\documentclass[journal, 12pt, onecolumn]{IEEEtran}
\usepackage{booktabs}

\ifCLASSINFOpdf
\else
\fi
\usepackage{xcolor}
\usepackage{booktabs}
\usepackage{array}
\usepackage{ragged2e}
\usepackage{mathtools}
\usepackage{amsmath,amssymb,amsthm}
\usepackage{tabularx}
\usepackage{graphicx}
\usepackage{float}
\usepackage{subcaption}
\usepackage{url}
\usepackage{xurl}
\usepackage{placeins}
\newcolumntype{Y}{>{\RaggedRight\arraybackslash}X}

\usepackage{algorithm}
\usepackage{algorithmic}

\usepackage{tikz}
\usetikzlibrary{arrows.meta,positioning,fit,calc,shapes.geometric,shapes.multipart}
\usetikzlibrary{backgrounds,decorations.pathreplacing}
\tikzset{
  arrow/.style={-Latex, line width=0.8pt},
  block/.style={draw, rounded corners=2pt, align=center, minimum height=6mm, inner sep=2pt, font=\small},
  op/.style={block, fill=gray!10},
  var/.style={block, fill=blue!5},
  gate/.style={block, fill=orange!12},
  delay/.style={block, fill=yellow!15},
  legendbox/.style={draw, rounded corners=2pt, inner sep=2pt, font=\scriptsize, fill=white},
  lbl/.style={font=\scriptsize, inner sep=1pt}
}

\makeatletter
\def\input@path{{Figures/}}
\makeatother

\usepackage{pgfplots}
\usepackage{pgfplotstable}
\pgfplotsset{compat=1.18}

\definecolor{colNOS}{RGB}{31,119,180}
\definecolor{coltGNN}{RGB}{255,127,14}
\definecolor{colGRU}{RGB}{44,160,44}
\definecolor{colRNN}{RGB}{214,39,40}
\definecolor{colMLP}{RGB}{148,103,189}

\usepackage{blindtext}

\usepackage{orcidlink}

\makeatletter
\g@addto@macro\normalsize{%
  \setlength{\abovedisplayskip}{1pt}%
  \setlength{\belowdisplayskip}{1pt}%
  \setlength{\abovedisplayshortskip}{0pt}%
  \setlength{\belowdisplayshortskip}{0pt}%
}
\makeatother

\begin{document}
%
\title{Large Language Models for Agentic NetOps and AIOps: Architectures, Evaluation, and Safety}
\newcommand\mdoubleplus{\mathbin{+\mkern-10mu+}}

\author{ Muhammad Bilal\,\orcidlink{0000-0003-4221-0877},~\IEEEmembership{~Senior Member, IEEE},
        Jon Crowcroft\,\orcidlink{0000-0002-7013-0121},~\IEEEmembership{~Fellow, IEEE},
        Ruizhi Wang,
        Xiaolong Xu,~\IEEEmembership{Senior Member,~IEEE,}
        and Schahram Dustdar,\orcidlink{0000-0001-6872-8821},~\IEEEmembership{~Fellow, IEEE}
\thanks{M. Bilal is with School of Computing and Communications Lancaster University, United Kingdom. }
\thanks{J. Crowcroft is with the Department of Computer Science and Technology, University of Cambridge, United Kingdom}
\thanks{R. Wang and X. Xu are with the School of Software, Nanjing University of Information Science and Technology, China}
\thanks{Schahram Dustdar is with the Distributed Systems Group of TU Wien, Austria and ICREA, Barcelona, Spain. }
\thanks{Corresponding author: Muhammad Bilal (e-mail: m.bilal@ieee.org)}
\thanks{Cite as: M. Bilal, J. Crowcroft, R. Wang, X. Xu, and S. Dustdar, “Large language models for agentic NetOps and AIOps: Architectures, evaluation, and safety,” Computer Science Review, vol. 63, Art. no. 101075, Feb. 2027, doi: 10.1016/j.cosrev.2026.101075.}}

\maketitle

\begin{abstract}
Large language models (LLMs) are increasingly being used in network operations (NetOps) and artificial intelligence for IT operations (AIOps) for tasks ranging from telemetry retrieval and incident diagnosis to configuration planning and bounded remediation. As these systems acquire greater access to operational tools, a central question is whether operational assurance develops commensurately with the authority granted to them. This survey examines that question through a structured, evidence-stratified review of agentic NetOps and AIOps, spanning directly evaluated LLM systems, operational mechanisms, transferred evidence, and normative guidance. We organise the field around autonomy, tool scope, evidence traces, assurance controls, evaluation, security, and governance, and introduce an operational assurance contract that links each autonomy level to permitted tools, required evidence, independent gates, execution budgets, rollout and rollback duties, and audit requirements. The synthesis indicates a capability–assurance gap: evidence is comparatively strong for read-oriented assistance and tool-grounded diagnosis, but becomes substantially less complete as systems approach configuration change, bounded execution, and closed-loop operation. We therefore develop a workflow-level evaluation framework covering evidence quality, tool use, policy and invariant compliance, staged execution, recovery, calibration, cost, and human intervention, together with a threat model for prompt-borne attacks, compromised operational evidence, excessive agency, privilege boundaries, and auditability. The resulting view treats agentic NetOps and AIOps as constrained operational control, in which increasing autonomy requires correspondingly stronger evidence, independently enforced safeguards, and recovery mechanisms.

\end{abstract}


\begin{IEEEkeywords}
Agentic AI; AIOps; Autonomous networks; Closed-loop control; Large language models; LLM agents; Network automation; Network operations; Operational assurance; Reliability; Tool-augmented agents; Root-cause analysis; Closed-loop control; Safety 
\end{IEEEkeywords}

\maketitle
\section{Introduction}
Modern networks and cloud services are operated through a dense layer of telemetry streams, tickets, runbooks, dashboards, and configuration artefacts. Over the last decade, this layer has grown faster than the teams that maintain it, largely because service graphs, deployment pipelines, and network policies now change continuously. In parallel, operators have accumulated powerful tools for reasoning about configuration and reachability (for example, offline analysis of routing and policy intent, and data-plane invariant checking) \cite{fogel2015batfish,Kazemian2012HSA,Khurshid2013VeriFlow}. What is new is not that we automate operations, but that large language models (LLMs) can sit across this entire surface and turn fragmented operational work into a single loop that reads, reasons, and acts.

This operational burden is being intensified by three related trends: the \emph{breadth} of the operational tool surface (more systems to query and more sources of truth), the \emph{velocity} of change (deployments, policy edits, and topology churn), and the \emph{volume} of interruption (alerts and incident handovers). In the field of SRE (Site Reliability Engineering), organizations aim to minimise detection and recovery times amidst increasing change frequency, making evidence collection and change security critical constraints \cite{Beyer2016SRE,Forsgren2018Accelerate}. The main advantage of LLMs lies in their ability to handle both procedural and linguistic tasks, both of which can be aided by tools. This perspective places agentic NetOps and AIOps within the context of autonomic computing. Kephart and Chess described autonomous systems as self-managing systems focused on configuration, remediation, and protection. LLM-based operational agents revisit this goal with modern tools, including natural-language interfaces, planning, and code generation. However, these tools also introduce new failure types, such as incorrect diagnoses, outdated evidence, unsafe tool use, prompt manipulation, and unrestrained execution \cite{Kephart2003Autonomic}.

LLMs transform the user interface by introducing natural language as the control layer for state queries and tool invocations. In early deployments, these models were primarily used for event aggregation, post-event analysis, or query building. As systems have evolved into agents that can plan multiple steps, call APIs or command-line tools, and change configurations directly, this shift has become more significant. Recent research demonstrates that by combining language inference with operational tools, language models can support event workflows, including monitoring and diagnostics \cite{Yu2024MonitorAssistant,xpert2024,Wang2024NetAssistant,chen2024autosys}. This survey focuses on this shift and its impact on architecture, assessment, and security. However, LLMs should not be considered the only way to achieve intelligent operations. In various NetOps and AIOps scenarios, causal reasoning and dependency graph methods have provided interpretable and less computationally intensive fault location and hypothesis generation mechanisms; therefore, a core design question is: how should LLM agents complement rather than mask or replace these causal mechanisms?

Consider an on-call incident where a latency SLO breaches for a microservice after a routine deployment. An agent must (i) retrieve recent changes and ownership context, (ii) issue a small number of discriminating log/metric/trace queries, (iii) propose a mitigation such as a feature-flag rollback or traffic shift, and (iv) route any write action through policy checks, invariant validation, and canary rollout with rollback triggers. The same example later illustrates the tool loop, the verification wall, the evaluation ladder, and the threat model.

The literature relevant to agentic operations is scattered across networking, systems, software engineering, and security. Networking venues emphasise correctness and invariants under policy and topology dynamics \cite{fogel2015batfish,Kazemian2012HSA,Khurshid2013VeriFlow,Reitblatt2012Abstractions}. Systems and SRE practice emphasise incident response latency, operational load, and the brittleness that comes from complex dependencies \cite{Beyer2016SRE}. Software engineering venues increasingly study how LLMs help operators express intent and retrieve the right operational evidence \cite{Yu2024MonitorAssistant,xpert2024}. Meanwhile, foundation-model research has produced patterns for tool use and iterative reasoning that are directly relevant to operations, but are rarely discussed with operational risk in mind \cite{Lewis2020RAG,schick2023toolformer,yao2023react}. The survey is timely because, although researchers are exploring agentic workflows, the field lacks shared terms for autonomy, a stable set of evaluation tasks, and a clear threat model that matches real operational tool surfaces.

\paragraph{Review questions and evidence discipline}
The review asks an overarching question: \emph{does operational assurance increase commensurately with the authority granted to an agent?} Three questions operationalise it: (RQ1) which operational tasks and tool interfaces are currently supported by LLM-based NetOps and AIOps systems; (RQ2) what evidence and independently enforced controls support claims at different autonomy levels; and (RQ3) how should workflow-level evaluation expose diagnosis quality, action safety, recovery, cost, and human intervention? Throughout the article, we distinguish four corpus statuses: \emph{direct LLM-facing evidence} from evaluated operational components or agents; \emph{domain-operational technical evidence} from NetOps, AIOps, cloud, SRE, and security-operations mechanisms; \emph{cross-domain transferred evidence} from agent, verification, safety, and human--automation research outside those operational domains; and \emph{normative infrastructure} from standards and official guidance. Author synthesis, including the contract schema and proposed metrics, is labelled separately and is not counted as an evidential status. This discipline prevents a transferred design principle from being reported as though it had already been validated in a write-capable LLM deployment.

\subsection{Why agentic operations now}

The structure of a running task is just as important as the functionality of the underlying model. In most cases, tasks involve a chain of queries and actions, often with missing details. An operator may need to check the blast radius, look for evidence in logs or counters, suggest ways to fix issues, make changes, and then watch for any problems that return. These steps follow a clear procedure. Much of the important context appears in work orders, manuals, commit messages, or handover notes. LLMs (Large Language Models) fit well here because they can work with both the process and the language found in these records.

\paragraph{(1) Running tasks require extensive retrieval}
Much of the work in these settings comes down to searching for evidence and correlating information from several tools.

Retrieval-enhanced designs address this need by joining model reasoning with dependable access to documents and system data \cite{Lewis2020RAG}.
In practice, the information we need is often present, but it can be hard to find. This is usually because of fragmented tools, changes over time, inconsistent names, or scattered records from past events.

\paragraph{(2) Tool invocation translates language into an executable plan}

Modern LLM agents can select tools, maintain state across steps, and modify the plan as new evidence emerges \cite{schick2023toolformer,yao2023react}.
This approach is similar to the workflow of an experienced operator, but agents can repeat the process more quickly on different interfaces. In incident management, this helps in developing systems capable of recommending investigative queries, summarising monitoring signals, and identifying potential root causes \cite{Yu2024MonitorAssistant,xpert2024,chen2024autosys}.

\paragraph{(3) Networks and clouds already have a correctness tradition}
NetOps has mature ways to check whether a change violates invariants, breaks reachability, or conflicts with policy intent \cite{Kazemian2012HSA,Khurshid2013VeriFlow,fogel2015batfish}.
This tradition introduces a beneficial constraint on agent-based systems. An agent may propose action plans, but the entire system must check critical properties before any action is taken and keep changes within safe limits. As network operations have shown, finding new configurations addresses only one aspect of the problem. The real challenge lies in deploying these changes without causing temporary violations or disruptions \cite{Reitblatt2012Abstractions}.
However, these advantages also introduce new types of failures.
Agent-based systems integrate interpretation and execution into a single operational flow. Misunderstandings of dashboards, unclear operating manuals, or imprecise objectives can all lead to changes that impact system availability. Once languages become the programming means for operating tools, the previous assumption that languages were merely harmless interfaces no longer holds true.
\paragraph{Practical question: measurable operational objectives}
Alongside the authority--assurance question, this article examines the following practical design question:
\emph{How should we design, evaluate, and manage NetOps and AIOps agents to maintain or even maximise reliability and security while reducing operational burden?}
From an operational perspective, the goal is to improve outcomes such as diagnostic and mitigation times while meeting explicit security constraints. These constraints include policy compliance, limiting the scope of impact, and ensuring rollback readiness \cite{Beyer2016SRE,Forsgren2018Accelerate}.
For this reason, the later discussion on evaluation focuses on workflow performance as the primary subject, rather than on static question answering.

\subsection{Scope and definitions}

We use \textbf{NetOps} to mean operational workflows specific to communication networks, including intent, configuration, troubleshooting, traffic engineering, and safe rollout \cite{leivadeas2023survey,Falkner2022IBNEnterprise,ElHassany2018NetComplete}.
We use \textbf{AIOps} to mean AI-supported IT operations, including log and metric anomaly detection, incident triage, diagnosis, and remediation \cite{notaro2021mgt}.
We use \textbf{agentic} to describe a system that maintains state across steps, selects tools or actions, and executes multi-step plans, possibly with human approval at selected stages \cite{zhou2024web,li2023apibank,patil2024gorilla}.

\paragraph{Terminology and evidential boundary}
The scope is represented on two orthogonal axes. The first axis contains six analytical pillars: \emph{operational capability and autonomy}, \emph{architecture and tool grounding}, \emph{assurance and operational control}, \emph{evaluation and benchmarking}, \emph{security, privacy, and adversarial robustness}, and \emph{human factors, governance, and standards}. These pillars define which part of an LLM-enabled NetOps or AIOps system a record informs. Each record receives one primary pillar for reproducible distribution counts and may receive secondary pillar tags where its contribution spans concerns.

The second axis records evidential status. \emph{Direct LLM-facing evidence} evaluates an LLM-based component or agent on a NetOps or AIOps task. \emph{Domain-operational technical evidence} evaluates a relevant mechanism inside networking, cloud operations, AIOps, SRE, or security operations without necessarily using an LLM. \emph{Cross-domain transferred evidence} supplies a mechanism from general agent systems, machine learning, software engineering, security, or human--automation research. \emph{Normative infrastructure} defines interfaces, duties, or controls through a standard or official framework. The axes are intentionally independent. A classical network verifier can justify an assurance gate without demonstrating that an LLM agent has used that gate successfully, while an LLM benchmark can demonstrate task performance without establishing deployment safety.

We use \emph{LLM-based operational agent} as the umbrella term. \emph{NetOps agent} and \emph{AIOps agent} identify the operational domain; \emph{operational agent} is used only when a claim applies to both. An \emph{agentic NetOps/AIOps system} is the surrounding system that combines the agent with evidence sources, tools, gates, rollout, rollback, and audit mechanisms. An \emph{autonomy rung} describes the authority granted to that system. The \emph{operational assurance contract}, shortened to \emph{assurance contract} where unambiguous, is the review's rung-specific schema for permitted tools, required evidence, gates, budgets, rollout, rollback, and audit. An \emph{operational contract} is a deployment-specific instance of that schema. An \emph{agent card} is a reporting artefact that can document these obligations, but does not itself enforce them.

\paragraph{Autonomy ladder as an operational object}
We model autonomy as a ladder of rungs. Each rung is characterised by which tools the agent may use and which gates are mandatory:
\begin{equation}
\mathcal{A}_k = \big(\mathcal{T}^{\mathrm{read}}_k,\; \mathcal{T}^{\mathrm{write}}_k,\; \mathcal{G}_k \big),
\label{eq:autonomy_rung}
\end{equation}
where $\mathcal{T}^{\mathrm{read}}_k$ are read-only tools (queries and retrieval), $\mathcal{T}^{\mathrm{write}}_k$ are write-capable tools, and $\mathcal{G}_k$ are non-bypassable gates (policy checks, invariants, approvals, canary and rollback rules) \cite{Khurshid2013VeriFlow,Reitblatt2012Abstractions,Li2019Jinjing}. For clarity, write authority is separated in prose into proposal and execution capabilities, so a system may draft a diff or change request without being allowed to apply it. The formal tool partition is introduced once, with the assurance contract, in Section~\ref{sec:agentic-patterns}. Later sections use $\mathcal{A}_k$ to state what is being evaluated and what safety contract must hold.

\paragraph{Evidence traces for audit and scoring}
The agent's interaction with tools is treated as a first-class output. To avoid repeating the same definition, the evidence trace is formalised once in Section~\ref{sec:llm-foundations}, where it is used for budget, provenance, and scoring analysis. This makes the concept of ``tool grounding'' more concrete: the evaluation can determine whether the tracing results are discriminative, whether the cost is limited, and whether it is sufficient to justify the proposed operation \cite{zhou2024web,li2023apibank,patil2024gorilla}.

Therefore, a practical approach to interpreting the term "agentic" in this survey is to view it as a ladder of autonomy. At the bottom is a read-only assistant responsible for aggregating and retrieving information. Mid-level systems can recommend queries, hypotheses, and manual procedures. Systems with higher autonomy can develop mitigation measures, generate configuration diffs, and run security checks. At the top is a closed-loop controller that continuously submits changes and verifies results. \cite{xpert2024,chen2024autosys,Li2019Jinjing}. Fig.~\ref{fig:taxonomy} outlines the main roles, the tools each role uses, and the safeguards required at different levels.

We focus on the \emph{operations tool interface}: monitoring backends, log storage, tracing systems, configuration repositories, CI/CD pipelines, network controllers, and ticketing systems. This boundary matters because security and governance depend on the tools an agent can invoke, the permissions attached to those tools, and the checks enforced before action \cite{notaro2021mgt,xpert2024,chen2024autosys}.

\subsection{Survey Method and Corpus Construction}
\label{sec:survey-method}

We conducted a structured integrative review to make the survey scope, evidence roles, and analytical procedure explicit. It is an explanatory synthesis across several research traditions. The aim was not to count every paper mentioning LLMs, NetOps, or AIOps, but to identify work explaining how agentic operational systems observe evidence, propose actions, pass checks, execute changes, and remain auditable near network and cloud control surfaces. This sets a functional boundary: we include work that changes how operational agents are designed, evaluated, constrained, or governed, rather than work that only applies LLMs to operational text.

The historical search log records searches across 16 source families: Scopus, Web of Science Core Collection, ACM Digital Library, IEEE Xplore, Elsevier ScienceDirect, Springer Nature Link, Wiley Online Library, USENIX proceedings, ACL Anthology, arXiv, OpenReview/ICLR proceedings, MLSys proceedings, Google Scholar, IETF/RFC Editor sources, NIST publications, and official project or specification pages. The search covered work published from January 2020 through 28 August 2026, since most LLM-agent and tool-use work appears in this period. The principal search was completed through 31 May 2026, after which a targeted literature-status, gap, and forward-search update extended the final boundary to 28 August 2026. The supplementary update log records the refreshed query families, eligible additions, and publication-status checks. Earlier work was retained where it supplied a necessary foundation for the survey, including network verification, configuration analysis, safe network updates, distributed tracing, autonomic computing, SRE practice, and classical accounts of automation risk \cite{Kazemian2012HSA,Khurshid2013VeriFlow,fogel2015batfish,Reitblatt2012Abstractions,Beyer2016SRE,Kephart2003Autonomic,Bainbridge1983Ironies,Parasuraman1997HumansAutomation}. This earlier material is needed because agentic operations inherits many of its safety constraints from older systems work, even when the user interface is now a language model.

Searches were organised into six strands. The NetOps strand covered intent-based networking, configuration synthesis, network verification, safe updates, reachability, isolation, SDN, controllers, and telemetry interfaces. The AIOps and SRE strand covered incident management, observability, anomaly detection, root-cause analysis, triage, remediation, rollback, and post-incident learning. The LLM-agent strand covered retrieval, tool use, structured outputs, planning, API calling, and multi-step operational workflows. The evaluation strand covered tool-use benchmarks, IT automation tasks, log and trace datasets, sandbox evaluation, trace replay, and workflow-level scoring. The safety and security strand covered prompt injection, indirect prompt injection, retrieval poisoning, telemetry manipulation, excessive agency, privacy, audit, and least privilege. The standards and practice strand covered OpenTelemetry, gNMI, NETCONF, YANG, intent-based networking definitions, telemetry frameworks, policy-as-code, and change governance. Search expressions were adapted to the syntax of each source, since academic databases, standards repositories, and Google Scholar do not expose the same query model.

\paragraph{Search-query families and reproducibility boundary}
The historical search was not implemented as one fixed Boolean string applied unchanged to every source. Source-specific and targeted expressions were used across heterogeneous databases, publisher platforms, repositories, standards sources, and citation-search routes. To make the central LLM-era operational search logic repeatable, Supplementary Search Record and Reproducibility Strategy S1 provides two common concept blocks and six \emph{consolidated reproducibility queries} aligned with the final analytical pillars: Q1 operational capability and autonomy, Q2 architecture and tool grounding, Q3 assurance and operational control, Q4 evaluation and benchmarking, Q5 security, privacy, and adversarial robustness, and Q6 human factors, governance, and standards. These consolidated queries summarise the recurring concepts used across the review; they are not claimed to be verbatim strings executed unchanged on every source and are not used to reconstruct the historical hit counts. The supplement separately describes the additional domain-operational, cross-domain, foundational, normative, citation-chaining, and venue-specific discovery lanes required to recover the broader 185-record evidence base, and it preserves the exact source-specific routes used in the final 28 August update.

Candidate records were screened in two stages. First, titles, abstracts, venues, and keywords were checked for relevance to at least one search strand. Second, the full text was checked against the survey question. A record was included if it met one or more of the following conditions: it studied a NetOps, AIOps, SRE, cloud operations, or security operations task; it proposed or evaluated a tool-using or retrieval-grounded LLM system for operational work; it defined a benchmark, dataset, or metric relevant to operational agents; it provided a verification, rollout, rollback, or governance mechanism that can serve as a gate for agentic systems; or it supplied a named mechanism used in a specific later claim about operational autonomy or assurance. Generic relevance to computing, networking, or AI was not sufficient. Records were excluded when they addressed generic chatbots, general prompting, or broad machine-learning methods without a clear operational task, tool surface, evaluation setting, or safety implication. Duplicate records and superseded versions were intended to be removed; the revision audit identified and removed one exact duplicate that remained in the earlier reference set. Where both a preprint and a peer-reviewed version existed, the peer-reviewed version was preferred unless the preprint contained later system or benchmark details needed for this survey.

\paragraph{Screening and coding responsibility}
The review was conducted within the five-author team. Screening and coding were not independently duplicated, and the project did not preserve a per-record assignment log that would support a reliable count of independent screeners or coders. Borderline eligibility and coding decisions were discussed within the author team until agreement. No inter-rater coefficient was calculated, and no review protocol was prospectively registered.

The historical PRISMA search record identifies 4,886 records across the 16 source families. Deduplication removed 1,972 records, leaving 2,914 unique records. Title and abstract screening excluded 2,319 records, so 595 advanced for further assessment. Ten reports could not be retrieved with sufficient full-text detail, leaving 585 full texts assessed for eligibility. Full-text assessment excluded 394 records. The resulting bibliography contains 191 records, of which 185 form the claim-linked evidence base used for quantitative corpus analysis and six are contextual comparison surveys outside that denominator. Figure~\ref{fig:prisma-flow} reports the identification, duplicate-removal, screening, eligibility, and inclusion stages.


\begin{figure}[t]
\centering

\begin{tikzpicture}[
  font=\small,
  >=Latex,
  flowbox/.style={
    draw=#1!70!black,
    fill=#1!20,
    rounded corners=4pt,
    align=center,
    minimum height=8mm,
    text width=7.1cm,
    inner sep=5pt,
    line width=0.8pt,
    font=\scriptsize
  },
  splitbox/.style={
    draw=#1!70!black,
    fill=#1!20,
    rounded corners=4pt,
    align=center,
    minimum height=9mm,
    text width=3.15cm,
    inner sep=5pt,
    line width=0.75pt,
    font=\scriptsize
  },
  auditbox/.style={
    draw=#1!70!black,
    fill=#1!20,
    rounded corners=4pt,
    align=center,
    minimum height=9mm,
    text width=7.1cm,
    inner sep=5pt,
    line width=0.8pt,
    font=\scriptsize
  },
  flowarrow/.style={
    -Latex,
    line width=0.75pt,
    draw=black!60
  }
]

\definecolor{obsC}{RGB}{170,185,230}
\definecolor{planC}{RGB}{185,178,228}
\definecolor{verifC}{RGB}{240,208,168}
\definecolor{deployC}{RGB}{238,228,165}
\definecolor{learnC}{RGB}{178,222,188}

\node[flowbox=obsC] (id)
  {\textbf{Identification}\\
   4,886 records across 16 source families};

\node[flowbox=obsC, below=4.0mm of id] (dedup)
  {\textbf{Deduplication}\\
   1,972 removed, 2,914 unique records remained};

\node[flowbox=planC, below=4.0mm of dedup] (screen)
  {\textbf{Title and abstract screening}\\
   2,319 excluded, 595 records advanced};

\node[flowbox=planC, below=4.0mm of screen] (full)
  {\textbf{Full-text retrieval and assessment}\\
   10 not retrieved with sufficient detail, 585 assessed};

\node[flowbox=verifC, below=4.0mm of full] (elig)
  {\textbf{Full-text eligibility}\\
   394 excluded, 191 bibliography records retained};

\node[
  splitbox=learnC,
  below left=6mm and -0.15cm of elig
] (evid)
  {\textbf{Claim-linked evidence}\\
   185 records};

\node[
  splitbox=deployC,
  below right=6mm and -0.15cm of elig
] (context)
  {\textbf{Contextual surveys}\\
   6 records outside denominator};

\draw[flowarrow] (id) -- (dedup);
\draw[flowarrow] (dedup) -- (screen);
\draw[flowarrow] (screen) -- (full);
\draw[flowarrow] (full) -- (elig);

\draw[flowarrow]
  (elig.south) -- ++(0,-2mm) -| (evid.north);

\draw[flowarrow]
  (elig.south) -- ++(0,-2mm) -| (context.north);

\end{tikzpicture}

\caption{PRISMA-style search and inclusion flow. The lower box shows
the later corpus reconciliation separately from the historical
identification and screening stages.}
\label{fig:prisma-flow}
\end{figure}

The revision-stage reconciliation is distinct from the full search flow. The June 2026 version used a coded corpus of 181 records. A later claim-source audit removed 14 records, including one exact duplicate, leaving an audited base of 167. The final targeted update through 28 August 2026 added 18 eligible records, producing the final 185-record claim-linked evidence base. At the cut-off, these 185 records comprise 162 peer-reviewed conference or journal papers, 10 preprints, and 13 standards, specifications, books, technical reports, or official guidance documents. The public dataset supplies the 185-row evidence table, the 14 exclusions, the 18 update records, the operational-system subset, derived distributions, a data dictionary, and the update log \cite{Bilal2026LLMNetOpsDataset}.

Peer-reviewed work is used as the main evidence base. Grey literature, standards, and project specifications are used more narrowly, where they define an operational interface, control surface, benchmark, or governance requirement that is difficult to recover from academic papers alone. This treatment is important for agentic NetOps and AIOps because several relevant control surfaces are specified outside ordinary research articles, including telemetry standards, network-management interfaces, model-context protocols, and risk-management guidance \cite{OpenTelemetrySpec,gNMISpec,RFC9232Telemetry,RFC9315Intent,MCP2025Spec,nist2023airmf,autio2024genai}.

\paragraph{Evidence roles and permissible claims}
Each retained source is assigned one of four evidential statuses. \emph{Direct LLM-facing} records evaluate an LLM-based operational component or agent on a NetOps or AIOps task. \emph{Domain-operational technical} records evaluate a mechanism within networking, cloud operations, AIOps, SRE, or security operations. \emph{Cross-domain transferred} records evaluate a relevant mechanism outside those operational domains. \emph{Normative} records specify an interface, governance expectation, or security practice. A direct record can support a task-specific empirical claim. A domain-operational record can support feasibility or an operational mechanism within its evaluated setting. A cross-domain record can motivate a transfer hypothesis or design requirement but cannot demonstrate that an LLM NetOps or AIOps agent satisfies it. A normative record can define an obligation but cannot establish effectiveness. Proposed contract clauses, metric templates, and research priorities are identified as author synthesis unless a passage explicitly reports evaluated practice.

For source-type attribution, the direct LLM-facing evidence is further divided by the object evaluated: \emph{operational systems} exercise an agent within a NetOps or AIOps workflow; \emph{enabling frameworks} contribute an adaptable model, architecture, or orchestration method; \emph{benchmarks} define tasks, interfaces, scenarios, or scoring oracles; and \emph{safety evaluations} probe attacks, failure modes, or controls. Operational systems establish observed workflow behaviour, frameworks establish architectural feasibility, benchmarks support comparative measurement, and safety evaluations establish bounded risk or control findings. None of these objects alone demonstrates broad closed-loop autonomy.

Each included record was coded along eight dimensions: domain, operational task, autonomy role, tool surface, evidence role, evaluation setting, safety control, and contribution type. The domain codes were NetOps, AIOps, SRE, cloud operations, security operations, and general agent infrastructure. The task codes included triage, telemetry query generation, incident summarisation, root-cause analysis, configuration synthesis, change planning, remediation, rollback, and post-incident learning. The autonomy codes followed the ladder used in this survey: read-only copilot, tool-grounded analyst, write-limited planner, and closed-loop controller. Tool surfaces were coded as logs, metrics, traces, tickets, runbooks, configuration repositories, topology stores, controllers, CI/CD systems, policy engines, validators, simulators, and action APIs. Evaluation settings were coded as static question answering, offline classification, trace replay, sandbox execution, live testbed, human study, production report, or benchmark suite. Safety controls were coded as least privilege, provenance tracking, freshness checks, schema validation, approval gates, policy checks, invariant verification, canary rollout, rollback readiness, audit logging, and red-team testing. Each record was also assigned one primary analytical pillar for non-overlapping corpus distributions and any applicable secondary pillar tags for cross-cutting interpretation. Autonomy, write capability, evaluation setting, and safety-control distributions are calculated only for direct operational systems. Frameworks, benchmarks, safety evaluations, domain-operational mechanisms, cross-domain foundations, and normative sources are marked not applicable rather than forced into operational-agent categories.

This coding scheme supports the contract-centred taxonomy used later in the paper. For each work, we record what an agent may observe, which tools it may call, what evidence it must collect, which gates it must pass, how changes are rolled out or undone, and what budget or stopping rule limits the workflow. Model family and AIOps task name are useful labels, but they do not show whether a system is safe near an operational write boundary. For this survey, a read-only RCA assistant and a rollback-capable controller are different operational objects, even if both use an LLM.

\subsection{Corpus characteristics and evidence distribution}

The final claim-linked evidence base contains 185 records. By publication form of the screened artefact, 162 are peer-reviewed conference or journal papers (87.6\%), 10 are preprints (5.4\%), and 13 are standards, specifications, books, technical reports, or official guidance documents (7.0\%). The 162 peer-reviewed records comprise 130 conference papers and 32 journal articles, using the publication types recorded in the accompanying machine-readable bibliography. By evidential status, 28 records are direct LLM-facing evidence (15.1\%), 95 are domain-operational technical evidence (51.4\%), 53 are cross-domain transferred evidence (28.6\%), and 9 are normative infrastructure (4.9\%). Publication form and evidential status are orthogonal. A peer-reviewed network verifier, for example, remains domain-operational evidence rather than direct evidence of an LLM agent. Every retained row records its first manuscript location, claim context, inclusion rationale, and permissible claim boundary.

\paragraph{Year and primary-domain distributions}
Within the contemporary search window, the year counts are 7 records from 2020, 6 from 2021, 8 from 2022, 23 from 2023, 44 from 2024, 38 from 2025, and 5 from 2026 through the 28 August cut-off. The 54 earlier records are retained foundations from 1975--2019 and are not counted as direct contemporary LLM-agent studies. Using one primary-domain assignment per retained record, the distribution is NetOps 51, AIOps 28, cloud operations 14, SRE or software operations 15, security operations or agent security 19, general agent infrastructure 33, and transferred general infrastructure or general ML evaluation 25. These domain labels describe the role of each record in the synthesis and do not imply that every record evaluates an operational LLM system.

The six primary analytical pillars describe the review object more accurately than a division between a small LLM core and a large adjacent periphery. Operational capability and autonomy contains 48 records (25.9\%). Architecture and tool grounding contains 21 (11.4\%). Assurance and operational control contains 39 (21.1\%). Evaluation and benchmarking contains 31 (16.8\%). Security, privacy, and adversarial robustness contains 27 (14.6\%). Human factors, governance, and standards contains 19 (10.3\%). Percentages use one non-overlapping primary assignment per record and may not sum to exactly 100\% because of rounding. Secondary pillar tags remain available in the supplement for records that contribute across several concerns.

The direct LLM-facing subset contains 28 records with distinct evidentiary objects: 13 operational systems, 4 enabling frameworks or models, 8 benchmarks or task-evaluation studies, and 3 safety or assurance evaluations. Autonomy and evaluation-setting distributions are derived only for the 13 operational systems: one read-only copilot, seven tool-grounded analysts, and five write-limited planners or bounded executors. No retained system demonstrates broad, independently assured closed-loop control. The direct records are also unevenly distributed across the pillars, with 12 in operational capability, 4 in architecture, 5 in assurance, 6 in evaluation, 1 in security, and none whose primary contribution is a longitudinal human-factors evaluation. These small and heterogeneous counts do not justify evidence-base-wide prevalence claims. They support the more cautious conclusion that our qualitative synthesis indicates a capability--assurance gap as authority approaches operational write access.

\paragraph{Comparable operational-system profile}
For a limited descriptive cross-tabulation, the 13 operational systems were grouped by the evaluation setting reported by the source. Seven use production reports, production incidents, deployment reports, or operational case evidence. Five use controlled benchmark, testbed, sandbox, experimental, or fault-injection settings. One is an early live operational prototype. The single R1 system is production-connected. Among the seven R2 systems, five are production-connected, one is evaluated in a controlled setting, and one is the early live prototype. Among the five R3 systems, one uses an operational case and four use controlled settings. Eight systems remain read-side. All five R3 systems demonstrate write-limited proposal or bounded execution, and four exercise bounded execution in an operational case, testbed, sandbox, or experimental environment. None evaluates a complete chain of proposal, independent verification, execution, and end-to-end rollback, and none demonstrates an independently assessed rollback mechanism. Reported controls are source-specific and overlapping, including read-only or scoped tools, schema or DSL validation, container or testbed isolation, guide or SOP bounding, and explicit stopping logic. We therefore report raw counts for the comparable authority and setting fields, but do not estimate corpus-wide safety-control prevalence from incompatible evidence objects.

The 28 August 2026 update adds 18 eligible records to the earlier 167-row audit and integrates them into the final denominator. The update contributes recent NetOps monitoring and intent systems, AIOps diagnosis and remediation agents, domain benchmarks, calibrated-confidence work, an operational-trust pipeline, and a cross-domain agent-security evaluation \cite{Yoon2025OFCnetLLM,Xiao2026YANGAgent,Yu2024MonitorAssistant,Ahmed2023CloudIncidents,Wang2023NetworkMeetsChatGPT,Cao2024LinuxAgents,Roy2024RCAAgents,Sarda2024AutoRemediation,Zhang2024ICLRCA,Goel2024XLifecycle,Guo2024OWL,Chen2025AIOpsLab,Xu2025OpenRCA,Pei2025FlowOfAction,Jacobs2025TrustIBN,Zhang2024LMPACE,Wang2025TeleEvalOS,Gasmi2026AgentSecurity}. The complete 185-row evidence audit, 14-row exclusion audit, 28-record direct-evidence classification, 13-system operational coding, formula-derived pillar and evidential-status distributions, six contextual-review records, and final update log are deposited in the associated version-specific Mendeley Data record \cite{Bilal2026LLMNetOpsDataset}.

Backward and forward snowballing were then applied to the screened corpus. Backward snowballing identified foundational work on network correctness, tracing, incident practice, and benchmark construction. Forward snowballing identified newer work on LLM agents, operational assistants, tool-use evaluation, and prompt-injection risks. Snowballing stopped when an additional pass did not add a new operational task, tool surface, evaluation type, or safety control. This ties the review to saturation of the coding scheme rather than to an open-ended citation search.

The method has two limits. Some operational agent systems are deployed inside companies and are described only through engineering reports, product documents, or not at all. In addition, the LLM-agent literature is changing quickly, especially in benchmark design and agent security. We therefore separate stable claims from current observations. Stable claims are grounded in older work on verification, SRE, tracing, least privilege, rollback, and audit. Current observations describe the present state of LLM agents for NetOps and AIOps, where the evidence base is still thinner and less even. This distinction matters because operational reliability is a slow discipline. The model interface may change quickly, but the need for evidence, gates, rollback, and accountability does not.

\subsection{Positioning Against Prior Surveys}
\label{sec:prior-survey-positioning}

The survey is related to several mature bodies of review work, but it is not a replacement for any one of them. Existing surveys mainly organise the literature around AIOps tasks, failure-management pipelines, network automation, LLM-agent architectures, retrieval methods, or tool-use benchmarks. Those views are useful. They do not, however, answer the central question of this survey: what must be true before an LLM-based operational agent is allowed to move from evidence gathering to proposed or executed change?

Table~\ref{tab:prior-survey-comparison} summarises the distinction. Earlier AIOps surveys define the operational problem space well. Notaro et al. survey AIOps methods for failure management and organise work around failure-related tasks such as detection, diagnosis, and remediation \cite{notaro2021mgt}. Cheng et al. review AIOps on cloud platforms and give a useful account of incident detection, failure prediction, root-cause analysis, and automated actions \cite{Cheng2023AIOpsCloud}. More recently, Zhang et al. provide a broad LLM4AIOps survey, covering data sources, AIOps tasks, LLM-based methods, and evaluation practice in the LLM era \cite{Zhang2025LLM4AIOpsCSUR}. These studies are closest to the AIOps side of this paper. Their main unit of analysis is the AIOps task or the LLM method. Our unit of analysis is the operational contract: which evidence is required, which tools may be invoked, which gates must pass, and which actions may be proposed or executed.

Network automation surveys form the second related group. Leivadeas and Falkner survey intent-based networking, including intent expression, translation, assurance, and closed-loop network management \cite{leivadeas2023survey}. Xu and Zhou survey systems approaches to configuration errors, including detection, diagnosis, testing, and prevention \cite{XuZhou2015ConfigErrorsSurvey}. These works are essential for the NetOps side of the paper because they show that network operations already has a strong correctness tradition. The missing link is the agentic one. Existing network-automation surveys do not study LLM-mediated tool use, prompt-borne risk, provenance-aware retrieval, or rollback-aware evaluation for operational agents.

A third group surveys LLM agents and tool use. Wang et al. and Xi et al. review LLM-based autonomous agents from broad architectural and application perspectives \cite{Wang2024LLMAgentsSurvey,Xi2025LLMAgentsSurvey}. Qin et al. survey tool learning with foundation models, while Li reviews prominent LLM-agent paradigms around tool use, planning, and feedback learning \cite{qin2023tool,Li2025AgentParadigms}. These surveys are valuable for agent construction. Yet they usually treat tools as capability extensions. In NetOps and AIOps, tools are also trust boundaries. A log query, a configuration diff, a controller API call, and a rollback command cannot be treated as equivalent actions simply because they are all tool calls.

A fourth group covers retrieval and evaluation. RAG surveys explain how external knowledge can improve factual grounding and freshness \cite{Gao2023RAGSurvey}. Agent benchmarks such as AgentBench, WebArena, API-Bank, OSWorld, SafeToolBench, and ITBench show how agent behaviour can be tested beyond single-turn question answering \cite{liu2023agent,zhou2024web,li2023apibank,Xie2024OSWorld,Xia2025SafeToolBench,Jha2025ITBench}. These works help frame the evaluation problem. Still, operational evaluation has extra structure. In an incident workflow, success is not only whether an answer is correct. It also includes whether the evidence trace is adequate, whether tool use stayed within budget, whether the proposal passed independent checks, and whether rollout and rollback behaviour were scored.

The position of this survey is therefore narrower than a general LLM-agent survey and broader than an LLM4AIOps task review. It treats agentic NetOps and AIOps as constrained operational control. The survey asks how an agent observes, proposes, verifies, executes, rolls back, and leaves an audit trail. This contract-centred view is the main difference from prior work.

\paragraph{Relation to adjacent assurance artefacts}
The assurance contract is not a replacement for an assurance case, safety case, access-control profile, policy specification, model card, or agent card. An assurance or safety case argues why a system is acceptably safe; a policy or access-control profile determines which operations are permitted; and a model or agent card reports characteristics and limitations. The contract schema instead supplies a compact, rung-indexed interface among these artefacts: it binds permitted tools to required evidence, independent gates, budgets, rollout and rollback duties, and an auditable trace. Its analytical consequence is the \emph{capability--assurance gap}: our qualitative synthesis indicates that support is strongest for read-side querying and diagnosis, while evidence becomes markedly less complete as claims approach autonomous execution, canary management, and recovery.

\begin{table*}[h]
\centering
\caption{Positioning of this survey against prior surveys and adjacent reviews}
\label{tab:prior-survey-comparison}
\small
\renewcommand{\arraystretch}{1.14}
\begin{tabular}{p{0.18\linewidth}p{0.22\linewidth}p{0.24\linewidth}p{0.28\linewidth}}
\toprule
\textbf{Prior survey line} & \textbf{Main focus} & \textbf{What it contributes} & \textbf{Remaining gap addressed here} \\
\midrule
AIOps failure-management surveys \cite{notaro2021mgt} &
Failure detection, diagnosis, and remediation &
Defines AIOps failure-management tasks and organises earlier ML-based methods &
Does not treat LLM agents as tool-mediated operational actors with explicit write boundaries \\
\midrule
Cloud AIOps reviews \cite{Cheng2023AIOpsCloud} &
AIOps on cloud platforms &
Covers incident detection, failure prediction, RCA, and automated actions in cloud settings &
Does not give a contract view of evidence, permissions, verification gates, rollout, and rollback \\
\midrule
LLM4AIOps surveys \cite{Zhang2025LLM4AIOpsCSUR} &
LLMs applied to AIOps &
Surveys LLM-based data sources, AIOps tasks, methods, and evaluation &
Centred on AIOps rather than joint NetOps and AIOps control surfaces. The write-side safety contract remains secondary \\
\midrule
Intent-based networking surveys \cite{leivadeas2023survey} &
Intent expression, translation, assurance, and closed-loop networking &
Provides the NetOps foundation for intent, policy, assurance, and automation &
Does not address LLM-mediated reasoning, tool calls, prompt-borne risk, or agent audit trails \\
\midrule
Configuration-error surveys \cite{XuZhou2015ConfigErrorsSurvey} &
Configuration errors in complex systems &
Explains detection, diagnosis, prevention, and operational causes of misconfiguration &
Pre-agentic. It does not cover LLM-generated diffs, tool-grounded planning, or gated execution \\
\midrule
LLM-agent and tool-use surveys \cite{Wang2024LLMAgentsSurvey,Xi2025LLMAgentsSurvey,qin2023tool,Li2025AgentParadigms} &
Agent architectures, tool use, planning, memory, and feedback &
Gives the general agentic design vocabulary needed for LLM-based agents &
Usually treats tools as capabilities. In operations, tools are also permissions, trust boundaries, and audit objects \\
\midrule
RAG surveys \cite{Gao2023RAGSurvey} &
Retrieval, generation, augmentation, and grounding &
Explains how external evidence can reduce staleness and improve factual grounding &
Does not model operational provenance, freshness, authority level, or write-side evidence obligations \\
\midrule
Agent benchmarks and safety benchmarks \cite{liu2023agent,zhou2024web,li2023apibank,Xie2024OSWorld,Xia2025SafeToolBench,Jha2025ITBench} &
Task completion, tool use, web interaction, desktop interaction, IT automation, and tool safety &
Shows how agent evaluation can move beyond static question answering &
Operational evaluation also needs trace quality, budgeted tool use, verification outcomes, canary behaviour, rollback, and audit scoring \\
\midrule
\textbf{This survey} &
\textbf{Agentic NetOps and AIOps as constrained operational control} &
\textbf{Organises the field around autonomy rungs, tool surfaces, evidence traces, gates, rollout, rollback, budgets, and audit} &
\textbf{Provides a contract-centred taxonomy for agents placed near network and cloud control surfaces} \\
\bottomrule
\end{tabular}
\end{table*}

\subsection{Contributions of this survey}
This paper makes the following contributions.

\begin{enumerate}
\item \textbf{Workflow-grounded framing} We frame agentic NetOps and AIOps as operational workflows that move from observation and diagnosis to decision-making and action. This shifts the unit of analysis from model architecture alone to the operational role an agent plays, the tools it may use, and the failure modes it may introduce.

\item \textbf{Contract-centred taxonomy} We develop a review-informed, theory-led integrative taxonomy for agentic NetOps and AIOps, organised around autonomy rungs, tool surfaces, evidence obligations, safety gates, rollout constraints, and audit requirements.

\item \textbf{Architectural patterns and guardrails} We extract recurring design patterns from recent systems, including retrieval and evidence management \cite{Lewis2020RAG}, tool-use loops \cite{schick2023toolformer,yao2023react}, and operational assistants for monitoring, query recommendation, incident analysis, and network troubleshooting \cite{Yu2024MonitorAssistant,xpert2024,Wang2024NetAssistant,chen2024autosys}. We map these patterns to guardrails such as verification, privilege minimisation, staging, and human approval.

\item \textbf{Evaluation practice and benchmark guidance} We synthesise evaluation requirements into proposed, benchmark-instantiable task and metric templates that reflect operational work, including event classification, diagnosis under partial observability, safe change proposal, rollout, and rollback. We also discuss how benchmarks can measure end-to-end utility rather than isolated language quality, drawing on emerging agent evaluation practice \cite{Jimenez2024SWEbench}.

\item \textbf{Security, governance, and compliance analysis} We analyse risks specific to agentic operations, including unsafe tool invocation, privilege boundaries, audit integrity, and the tension between autonomy and accountability. We show how existing network verification and update abstractions can be repurposed as enforceable safety constraints \cite{Kazemian2012HSA,Khurshid2013VeriFlow,fogel2015batfish,Reitblatt2012Abstractions}.
\end{enumerate}

\begin{figure*}[t]
\centering
\includegraphics[width=\textwidth]{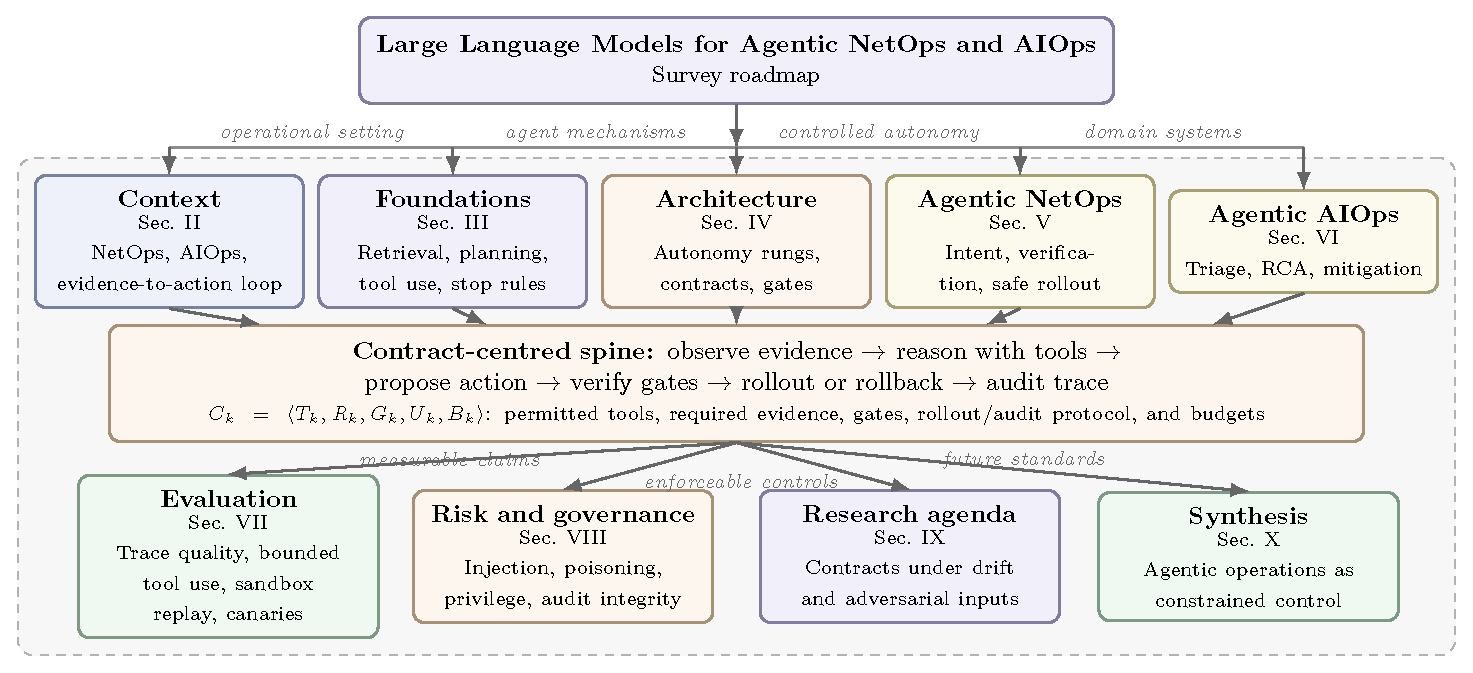}
\caption{Survey roadmap. Unlike the later control-loop, architecture, autonomy, and human-review figures, this figure shows only the article's argumentative progression from evidence and operational context to contracts, evaluation, governance, and synthesis}
\label{fig:survey-roadmap}
\end{figure*}

Fig.~\ref{fig:survey-roadmap} summarises the survey structure and its contract-centred progression from operational context to open problems.

\section{Operational background: NetOps and AIOps}
\label{sec:ops-background}

Operations is best understood as a control problem carried out under uncertainty and organisational constraint.
Most incidents are not solved by one brilliant query.
They are solved by repeatedly turning partial signals into safe, reversible actions while keeping the blast radius small.
That is why ``helpful text'' is only a small part of the story. Key challenges include the collection of evidence, the proposal and review of changes, and the documentation of decisions for subsequent audit and organizational learning. Early research on Internet operations demonstrated that even minor configuration errors or policy mistakes can result in unexpected and widespread failures \cite{Mahajan2002BGP}. As systems increase in scale and software updates occur more frequently, the likelihood of such issues escalates.

\subsection{A unified evidence-to-action control loop}
\label{subsec:unified-ops-loop}

Both NetOps and AIOps can be modelled as a partially observed control loop.
There is a latent system state \(x_t\) (network and service state), operators and automation observe it only through tool-mediated observations \(o_t\) (logs, metrics, traces, reachability checks), and they apply actions \(a_t\) (changes, mitigations) that alter future state \cite{fogel2015batfish,chen2024autosys,xpert2024,notaro2021mgt}.
A compact abstraction is:

\begin{equation}
x_{t+1} \sim P(x_{t+1}\mid x_t,a_t), \qquad o_t \sim O(o_t\mid x_t), \qquad a_t \in \mathcal{A}_k .
\label{eq:ops_pomdp}
\end{equation}

The key operational point is the rightmost term.
The agent does not act in an unconstrained action space.
It acts within an autonomy rung \(k\), which is defined by the tool permissions and the mandatory gates that must be satisfied before any write-side action is executed \cite{Khurshid2013VeriFlow,Reitblatt2012Abstractions}.
Following the introduction, we treat an autonomy rung as an operational object:
\begin{equation}
\mathcal{A}_k = \big(\mathcal{T}^{\text{read}}_k,\mathcal{T}^{\text{write}}_k,\mathcal{G}_k\big),
\label{eq:ops_ob}
\end{equation}
where \(\mathcal{T}^{\text{read}}_k\) are read-only tools (queries and retrieval), \(\mathcal{T}^{\text{write}}_k\) are write-capable tools, and \(\mathcal{G}_k\) are non-bypassable gates (policy checks, invariants, approvals, rollout constraints) \cite{Khurshid2013VeriFlow,Reitblatt2012Abstractions}.
This makes ``tools as a boundary of trust'' concrete: the trust boundary is precisely the interface where a proposal becomes an executed action, and where \(\mathcal{G}_k\) must hold.

A second shared primitive is mismatch between what is intended and what is actually realised \cite{leivadeas2023survey,fogel2015batfish}.
Let \(x_t^{\text{desired}}\) denote the declared desired state (intent, policy, change request), and let \(x_t^{\text{realised}}\) denote the realised operational state (what the system actually does, as evidenced by measurements and derived models) \cite{leivadeas2023survey,fogel2015batfish}.
We can express mismatch as:

\begin{equation}
\Delta_t = d\!\left(x_t^{\text{desired}},\, x_t^{\text{realised}}\right),
\label{eq:state_mismatch}
\end{equation}

where \(d(\cdot,\cdot)\) is a domain-appropriate distance (for example, a binary violation indicator for an invariant, a count of violated intent clauses, or a normalised diff magnitude for a configuration slice).
This is useful later because \(\Delta_t\) is an evaluation axis: sound operational practice reduces \(\Delta_t\) safely, or refuses to act when uncertainty is too high to reduce it without undue risk \cite{Khurshid2013VeriFlow,Reitblatt2012Abstractions}.

\begin{figure}[t]
\centering
\includegraphics[width=.8\linewidth]{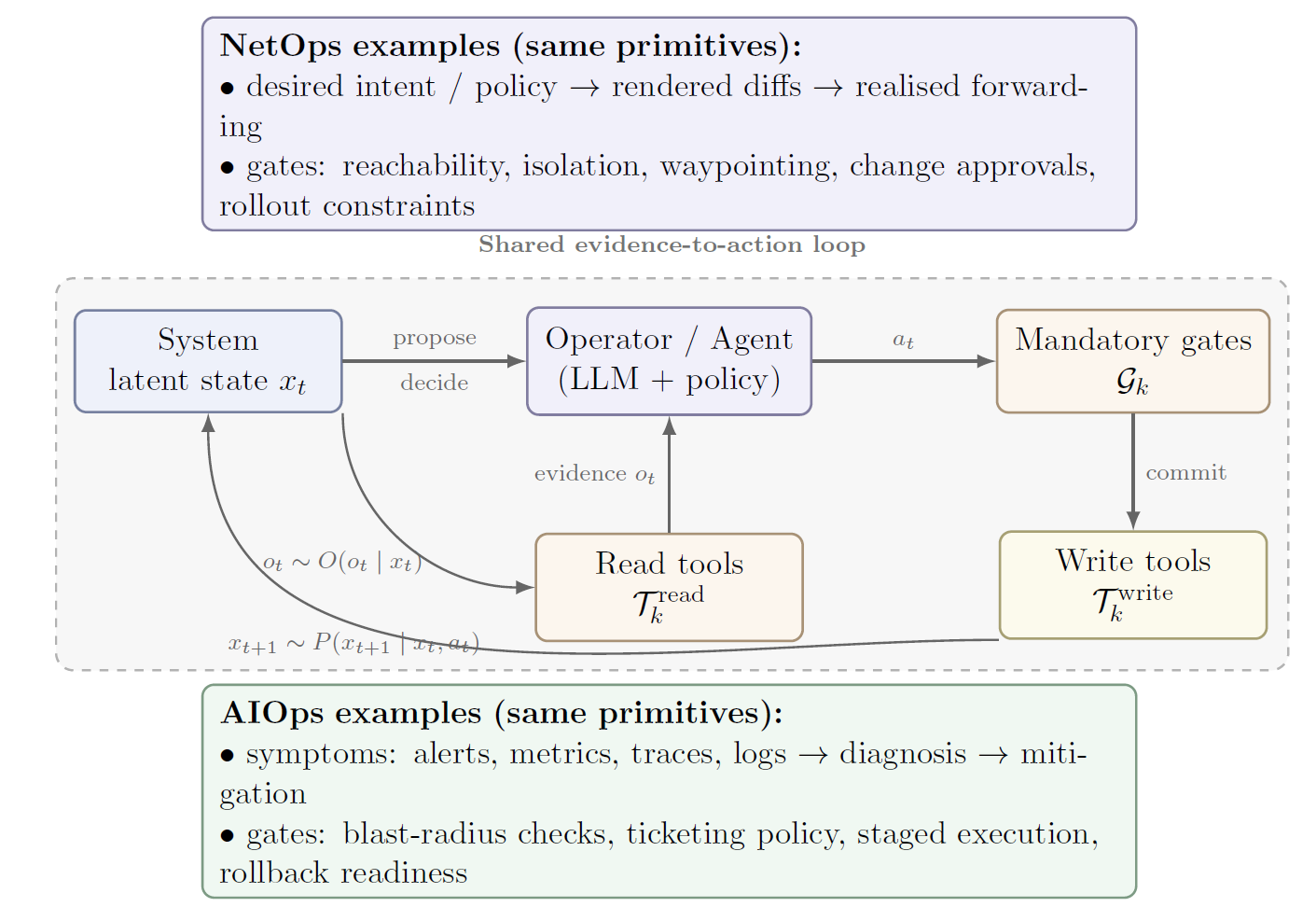}
\caption{Cross-domain analytical model. This figure aligns NetOps and AIOps as partially observed evidence-to-action loops: tools expose \(o_t\), rung \(\mathcal{A}_k\) bounds actions \(a_t\), and mandatory gates \(\mathcal{G}_k\) mediate write-side execution. It is distinct from the domain-specific workflow figures, which show task sequences}
\label{fig:aiops-workflow}
\end{figure}

\subsection{NetOps workflows and artefacts}

NetOps encompasses the operational lifecycle of a communication network, and its workflow typically consists of five stages: capturing intents, integrating device or controller changes, validating against invariants and operational policies, executing phased deployment, and conducting post-change monitoring with rollback as necessary. In programmable networks, the controller has a central role in this process. This change began with Software-Defined Networking (SDN), which separated the control plane from the data plane. This separation gave operators a clearer way to manage the control plane \cite{McKeown2008OpenFlow}. Network operations (NetOps) depend on artefacts at multiple levels: desired states, such as intents and policies; rendered states, including configurations and controller rules; and realised states, encompassing forwarding behaviour, counters, and alarms. Issues frequently occur when these layers are misaligned. In notation \eqref{eq:state_mismatch}, \(x_t^{\text{desired}}\) represents intents and policies, whereas \(x_t^{\text{realised}}\) denotes the resulting data-plane and control-plane behaviour. Intents are typically too abstract for simple verification, while vendor-specific configurations, although precise, are often difficult to interpret. Intent-based networking tackles this challenge through employing closed-loop transformation, execution, and monitoring, rather than relying on a single translation step \cite{leivadeas2023survey}.

\paragraph{Why configuration is operationally hard}
Configuration is not simply parameter setting. It is a distributed process across heterogeneous devices, with asynchronous convergence and emergent behaviour. Errors arise from local mistakes, policy conflicts, drift, or fragile change sequences. Empirical work on BGP configuration errors shows that even minor policy mistakes can propagate and cause system-wide effects, which is still a warning for automation \cite{Mahajan2002BGP}. Even when the final state is correct, a wrong update order may create unsafe intermediate states. This has motivated consistent update protocols and phased deployment designs for managing transient inconsistencies \cite{Reitblatt2011ConsistentUpdates}.

\paragraph{Causal diagnosis as a NetOps primitive}
Not all NetOps diagnoses are suited to language reasoning or pattern recognition. A complementary approach uses causal inference, dependency models, and event structures to explain observed symptoms \cite{Yan2012GRCA,Kobayashi2018MiningCausality,Kobayashi2019CausalNetworkLogs}. This approach is attractive in network operations because faults propagate through known relationships, including topology, protocol dependencies, control-plane interactions, service dependencies, and time-ordered event sequences \cite{Yan2012GRCA,Kobayashi2019CausalNetworkLogs}. Causal approaches can construct compact, interpretable fault graphs. When the underlying causal or dependency structure is known, the graph can reduce repeated resource-intensive inference over raw telemetry and operational manuals. For network operations, causal reasoning can therefore serve as a primary tool for $\mathcal{T}^{\mathrm{read}}_k$. An LLM can query or summarise the model, while the causal graph makes the assumptions and evidence relationships explicit. This structured division can improve practical efficiency \cite{Yan2012GRCA,Kobayashi2018MiningCausality,Kobayashi2019CausalNetworkLogs}.

\paragraph{Validation and verification in everyday NetOps}
In network operations, a change is normally checked before acceptance. Even a small configuration edit may alter reachability, isolation, waypointing, or policy. This need has produced tools that turn configurations into network models and check important properties before rollout. Batfish extracts the intended data plane from candidate configurations and supports pre-deployment what-if analysis \cite{fogel2015batfish}. Minesweeper broadens configuration verification by checking richer properties at scale while retaining operator-facing workflows \cite{Beckett2017Minesweeper}.
In dynamic, controller-managed networks, verification must run at the write boundary, not only offline. VeriFlow inserts a layer between the controller and devices to block unsafe rules before they reach the data plane \cite{Khurshid2013VeriFlow}. NoD adds a related lesson: operator assumptions about reachability, path selection, and service access must also be checked, since mismatches can expose policy gaps and security risks \cite{Lopes2015NoD}. NetOps governance defines who may review, approve, or apply changes through configuration records, CI checks, controller interfaces, and ticketing systems. For agentic systems, this layer sets the permitted action boundary.

\subsection{AIOps workflows and artefacts}

AIOps turns telemetry into incident evidence and response options. Its core workflow covers detection, triage, diagnosis, mitigation, and post-incident learning, often in a non-linear order.

\paragraph{Signals to decisions}
AIOps begins with noisy and incomplete alerts. Triage filters common causes, while diagnosis links symptoms to plausible failures. In a sound process, tests rule out weaker explanations instead of supporting a merely convincing narrative.

Early systems work on problem determination in large services already stressed that diagnosis must be grounded in measured behaviour, since modern services are dynamic and failures often have several contributing causes \cite{Chen2002Pinpoint}. As systems have grown in complexity, diagnostic practice has expanded to include provenance and the context of recent changes. Orca illustrates how operational debugging can be improved by tying incidents to code and deployment provenance, so that suspect changes can be ranked and inspected \cite{Bhagwan2018Orca}.

\paragraph{Artefacts: incident timelines, runbooks, and postmortems}
Compared with NetOps, AIOps produces a broader and more detailed set of narrative artefacts \cite{notaro2021mgt,Chen2020IncidentManagement}. Incidents generate timelines, chat transcripts, dashboards, on-call notes, runbooks, and postmortems \cite{Chen2020IncidentManagement,Jiang2020TSG}. These artefacts capture both policy and practice. They show what the organisation treats as safe, which actions are permitted, and what evidence must be gathered before mitigation is attempted \cite{Chen2020IncidentManagement,Jiang2020TSG}.

They are also a source of systematic bias. Runbooks may be stale, postmortems may omit sensitive details, and tickets may contain ambiguous or adversarial text \cite{Chen2020IncidentManagement,Li2022Observability}. For agentic systems, these artefacts are both a knowledge base and an attack surface. They should therefore be treated as governed inputs rather than informal documentation.

\paragraph{From anomaly detection to actionable RCA}
Early research in AIOps largely focused on incident detection. However, the true operational benefits lie in the actual mitigation of incidents \cite{notaro2021mgt,Chen2020IncidentManagement}. This early focus led researchers to pay more attention to root cause analysis, often combining different types of data to support actual fault localisation \cite{Wang2024MRCA,Zhu2024HeMiRCA}.
Recent work in automated software engineering argues for multi-modal RCA at finer granularity, combining metrics with other signals to localise causes more precisely in microservice settings \cite{Wang2024MRCA}.
This reinforces a practical point: operations data is incomplete, and single-modality answers are often too fragile \cite{Li2022Observability,Wang2024MRCA,Zhu2024HeMiRCA}.
Figure~\ref{fig:aiops-workflow} follows the standard incident-response loop used in large-scale services and observability stacks, informed by classic service diagnosis and tracing systems \cite{Chen2002Pinpoint,Fonseca2007XTrace,Kaldor2017Canopy,Chen2020IncidentManagement}.

\subsection{Operational artefacts as data types: trust and freshness}
\label{subsec:artefacts-trust-freshness}

\begin{figure}[t]
\centering
\includegraphics[width=.6\linewidth]{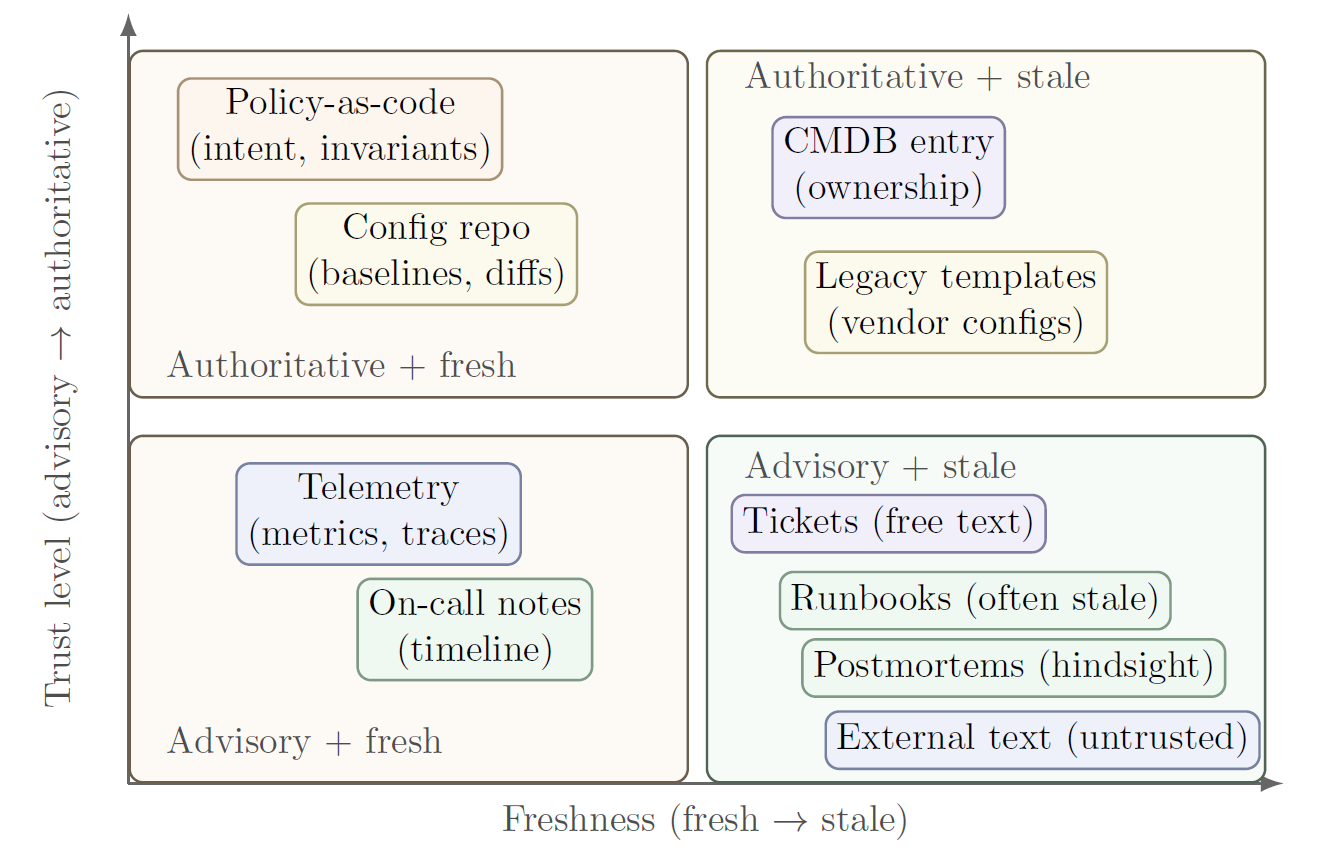}
\caption{Operational artefacts by trust level and staleness risk. The same artefact can move over time. For example, a runbook can become stale after a logging change. Retrieval therefore needs provenance and freshness signals when actions approach the write boundary}
\label{fig:artefacts-quadrant}
\end{figure}

Agentic operations systems inevitably consume and produce artefacts \cite{Chen2020IncidentManagement,notaro2021mgt}.
Treating these artefacts as data types, instead of as ``just text'', makes the trust boundary visible. It also reduces accidental over-trust during incidents, when people and systems are already working under pressure.
Figure~\ref{fig:artefacts-quadrant} summarises the two axes used in this subsection: trust level, from advisory to authoritative, and freshness, from fresh to stale.
The figure also shows why the same artefact can change its operational status over time, which matters when retrieved evidence is used near the write boundary.

\paragraph{Trust level}
Some artefacts are authoritative, such as policy-as-code, signed configuration baselines, and CMDB entries with ownership.
Some are advisory, such as runbooks, postmortems, and tickets.
Other inputs are considered untrusted, including free-form chat logs, externally provided documents, and user-influenced log strings. The operating rule is simple: untrusted artefacts can provide information for hypotheses, but without independent checks in \(\mathcal{G}_k\), they must not be used to drive write operations.

\paragraph{Freshness and staleness risk}
Operational knowledge ages \cite{Chen2020IncidentManagement,Jiang2020TSG}.
Runbooks lag behind deployments, ownership rotates, and instrumentation coverage changes \cite{Jiang2020TSG,Li2022Observability}.
Freshness should therefore be tracked at the artefact level through timestamp, version, and provenance.
When a write action is under consideration, retrieval should favour recent and authoritative sources.

\subsection{Observability modalities and tool surface}

Agentic operations must connect different kinds of evidence, from live telemetry to standing policy, to the tools that expose or act on them. Table \ref{tab:ops-artefacts} summarises these artefacts by trust level and common failure mode. Each modality is a partial sensor for a distributed system. Each tool is either a read-side interface that produces observations \(o_t\), or a write-side actuator that proposes or executes actions \(a_t\), thereby triggering the gates \(\mathcal{G}_k\).

\paragraph{Logs: rich context, weak structure}
Logs carry useful local context, but they are often unstructured, noisy, and uneven across components.
Log analysis therefore usually starts with parsing and template extraction, since later correlation depends on shared structure.
Drain is a widely used example of online log parsing for streaming environments \cite{He2017Drain}.
DeepLog shows that systems can learn usual log patterns and flag departures from them, although reliability still depends on timing robustness and parser accuracy \cite{Du2017DeepLog}. Loghub supports evaluation across varied real-world logs and helps avoid overfitting to one system's logging habits \cite{Zhu2023Loghub}.

\paragraph{Metrics: regular, scalable, and ambiguous}
Metrics remain common in alerting because they are easy to collect, combine, and compare. Their weakness is ambiguity. Different failures can leave much the same numerical trace. Pipelines therefore attach labels and dimensions for service layout, deployment state, and other operating details. Standards such as OpenTelemetry aim to align these conventions, reducing the translation needed before diagnosis can begin \cite{OpenTelemetrySpec}.

\paragraph{Traces: causal hints with coverage gaps}
Distributed tracing is valuable because it exposes request paths and latency contributors. By linking events through request context, tracing lets engineers ask causal questions rather than search raw logs alone \cite{Fonseca2007XTrace}. At scale, traceability remains both a systems problem and a data-management problem. Canopy addresses this through end-to-end traceability and analysis for high-volume production settings \cite{Kaldor2017Canopy}. The limit is coverage. Traces are sampled, instrumentation is incomplete, and context propagation can fail across service boundaries.

\paragraph{Tickets and change records: the policy layer in disguise}
Tickets and change records capture intent, approval, and accountability \cite{Chen2020IncidentManagement,Wu2023ChangeInducedIncidents}.
They are not telemetry, but they often hold the constraints that decide whether an action is safe \cite{Jiang2020TSG,Chen2020IncidentManagement}.
They also help define operational ground truth, including whether mitigation was accepted, whether a change was rolled back, and whether a similar incident later recurred \cite{Wu2023ChangeInducedIncidents,Chen2020LinkedIncidents}.
For agentic systems, incident-management and change-control interfaces should therefore be treated as first-class tool surfaces \cite{Chen2020IncidentManagement,LasCasas2024LLexus}.

\paragraph{Tools as a boundary of trust}
Operational tools determine what an agent can observe and what it can safely do. Query tools expose evidence. Change tools alter the system. This asymmetry is one reason operations differ from many general agent benchmarks. Read-only assistants may suit moderate-risk settings, while write-enabled agents require least privilege, staged deployment, and independent verification. Without these controls, the agent may simply automate failure, which is efficiency of the least helpful kind. In the unified model \eqref{eq:ops_pomdp}, this is the distinction between tools in \(\mathcal{T}^{\text{read}}_k\) and tools in \(\mathcal{T}^{\text{write}}_k\). It is also where the gate \(\mathcal{G}_k\) must sit. Work on network verification and secure updates provides concrete precedents for this trust boundary \cite{fogel2015batfish,Beckett2017Minesweeper,Khurshid2013VeriFlow,Reitblatt2011ConsistentUpdates}. Since tools cross these boundaries, the artefacts they consume and produce must be weighted according to whether they are authoritative or advisory, as shown in Table \ref{tab:ops-artefacts}.

\begin{table}[t]
\centering
\footnotesize
\setlength{\tabcolsep}{4pt}
\renewcommand{\arraystretch}{1.08}
\caption{Operational artefacts as data types, with typical trust and freshness concerns}
\label{tab:ops-artefacts}
\begin{tabularx}{\linewidth}{@{}
>{\RaggedRight\arraybackslash}p{0.20\linewidth}
>{\RaggedRight\arraybackslash}p{0.22\linewidth}
>{\RaggedRight\arraybackslash}p{0.15\linewidth}
>{\RaggedRight\arraybackslash}X
@{}}
\toprule
Artefact type & Typical source & Trust class & Common staleness or failure mode \\
\midrule

Policy / intent
& policy-as-code, change requests \cite{Opdebeeck2026PaC,Sokolowski2024IaCTest}
& Authoritative
& Policy drift, incomplete translation into device or service configurations \cite{Opdebeeck2026PaC,Sokolowski2024IaCTest}. \\

Config baselines / diffs
& repositories, controllers \cite{Reitblatt2011ConsistentUpdates,AlFares2023ChangeManagement}
& Authoritative
& Drift from live state, unsafe intermediate states during rollout \cite{Reitblatt2011ConsistentUpdates,AlFares2023ChangeManagement}. \\

Telemetry: logs / metrics / traces
& observability stack \cite{Chen2002Pinpoint,Fonseca2007XTrace,Kaldor2017Canopy,Li2022Observability}
& Mixed
& Missing coverage, misleading signals, sampling bias \cite{Li2022Observability,notaro2021mgt}. \\

Tickets / runbooks
& ITSM, wikis \cite{Jiang2020TSG,Chen2020IncidentManagement}
& Advisory
& Stale steps, unclear intent, injection through free-form text \cite{Jiang2020TSG,Chen2020IncidentManagement}. \\

Postmortems / timelines
& incident documents \cite{Chen2020IncidentManagement,Sillito2024LearningFromFailure}
& Advisory
& Incomplete detail, hindsight bias, assumptions that no longer hold \cite{Chen2020IncidentManagement,Sillito2024LearningFromFailure}. \\

\bottomrule
\end{tabularx}
\end{table}

\section{LLM foundations for operations}
\label{sec:llm-foundations}

\begin{figure}[t]
\centering
\includegraphics[width=.7\linewidth]{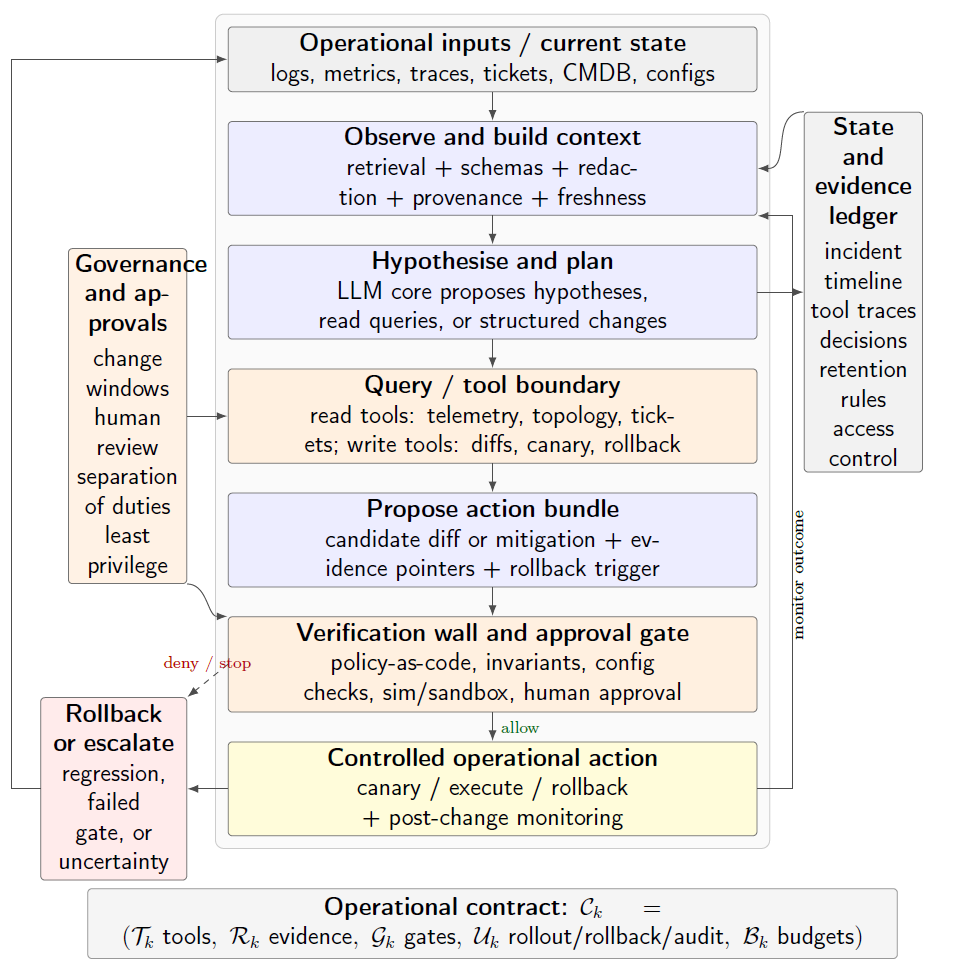}
\caption{Reference component stack. The model coordinates evidence, hypotheses, and proposals; typed tools, provenance logging, verification gates, approval paths, and rollback hooks define the enforceable system boundary. This component view is distinct from the autonomy taxonomy and task-specific workflow figures}
\label{fig:ops-stack}
\end{figure}

LLMs matter in NetOps and AIOps because they can coordinate evidence, hypotheses, tool calls, and change proposals across fragmented operational artefacts.
They do not remove the older constraints of operations.
Queries, dashboards, validators, runbooks, ticketing systems, change windows, approvals, and audit trails still define what can be safely done \cite{Beyer2016SRE,Forsgren2018Accelerate,Humble2010ContinuousDelivery,AlFares2023ChangeManagement,Alipourfard2019RiskBasedPlanning,Chen2020IncidentManagement}.
For this reason, the useful foundation is not a general account of prompting or language modelling.
It is the interface between the model and the operational system: what the model may observe, which tools it may call, what evidence it must collect, which checks it must pass, and how unsafe or uncertain actions are stopped \cite{Wang2024NetAssistant,Yu2024MonitorAssistant,xpert2024,LasCasas2024LLexus,autio2024genai,phiri2025auditable}. Figure~\ref{fig:ops-stack} sketches the resulting stack.
The LLM sits between operational evidence and controlled action.
Our synthesis suggests that operational reliability depends critically on the control system surrounding the model, especially as authority approaches write-capable interfaces. Typed tool interfaces, provenance-aware retrieval, explicit budgets, independent gates, and rollback-aware execution are therefore treated here as assurance mechanisms alongside model capability \cite{Lewis2020RAG,qin2023tool,schick2023toolformer,mondal2023routerconfigs,zhou2025meshagent}.
This section therefore keeps only the primitives needed by later sections: evidence traces, budgeted planning, freshness and provenance, controlled tool interfaces, the verification wall, human approval, and rollback.

\paragraph{Status of the mathematical notation}
The equations in this review provide a formal conceptual vocabulary for comparing operational agents; they are not theorem claims or evidence that a particular deployment satisfies a bound. The autonomy object, trace, budgets, gates, and score components make assumptions and reporting obligations explicit. Benchmark-specific quantities---including risk estimates, calibration bins, cost weights, stopping thresholds, and aggregation rules---must be instantiated and calibrated against a declared task distribution, oracle, horizon, and consequence model before numerical comparison. Where the article introduces a score or contract component rather than reporting an established metric, it is labelled as a proposed template. A further notation audit also removed repeated displayed definitions and re-statements that were not needed by later analysis. The remaining displayed equations are retained only where the same object is reused across the contract, evaluation, security, or research-agenda sections.

\subsection{Model-level synthesis and attribution}
Model choice matters, but operational results belong to the compound system unless an ablation isolates the model from retrieval, prompting, adaptation, tools, and control logic. Table~\ref{tab:systems} therefore retains the operational comparison as Panel A and adds a continuation, Panel B, recording the reported backbone and scale, prompting or agent strategy, retrieval-augmented generation, fine-tuning or domain adaptation, tool interface, and benchmark or evaluation setting. Missing implementation details are reported as not stated rather than inferred. This separation answers model-level comparison questions without attributing an agent's workflow result to its backbone alone.
NetLLM is classified separately as an enabling framework because it evaluates networking-oriented model adaptation without, by itself, demonstrating operational authority or production assurance \cite{Wu2024NetLLM}.
The update sharpens this attribution boundary. OWL adapts a LLaMA2-13B backbone with operations-specific instruction data, long-context processing, and a mixture of LoRA adapters, then evaluates the resulting model on Owl-Bench and log-analysis tasks \cite{Guo2024OWL}. OFCnetLLM reports an early multi-agent monitoring prototype on the OFC conference network, which supports system-feasibility and architecture claims but does not provide a mature comparative evaluation \cite{Yoon2025OFCnetLLM}. 
Network Meets ChatGPT proposes an intent-oriented management architecture, while Confucius provides production-facing evidence for multi-agent network management by representing operational workflows as DAGs, integrating LLM agents with existing management tools and retrieval, and coupling critical operations to validation mechanisms \cite{Wang2023NetworkMeetsChatGPT,Wang2025Confucius}. AIOpsLab and OpenRCA complement these systems with controlled evaluations of compound agents, operational tasks, and telemetry interfaces \cite{Chen2025AIOpsLab,Xu2025OpenRCA}. These records support model, architecture, system, or benchmark claims at their evaluated level. They do not by themselves establish independently assured production authority.

\subsection{Evidence traces, budgets, and stop rules}
\label{sec:workflow-contract}

An operational agent should be evaluated through its trace, not only through its final answer.
We write the evidence trace as
\begin{equation}
E =
\big((\tau_1,y_1),\ldots,(\tau_n,y_n)\big),
\label{eq:evidence_trace_sec3}
\end{equation}
where \(\tau_i\) is a typed tool call, including its arguments, and \(y_i\) is the corresponding output.
This makes the agent's reasoning inspectable: a reviewer can see which logs, metrics, traces, tickets, configurations, validators, or simulators were queried, and whether the resulting evidence was sufficient for the proposed diagnosis or action \cite{Sigelman2010Dapper,Fonseca2007XTrace,Mace2015PivotTracing,Kaldor2017Canopy,OpenTelemetrySpec}.

Operational loops are also budgeted.
Budgets are not only a cost-control device.
They prevent open-ended tool churn during incidents and provide a mechanical point at which an agent must stop, escalate, or roll back \cite{Jha2025ITBench,liu2023agent,yao2025taubench}.
We use
\begin{equation}
B = (B^{\text{tool}},B^{\text{tok}},B^{\text{time}},B^{\text{risk}}),
\label{eq:budget_vector_sec3}
\end{equation}
where \(B^{\text{tool}}\) bounds tool calls, \(B^{\text{tok}}\) bounds context and generation, \(B^{\text{time}}\) bounds latency, and \(B^{\text{risk}}\) bounds the risk of committing a violating write action.

A compact planning constraint is
\begin{equation}
\sum_{i=1}^{n} c(\tau_i) \le B^{\text{tool}},
\qquad
\sum_{i=1}^{n} \ell_i \le B^{\text{tok}},
\qquad
\sum_{i=1}^{n} \Delta t_i \le B^{\text{time}},
\label{eq:budget_constraints_sec3}
\end{equation}
where \(c(\tau_i)\) is a per-call cost, \(\ell_i\) is token consumption at step \(i\), and \(\Delta t_i\) is wall-clock time.
For write-side risk, we use
\begin{equation}
\Pr\!\big(V_{\Pi,\mathcal{I}}(a_t,E)=1\big) \le \epsilon,
\qquad
\epsilon \doteq B^{\text{risk}},
\label{eq:risk_constraint_sec3}
\end{equation}
where \(a_t\) is a candidate action, usually a diff plus rollout plan, and \(V_{\Pi,\mathcal{I}}\) is a violation indicator under the relevant policy \(\Pi\) and invariants \(\mathcal{I}\).
This makes refusal and escalation measurable.
A system can report when it stopped because evidence was missing, the budget was exhausted, the gate failed, or the estimated risk exceeded \(\epsilon\) \cite{nist2023airmf,autio2024genai,NCSC2025Integrity,greshake2023indirect,chen2025struq,pasquini2025aioopsdoom}.
In a benchmark, the probability in Eq.~\eqref{eq:risk_constraint_sec3} must be estimated over a declared incident and perturbation distribution using an independent violation oracle. The threshold \(\epsilon\) is consequently a calibrated deployment parameter, not a universal constant supplied by the LLM.

MonitorAssistant provides complementary operational evidence from cloud-service monitoring. It uses LLMs to assist monitoring configuration, knowledge reuse, and alarm interpretation within an operational workflow \cite{Yu2024MonitorAssistant}. The study is useful here because it places the language model inside an established monitoring process rather than treating model output as an independent operational decision. Its evidence supports assisted monitoring and diagnosis, but not autonomous remediation or independently verified write-capable control.

Stop rules should be explicit and mechanically enforceable.
Common examples are budget exhaustion, missing required evidence, a failed gate, repeated contradictory observations, non-diagnosable ambiguity, and any request that would move the system outside the permitted autonomy rung.
In operations, knowing when to stop is part of the safety case \cite{Sillito2020FailuresFixes,Widder2021TrustAutomation,Sillito2024LearningFromFailure}.

\subsection{Tool-first evidence and controlled interfaces}
\label{sec:tool-first-foundation}

Two operational facts explain why fluent text is a weak success criterion.
First, evidence is produced by tools.
In incident response, the system status is best determined by evidence provided by tools, such as logs, metrics, topology status, ticket records, and validators. Recent operational assistants therefore collect and structure diagnostic artefacts before asking the model to draw conclusions \cite{chen2024autosys,xpert2024,Wang2024NetAssistant,Yu2024MonitorAssistant,Wang2024RCAgentCIKM}.
Second, actions must be reviewable and reversible.
A diagnosis often implies a change, but a change must be expressed as a structured proposal, checked outside the model, and applied with limited permissions \cite{acto_sosp23,conveyor_osdi23,AlFares2023ChangeManagement,Opdebeeck2026PaC,Sokolowski2024IaCTest}. Work on router configuration synthesis shows the same pattern. Direct LLM output is brittle when used alone. Correctness comes from verifiers, repair loops, and explicit checks around the generator \cite{mondal2023routerconfigs,Nazari2024LocalizedExplanationsHotNets,Liu2024AutoConfigRepair}.

Operational retrieval is useful only when information remains fresh, attributable, and authoritative. Runbooks change after incidents, ownership shifts, and baselines drift. Stale instructions can be worse than none when they guide operational changes. The retrieval layer should therefore record source, validity period, trust level, and the presence of newer evidence \cite{Lewis2020RAG,zhang2025mrag,qi2024attributionrag}.

Sources should be classified as authoritative, such as policies as code, CMDB entries, verified topologies, and configuration baselines; consultative, such as operating manuals, tickets, and post-event analyses; or untrusted, such as chat logs and externally supplied text. The class determines how the evidence may be used \cite{greshake2023indirect,zou2025poisonedrag,shafran2025machineRAG,NCSC2025Integrity,Chismon2025PromptInjection}.

Operational tools should be typed, versioned, access-controlled, and scope-limited. Broad shell access should not be the default. Privileged tasks should use restricted APIs for validation, tracing, change requests, sandbox tests, canary deployment, and rollback. Tool failures should be expected, including poor tool choice, malformed parameters, dependence on output wording, and loops that continue after the task is complete \cite{qin2024toolllm,liu2023agent,zhou2024web,cai2024latm,patil2024gorilla,li2023apibank}.

Structured output links language to operational control. Typed assumptions, action proposals, configuration discrepancies, approval packages, and rollback plans can be parsed, replayed, checked against policy, and reused in regression tests from past incidents. This is important in network operations, where ambiguous intent may conceal conflicts in reachability, isolation, or priority. Ambiguities in configuration synthesis should therefore be resolved or rejected before a change is generated.

Algorithm~\ref{alg:planner-executor} is normative pseudocode for the proposed contract, not an implementation evaluated in the retained studies. It identifies where independent policy, budget, and gate checks must occur. An implementation must also define concurrency, retry semantics, partial tool failure, idempotence, and transaction or compensation behaviour.

Termination is guaranteed only if all relevant budget counters are monotone, finite bounds are imposed, and the stop rules fire when those bounds or a finite retry limit is reached. If a planner may emit an empty or repeated plan without advancing a counter or invoking a stop rule, Algorithm~\ref{alg:planner-executor} gives no termination guarantee.

\begin{algorithm}[!ht]
\caption{Normative planner--executor procedure under operational gates}
\label{alg:planner-executor}
\begin{algorithmic}[1]
\STATE \textbf{Input:} incident signal \(s\), autonomy rung \(\mathcal{A}_k=(\mathcal{T}^{\mathrm{read}}_k,\mathcal{T}^{\mathrm{write}}_k,\mathcal{G}_k)\),
        policy \(\Pi\), invariants \(\mathcal{I}\), required evidence \(\mathcal{R}\), budgets \(B\), stop rules \(\mathcal{S}\)
\STATE \(E \leftarrow \emptyset\), \(H \leftarrow \emptyset\), \(b \leftarrow 0\) \hfill evidence, hypotheses, budget counters
\WHILE{not resolved}
  \IF{\(\mathcal{S}\) triggers on \((E, H, s, B)\)}
    \STATE \textbf{return} escalate or rollback with evidence trace \(E\)
  \ENDIF
  \STATE \(p \leftarrow \textsc{Plan}(s, H, E, \Pi, \mathcal{I}, B)\)
  \IF{not \(\textsc{PolicyCheck}(p, \Pi)\)}
    \STATE \textbf{return} escalate with \(E\)
  \ENDIF
  \FOR{each \((\tau_i, \mathit{args}_i)\) in \(p\)}
    \IF{\(\tau_i \notin \mathcal{T}^{\mathrm{read}}_k \cup \mathcal{T}^{\mathrm{write}}_k\)}
      \STATE \textbf{return} deny action and escalate with \(E\)
    \ENDIF
    \STATE update budget counters \(b\)
    \IF{\(b\) exceeds \(B\)}
      \STATE \textbf{return} stop and escalate with \(E\)
    \ENDIF
    \IF{\(\tau_i \in \mathcal{T}^{\mathrm{write}}_k\) and not \(\textsc{GateCheck}(\tau_i, \mathit{args}_i, E, \Pi, \mathcal{I}, \mathcal{G}_k, B^{\mathrm{risk}})\)}
      \STATE \textbf{return} reject action, propose safe alternatives, and escalate with \(E\)
    \ENDIF
    \STATE \(y_i \leftarrow \tau_i(\mathit{args}_i)\)
    \STATE append \((\tau_i, y_i)\) to \(E\)
    \STATE \(H \leftarrow \textsc{UpdateHypotheses}(H, E)\)
    \STATE \(s \leftarrow \textsc{UpdateSignal}(s, E)\)
  \ENDFOR
\ENDWHILE
\STATE \textbf{return} diagnosis, next safe actions, and evidence trace \(E\)
\end{algorithmic}
\end{algorithm}

\subsection{Verification wall and gated planning}

\label{sec:verification-wall-prop}

The verification wall is the non-bypassable boundary between proposal and execution. It turns agentic operation from free-form text generation into a controlled change process. Let \(\mathcal{A}_k=(\mathcal{T}^{\text{read}}_k,\mathcal{T}^{\text{write}}_k,\mathcal{G}_k)\). Every candidate write action (a), or equivalently every tool call \(\tau\in\mathcal{T}^{\text{write}}_k\), must pass a gate
\begin{equation}
g:\ (a,E,\Pi,\mathcal{I})\mapsto \{\texttt{allow},\texttt{deny}\},\qquad g\in\mathcal{G}_k,
\label{eq:gate_sec3}
\end{equation}
Here, \(E\) is the evidence trace, \(\Pi\) is policy, and \(\mathcal{I}\) is the set of invariants. If \(g\) is sound for its declared model and checked property set, an allowed write satisfies only those modelled policies and invariants; this does not guarantee correctness under unmodelled state, stale evidence, checker defects, or actuator failure \cite{Kazemian2012HSA,Khurshid2013VeriFlow,fogel2015batfish,Beckett2017Minesweeper,gember-jacobson2017integrating}. Planning must consider both tool limits and gate rules. Algorithm~\ref{alg:planner-executor} shows this loop, which we use throughout the survey.

\paragraph{Safety risks}
Safety matters because agentic operations connect untrusted text and telemetry to privileged tools. Prompt injection may enter through tickets, runbooks, chat logs, postmortems, or customer-provided text, so retrieved content should remain separate from system instructions, with narrow permissions, provenance records, and independent checks before action \cite{Chismon2025PromptInjection,greshake2023indirect,chen2025struq}. Telemetry manipulation can also mislead the agent by corrupting logs, metrics, traces, alerts, or topology. This is an evidence-integrity failure, not only a reasoning error. Defences include cross-checking telemetry, validating formats, tracking provenance, and refusing action when evidence is too weak \cite{pasquini2025aioopsdoom,nist2023airmf,autio2024genai}. The main operational safeguards are least privilege, independent verification, approval gates, canary rollout, and rollback plans \cite{saltzer1975protection,acto_sosp23,AlFares2023ChangeManagement,Alipourfard2019RiskBasedPlanning}. An agent may identify the right cause and still fail operationally if it cannot show evidence, pass checks, prepare rollback, or leave an audit trail \cite{Widder2021TrustAutomation,Jha2025ITBench,phiri2025auditable}.


\section{Agentic architecture patterns for operations}
\label{sec:agentic-patterns}

Agentic NetOps and AIOps systems differ mainly in their permitted authority \cite{Chen2020IncidentManagement,LasCasas2024LLexus,zhou2025meshagent}. We describe this through a ladder of autonomy, where each rung adds capability and risk. Lower rungs read, filter, and organise evidence. Higher rungs propose or execute operational changes. As autonomy increases, permissions, verification checks, approval paths, and rollback plans must be part of the architecture from the start \cite{Wang2024NetAssistant,Yu2024MonitorAssistant,xpert2024,zhou2025meshagent}.

\begin{figure}[t]
\centering
\includegraphics[width=.8\linewidth]{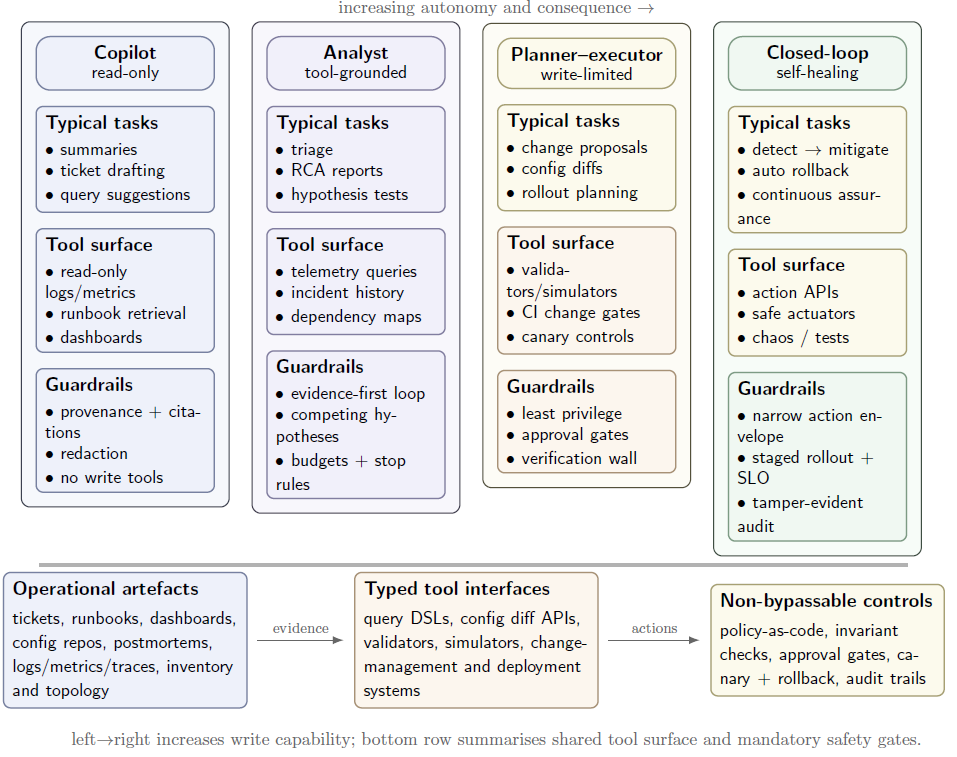}
\caption{Authority taxonomy for LLM-based operational agents. The figure classifies roles by read, proposal, and execution authority; Table~\ref{tab:autonomy-ladder} separately specifies the evidence and assurance obligations associated with each role}
\label{fig:taxonomy}
\end{figure}

Figure~\ref{fig:taxonomy} shows the role-level taxonomy used here. It separates read-only copilots, tool-based analysts, write-limited planner-executors, and closed-loop controllers by tool surface and guardrails. Table~\ref{tab:autonomy-ladder} then makes the same ladder operational by writing each rung as a contract over allowed tools, required evidence, gates, rollout and audit obligations, and budgets or success criteria.
The figure explains the roles, while Table~\ref{tab:autonomy-ladder} defines the controls that make those roles safe enough to evaluate.

\subsection{Autonomy rungs as operational assurance contracts}

We make the autonomy rung operational by treating it as a contract between the agent and the surrounding operations system.
At rung \(k\), the agent is not defined only by the model it uses, but by what it may observe, which tools it may call, what evidence it must collect, which checks it must pass, how changes are rolled out or undone, and which budget limits the workflow.
We write this rung-indexed assurance contract as
\begin{equation}
\mathcal{C}_k
=
\big(
\mathcal{T}_k,\,
\mathcal{R}_k,\,
\mathcal{G}_k,\,
\mathcal{U}_k,\,
\mathcal{B}_k
\big),
\label{eq:assurance_contract_k}
\end{equation}
where \(\mathcal{T}_k\) is the permitted tool surface, \(\mathcal{R}_k\) is the required evidence to collect before a conclusion or action, \(\mathcal{G}_k\) is the set of non-bypassable gates, \(\mathcal{U}_k\) is the rollout, rollback, and audit protocol, and \(\mathcal{B}_k\) denotes budgets for tool cost, latency, uncertainty, and action risk.
The permitted tool surface is partitioned as
\begin{equation}
\mathcal{T}_k
=
\mathcal{T}_k^{\text{read}}
\cup
\mathcal{T}_k^{\text{propose}}
\cup
\mathcal{T}_k^{\text{execute}},
\label{eq:tool_partition_k}
\end{equation}
where read tools retrieve or query evidence, propose tools generate reports, hypotheses, plans, or configuration diffs, and execute tools apply changes to an operational system.

The evidence trace remains a first-class output:
\begin{equation}
\label{eq:evidence_trace_sec4}
E =
\big((\tau_1,y_1),\ldots,(\tau_n,y_n)\big),
\end{equation}
where \(\tau_i\) is a typed tool call and \(y_i\) is its output.
For any candidate action \(a\), gating is modelled as a non-bypassable decision function
\begin{equation}
\label{eq:gate_function}
g_k(a,E,\Pi,\mathcal{I}) \in \{0,1\},
\end{equation}
which returns \(1\) only when the action is permitted under policy \(\Pi\), supported by the required evidence, satisfies the relevant invariants \(\mathcal{I}\), and passes the gate conditions in \(\mathcal{G}_k\) \cite{Khurshid2013VeriFlow,fogel2015batfish,Kazemian2013NetPlumber,Ye2020Hoyan,Sokolowski2024IaCTest}.
The shift from advisory operation to operational autonomy is therefore marked by the regime
\(\mathcal{T}_k^{\text{execute}}\neq\emptyset\), where verification, rollout, rollback, and audit become binding constraints rather than optional safeguards \cite{saltzer1975protection,autio2024genai,phiri2025auditable}.

Contract satisfaction is a trace property.
For a trace \(\tau\), we write \(\tau \models \mathcal{C}_k\) when the trace contains the mandated evidence, respects the relevant budgets, never bypasses the gate for executed actions, and follows the rollout and audit protocol:
\begin{align}
&\textsf{Evidence}(\tau) \supseteq \mathcal{R}_k
\quad\wedge\quad
\textsf{Budget}(\tau)\le \mathcal{B}_k,
\label{eq:contract_evidence_budget_sec4}
\\
&\wedge\ \forall a \in \tau \cap \mathcal{T}_k^{\text{execute}}:\ 
g_k(a,E,\Pi,\mathcal{I})=1,
\label{eq:contract_gate_sec4}
\\
&\wedge\ {\textsf{Rollout}(\tau),\ \textsf{Rollback}(\tau),\ \textsf{Audit}(\tau)
\ \text{satisfy}\ \mathcal{U}_k} .
\label{eq:contract_rollout_audit_sec4}
\end{align}
This lets later sections compare agentic NetOps and AIOps systems by their operational contract, rather than by model family alone.
Equations~\eqref{eq:assurance_contract_k}--\eqref{eq:contract_rollout_audit_sec4} are a proposed reporting and design template. They do not imply that every component is presently implemented or empirically validated at every rung. A benchmark must specify the concrete evidence oracle, gate implementation, budget units, rollout horizon, audit predicate, and failure handling before reporting contract satisfaction.

\subsection{Ladder of autonomy with explicit tools, gates, artefacts, and success}
Table~\ref{tab:autonomy-ladder} summarises the same four autonomy levels as operational contracts.
For each rung, it lists the allowed tools \(\mathcal{T}_k\), required evidence \(\mathcal{R}_k\), gates \(\mathcal{G}_k\), rollout and audit obligations \(\mathcal{U}_k\), and budgets or success criteria \(\mathcal{B}_k\).
The purpose is to ensure that later architecture and evaluation sections refer to the same contract objects, rather than restating autonomy in prose.

\begin{table}[t]
\centering
\footnotesize
\setlength{\tabcolsep}{2.2pt}
\renewcommand{\arraystretch}{1.08}
\caption{A ladder of autonomy for agentic NetOps/AIOps written as operational contracts. Higher autonomy shifts the engineering burden from answer quality to evidence obligations, gated action, least privilege, rollout, rollback, and audit \cite{saltzer1975protection,Alipourfard2019RiskBasedPlanning,acto_sosp23,conveyor_osdi23,weber2013undoability}}
\label{tab:autonomy-ladder}
\begin{tabularx}{\linewidth}{@{}
>{\RaggedRight\arraybackslash}p{0.12\linewidth}
>{\RaggedRight\arraybackslash}p{0.18\linewidth}
>{\RaggedRight\arraybackslash}p{0.16\linewidth}
>{\RaggedRight\arraybackslash}p{0.15\linewidth}
>{\RaggedRight\arraybackslash}p{0.16\linewidth}
>{\RaggedRight\arraybackslash}X
@{}}
\toprule
Rung $k$ & $\mathcal{T}_k$: allowed tools & $\mathcal{R}_k$: required evidence & $\mathcal{G}_k$: gates & $\mathcal{U}_k$: rollout and audit & $\mathcal{B}_k$: budgets and success criteria \\
\midrule

Copilot (read-only)
& Search, retrieval, summarisation, and query suggestion; no propose or execute tools \cite{xpert2024,Yu2024MonitorAssistant,LasCasas2024LLexus}
& Provenance-linked observations; cited source snippets; freshness markers where available \cite{xpert2024,Yu2024MonitorAssistant,OpenTelemetrySpec}
& Redaction, provenance rules, citation requirement, and rate limits \cite{NCSC2025Integrity,Chismon2025PromptInjection,qi2024attributionrag}
& Evidence trace $E$; incident summary; query candidates; no rollout obligation
& Faster evidence location; fewer missed signals; low false citations; bounded tool calls \cite{xpert2024,Yu2024MonitorAssistant,yao2025taubench}. \\

\addlinespace
Analyst (read-mostly)
& Read tools plus diagnosis tools, including causal graphs, dependency models, event-correlation engines, and telemetry queries \cite{Wang2024NetAssistant,chen2024autosys,Wang2024RCAgentCIKM}
& Evidence trace $E$; hypothesis set $H$; RCA evidence pointers; contradiction checks \cite{chen2024autosys,Wang2024MRCA,Zhu2024HeMiRCA,changerca2024}
& Citation to tool outputs; stop rules; escalation policy; no unsupervised write execution \cite{chen2024autosys,LasCasas2024LLexus,phiri2025auditable}
& RCA report; audit-ready trace of queries and discarded hypotheses; human handoff where uncertainty remains
& Correct top-$k$ localisation; low tool churn; calibrated uncertainty; bounded diagnostic latency \cite{Wang2024MRCA,Zhu2024HeMiRCA,Guo2017Calibration,Jha2025ITBench}. \\

\addlinespace
Planner--executor (write-limited)
& Read tools, propose tools for diff synthesis, and execute tools limited to approved safe actuators \cite{mondal2023routerconfigs,zhou2025meshagent,Liu2024AutoConfigRepair}
& Evidence trace $E$; candidate change diff $\Delta$; preconditions; risk estimate; rollback evidence \cite{conveyor_osdi23,acto_sosp23,weber2013undoability,AlFares2023ChangeManagement}
& Non-bypassable verification wall; approvals; canary constraints; rollback triggers \cite{fogel2015batfish,Khurshid2013VeriFlow,Kazemian2013NetPlumber,Ye2020Hoyan,acto_sosp23,Alipourfard2019RiskBasedPlanning}
& Staged rollout plan; rollback plan; signed change record; post-change observation window
& No invariant violation in the checked domain; safe rollout; quick rollback on regression; bounded risk and time budget \cite{fogel2015batfish,Khurshid2013VeriFlow,Beckett2017Minesweeper,conveyor_osdi23,changerca2024}. \\

\addlinespace
Closed-loop (self-healing)
& Continuous read plus execute tools inside a narrow action envelope \cite{acto_sosp23,conveyor_osdi23,Basiri2019ChaosEngineering}
& Continuous evidence trace $E$; monitor state; trigger condition; recovery evidence; periodic review artefacts \cite{OpenTelemetrySpec,RFC9232Telemetry,crosby2009tamper}
& Strict action envelope; automated rollback; continuous monitors; periodic audits \cite{saltzer1975protection,autio2024genai,phiri2025auditable,crosby2009tamper}
& Signed action log; recovery certificate where possible; rollback record; scheduled audit of repeated actions
& Improved MTTR with bounded regressions; stable long-run drift $\Delta_t$; bounded action frequency and risk \cite{acto_sosp23,conveyor_osdi23,Basiri2019ChaosEngineering,Forsgren2018Accelerate}. \\

\bottomrule
\end{tabularx}
\end{table}

\subsection{Reference architecture and evaluation observables}
Figure~\ref{fig:ops-stack} gives the reference architecture used throughout this survey.
It condenses the safe-autonomy commitments that recur across incident assistants, tool-grounded diagnosis systems, and verification-heavy NetOps workflows \cite{xpert2024,Yu2024MonitorAssistant,chen2024autosys,Wang2024NetAssistant,fogel2015batfish,Khurshid2013VeriFlow,Reitblatt2011ConsistentUpdates,Beckett2017Minesweeper,zhou2025meshagent,LasCasas2024LLexus,autio2024genai,phiri2025auditable}.

The practical value of this architecture is that it makes later evaluation concrete.
It exposes observables that can be measured across tasks, traces, and deployment stages \cite{Jha2025ITBench,yao2025taubench,Liang2022HELM}.
Examples include:
\begin{itemize}
  \item \textbf{Evidence quality:} completeness and correctness of \(E\), including missing tool calls, incorrect citations, and stale artefacts \cite{qi2024attributionrag,zhang2025mrag}.
  \item \textbf{Gate behaviour:} pass and fail rates of \(g_k(\cdot)\), together with the reasons for rejection, such as policy violation, invariant failure, missing approval, or budget exhaustion \cite{fogel2015batfish,Khurshid2013VeriFlow,Kazemian2013NetPlumber,Ye2020Hoyan,autio2024genai}.
  \item \textbf{Proposal quality:} validity of diffs \(\Delta\), satisfaction of preconditions, and behaviour under staged rollout \cite{mondal2023routerconfigs,Liu2024AutoConfigRepair,conveyor_osdi23,weber2013undoability}.
  \item \textbf{Operational outcomes:} time to mitigate, rollback frequency, and post-change drift or mismatch metrics, for example \(\Delta_t\) from Section~\ref{sec:ops-background} \cite{Forsgren2018Accelerate,Chen2020IncidentManagement,changerca2024,Wu2023ChangeInducedIncidents}.
\end{itemize}

\subsection{Human review as a concrete object}
In guarded planner--executor systems, the human does not review a chat transcript.
They review a \emph{bundle} that can be scored, archived, and replayed \cite{Widder2021TrustAutomation,AlFares2023ChangeManagement,weber2013undoability,phiri2025auditable}.
We model the review bundle as
\begin{equation}
\label{eq:review_bundle}
\mathcal{R} \;=\; \big(\Delta,\; \mathsf{Pre},\; \mathsf{Checks},\; \mathsf{Canary},\; \mathsf{Rollback},\; \mathsf{Ptrs}(E)\big),
\end{equation}
where $\Delta$ is the proposed change (diff or action plan), $\mathsf{Pre}$ are preconditions, $\mathsf{Checks}$ are required validations (policy-as-code, invariants, sandbox), $\mathsf{Canary}$ is the staged rollout plan, $\mathsf{Rollback}$ defines triggers and steps, and $\mathsf{Ptrs}(E)$ are evidence pointers into the trace \cite{fogel2015batfish,Khurshid2013VeriFlow,acto_sosp23,conveyor_osdi23,weber2013undoability}.

\begin{figure}[t]
\centering
\includegraphics[width=.65\linewidth]{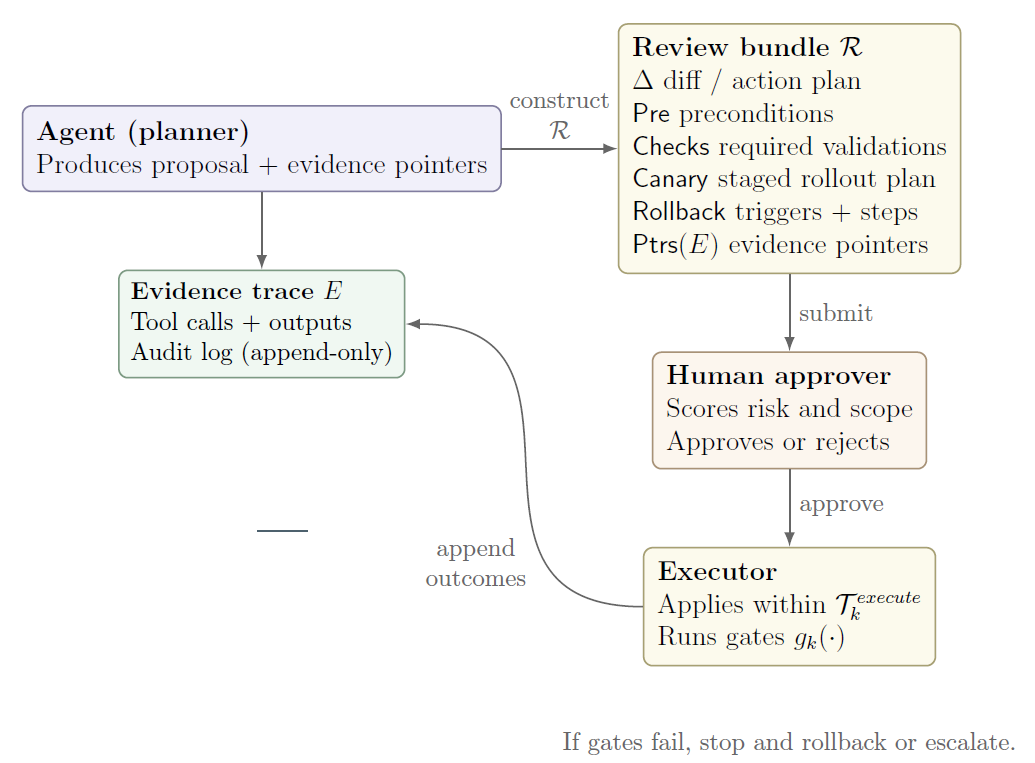}
\caption{Human decision interface for a guarded planner--executor. The review bundle $\mathcal{R}$ contains the diff, preconditions, completed checks, canary plan, rollback triggers, and evidence pointers into $E$; it does not imply that approval quality is automatic}
\label{fig:review-bundle}
\end{figure}

Figure~\ref{fig:review-bundle} sketches this review flow, which is the practical interface for accountability \cite{phiri2025auditable,autio2024genai}.

\subsection{Failure modes by rung}
Autonomy changes which failures dominate \cite{Widder2021TrustAutomation,Jha2025ITBench,autio2024genai}.
At low rungs, the main damage is wasted time, unsupported advice, and misleading explanations.
At high rungs, the main damage is unsafe change, delayed rollback, and hard-to-audit action \cite{saltzer1975protection,phiri2025auditable}.

\noindent\fbox{%
\centering
\begin{minipage}{\linewidth}
\textbf{Design rule: autonomy raises failure severity}
As an agent moves from copilot to analyst, planner--executor, and closed-loop controller, its capability set changes from read-only evidence support to proposal generation and, eventually, bounded execution.
Hallucination or story-mode reasoning can affect every rung, but tool misuse, prompt injection through operational artefacts, drift between verified and realised state, and rollback failure become more serious once the agent is allowed to propose or execute actions \cite{greshake2023indirect,Chismon2025PromptInjection,pasquini2025aioopsdoom,Wu2023ChangeInducedIncidents,weber2013undoability}.
Higher autonomy is therefore acceptable only when the operational contract strengthens the tool boundary, evidence obligations, gates, rollout controls, rollback triggers, budgets, and audit record.
\end{minipage}}

\vspace{0.5em}
The design rule above is qualitative rather than an empirical measurement.
It follows from each rung's capability set
\(\mathcal{T}_k=\mathcal{T}_k^{\text{read}}\cup\mathcal{T}_k^{\text{propose}}\cup\mathcal{T}_k^{\text{execute}}\)
and gate strength \(\mathcal{G}_k\).
Risks driven by write capability increase when
\(\mathcal{T}_k^{\text{execute}}\neq\emptyset\), while strong, non-bypassable gates reduce the likelihood that these risks propagate into executed actions \cite{fogel2015batfish,Khurshid2013VeriFlow,Kazemian2013NetPlumber,Ye2020Hoyan,saltzer1975protection}.
Table~\ref{tab:autonomy-ladder} operationalises this autonomy--risk coupling by specifying, for each rung, the allowed tools, required evidence, gates, rollout and audit duties, and budgets.
The qualitative takeaway is that gating and verification are not optional extras at higher rungs.
They are binding constraints once \(\mathcal{T}_k^{\text{execute}}\) is non-empty \cite{autio2024genai,phiri2025auditable}.

\subsection{Pattern summaries tied back to the contract}
The following patterns are designed to align with the contract object \((\mathcal{T}_k,\mathcal{G}_k,E,\Pi,\mathcal{I})\).

\paragraph{Pattern A: Co-pilot (Read-only)}
The Co-pilot pattern improves operator throughput by compressing and navigating evidence. Its security depends primarily on the source, editing, and faithful referencing of the tool's output. The measure of success should be faster discovery of relevant evidence and fewer missed signals, rather than a more polished incident summary \cite{xpert2024,Yu2024MonitorAssistant,LasCasas2024LLexus}.

\paragraph{Pattern B: Analyst (tool-grounded diagnosis)}
The Analyst agent runs read-only queries, keeps several hypotheses in view, and produces a structured root cause analysis (RCA) report with traceable evidence clues. Its main failure mode is narrative drift: a coherent explanation that is only lightly attached to the system under investigation. The design should therefore require tool-based citations, checks against contrary evidence, and clear stopping conditions. If uncertainty remains, the agent should escalate rather than close the case on a tidy but weak explanation \cite{Wang2024NetAssistant,chen2024autosys,Wang2024RCAgentCIKM,Wang2024MRCA,Zhu2024HeMiRCA,changerca2024}.

\paragraph{Pattern C: Planner--executor (write-limited)}
This is the first rung where safe autonomy becomes a property of the whole system.
The agent proposes diffs and plans, while execution is constrained by least privilege and by a verification wall that checks policy and invariants \cite{fogel2015batfish,Beckett2017Minesweeper,Khurshid2013VeriFlow,Kazemian2013NetPlumber,Ye2020Hoyan,zhou2025meshagent}.
Success is not only whether the change was applied.
It is whether the change was applied without violating checked properties, and whether the system could return quickly to a known-good state if reality objected \cite{conveyor_osdi23,weber2013undoability,changerca2024}.

\paragraph{Pattern D: Closed-loop (self-healing)}
A closed-loop system aims to integrate detection, diagnosis, mitigation, and recovery verification into a continuous process.
At this rung, since the system is part of the control loop, the range of action must be very limited, and rollback needs to happen automatically.
Thus, the main evaluation metric should be the system's long-term stability in the face of faults and changes, rather than the outcome of any single event \cite{acto_sosp23,conveyor_osdi23,Basiri2019ChaosEngineering,autio2024genai}.


\begin{figure}[t]
\centering
\includegraphics[width=.7\linewidth]{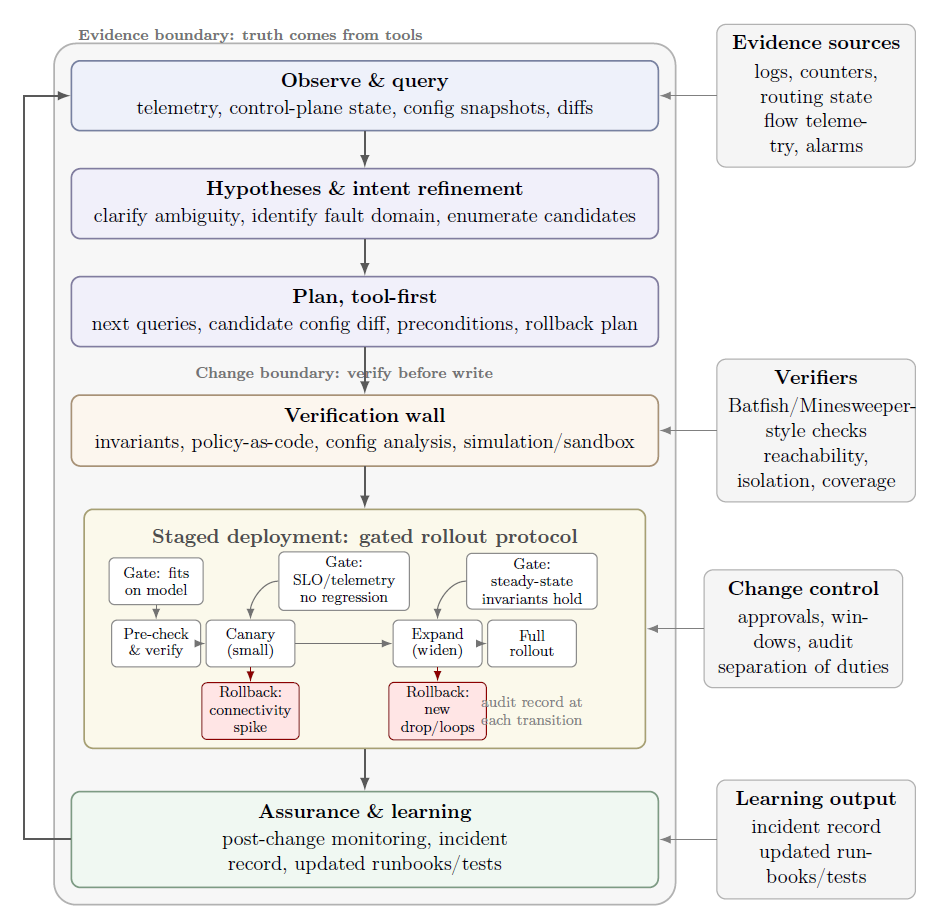}
\caption{Agentic NetOps loop specialised to high-consequence change. The LLM is useful as a workflow controller: it gathers evidence, clarifies intent, proposes minimally invasive diffs, and routes every write through an explicit verification wall, staged rollout, and rollback path}
\label{fig:agentic-netops-loop}
\end{figure}

\section{Agentic NetOps}
\label{sec:agentic-netops}

NetOps is a natural target for agentic systems because many operational tasks already follow a repeatable loop:
(i) gather evidence from measurement and control-plane state, (ii) form a hypothesis about a failure or a policy violation, (iii) propose a change, and (iv) validate the change before rollout \cite{RFC9315Intent,Falkner2022IBNEnterprise,leivadeas2023survey,Chen2020IncidentManagement,AlFares2023ChangeManagement}. At the same time, NetOps has a higher safety bar than many AIOps settings because a single incorrect change can propagate quickly, trigger large blast-radius outages, or create hard-to-debug transient behaviours during convergence \cite{Reitblatt2011ConsistentUpdates,Reitblatt2012Abstractions,Turner2010CalFaultLines,Gill2011DCFailures}. Empirical studies show that operational failures in networks often arise from simple causes, including component faults, misconfigurations, and side effects from other configuration changes. Although these causes may appear simple, they are difficult to predict without systematic testing \cite{Turner2010CalFaultLines,Gill2011DCFailures,XuZhou2015ConfigErrorsSurvey,Mahajan2002BGP,Wu2023ChangeInducedIncidents}.

This combination of structured workflows and high consequence makes NetOps a useful stress test for agentic LLM design.
The LLM is rarely the correct place to ``decide truth'' about network state. Instead, the LLM is most valuable as a controller of \emph{workflow}: it translates intent into queries, routes evidence through verifiers, proposes minimally invasive diffs, and documents preconditions and rollback \cite{Wang2024NetAssistant,mondal2023routerconfigs,zhou2025meshagent}.
Figure~\ref{fig:agentic-netops-loop} summarises this workflow view: evidence is gathered through tools, intent is refined before planning, and every write-side action is routed through verification, staged rollout, rollback triggers, and assurance.

In other words, NetOps rewards agentic systems that are tool-first and verification-first, rather than conversationally fluent \cite{fogel2015batfish,Khurshid2013VeriFlow,Ye2020Hoyan,Brown2023BatfishLessons}. This separation also leaves room for specialised, network-native control mechanisms such as NOS \cite{bilal2025nos}, where fast adaptation is handled by the control layer rather than by the language model itself.  For example, in O-RAN scheduling, the lower-layer  manages latency tails and spectrum targets. Meanwhile, higher-level systems interpret intent, select policies, and oversee auditable changes \cite{bilal2025nosoran}.
This positioning is consistent with practical NetOps verification and safe-change practice, where the hard part is not proposing a change but justifying and checking it before it is trusted \cite{fogel2015batfish,Beckett2017Minesweeper,Khurshid2013VeriFlow,Kazemian2013NetPlumber,DeltaNet2017NSDI}.

\subsection{NetOps property definitions (targets for the verification wall)}
To make ``verification'' concrete, we fix a compact model and a small palette of properties that recur across the NetOps literature \cite{Kazemian2012HSA,Kazemian2013NetPlumber,fogel2015batfish,Beckett2017Minesweeper}.
Let the network be a directed graph \(G=(V,E)\).
Let \(x\) denote the realised control/forwarding state (routing adjacencies, FIB entries, ACL tables, tunnel state, etc.).
For a traffic class \(c\) (e.g., 5-tuple predicate, VRF, DSCP class), let \(F_x(v,c)\) denote the next-hop relation induced by \(x\).
This induces a (possibly empty) forwarding path \(\textsf{Path}_x(s,c)\) obtained by iterating \(F_x\) from source \(s\) until termination (deliver, drop) or a loop.
These abstractions match the intent of data-plane and control-plane analysis systems that operationalise reachability and policy reasoning from configurations \cite{Kazemian2012HSA,fogel2015batfish,Kazemian2013NetPlumber,Lopes2015NoD}.

\begin{figure}[t]
\centering
\includegraphics[width=.8\linewidth]{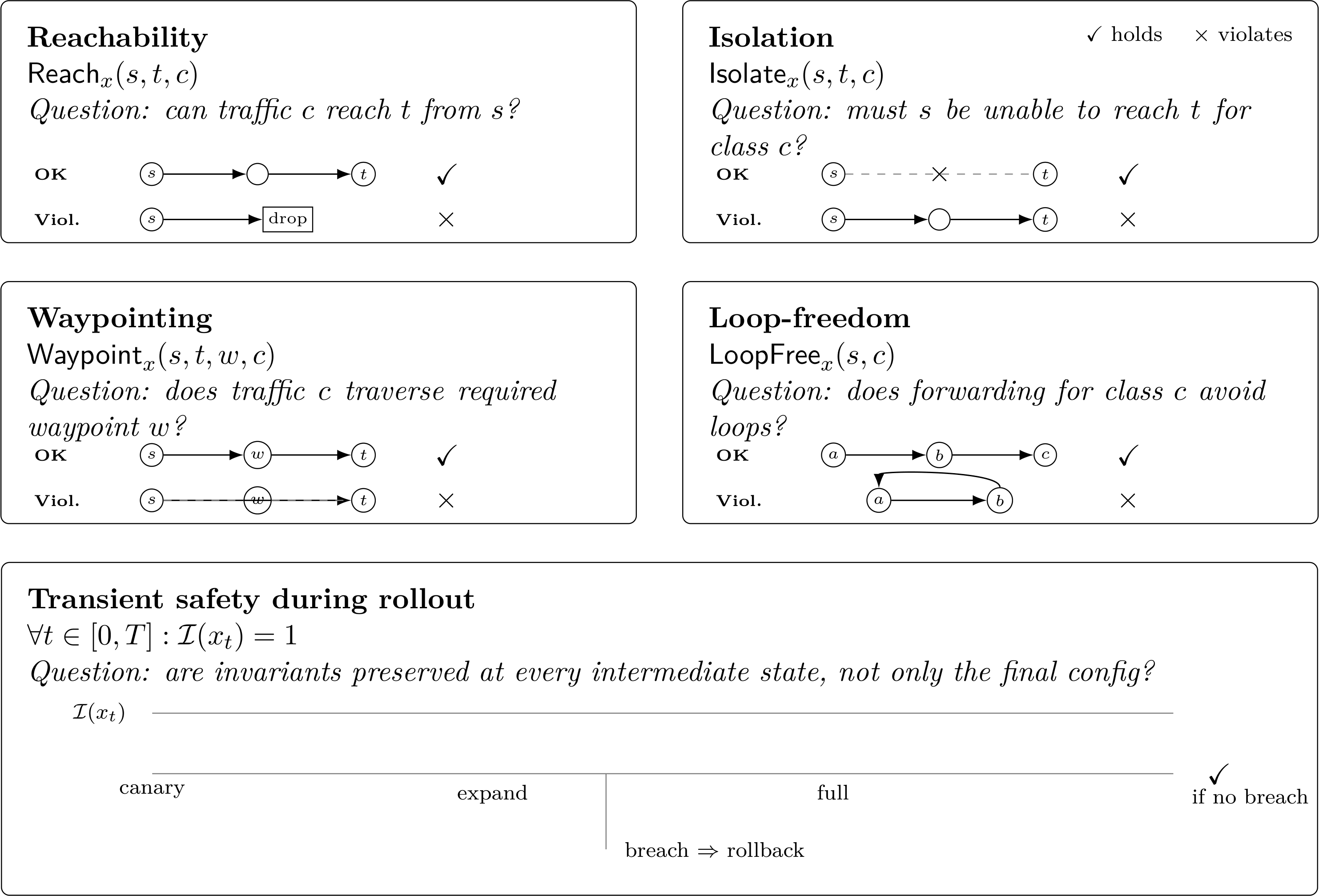}
\caption{Common NetOps invariants that can form the verification wall. The point is not the notation; it is to make the targets explicit and reusable across synthesis, rollout safety, and autonomy gating}
\label{fig:netops-invariants}
\end{figure}

We use the following predicates as reusable invariants, illustrated in Figure~\ref{fig:netops-invariants}:

\begin{align}
\textsf{Reach}_x(s,t,c)
&\equiv \textsf{Path}_x(s,c)
   \text{ terminates at } t \notag\\
&\quad \text{without drop or loop}, \\
\textsf{Isolate}_x(s,t,c)
&\equiv \neg \textsf{Reach}_x(s,t,c), \\
\textsf{Waypoint}_x(s,t,w,c)
&\equiv \textsf{Reach}_x(s,t,c)
   \wedge w \in \textsf{Path}_x(s,c), \\
\textsf{LoopFree}_x(s,c)
&\equiv \textsf{Path}_x(s,c)
   \text{ contains no repeated node}
\end{align}

These cover the most common operational questions: ``can A reach B'', ``must A never reach B'', ``must traffic pass through a firewall/proxy'', and ``are we creating transient loops''.
Figure~\ref{fig:netops-invariants} also makes explicit why rollout safety is stricter than final-state correctness: the invariants should hold not only after the final configuration is installed, but across intermediate states during canary, expansion, and full rollout.
They align with what configuration analysers and invariant checkers are typically asked to prove or refute at scale \cite{fogel2015batfish,Beckett2017Minesweeper,Khurshid2013VeriFlow,Ye2020Hoyan,ElHassany2018NetComplete,DeltaNet2017NSDI}.

\subsection{Intent, ambiguity, and when clarification is mandatory}
Natural language intent is typically underspecified, even when expressed by experienced operators \cite{RFC9315Intent,Falkner2022IBNEnterprise,leivadeas2023survey,mondal2025ambiguity}.
Intent often omits operational constraints that matter for safety, such as acceptable blast radius, maintenance window, traffic class priorities, failure-domain boundaries, or whether a waypoint is ``preferred'' versus ``mandatory''.
Intent-based networking is typically conceptualised as a closed-loop system that provides continuous assurance, rather than as a single compilation step \cite{leivadeas2023survey,RFC9315Intent,Falkner2022IBNEnterprise}.
Prior to the advent of LLMs, research approached configuration as a synthesis problem defined by explicit correctness criteria, which typically focused on translating high level policies into device-level configurations, ensuring invariants are maintained, and protocol semantics are respected \cite{Beckett2016DontMindGap,ElHassany2017AbstractTopologies,ElHassany2018NetComplete,Synet2017SIGCOMM}.

For agentic NetOps, this suggests a safe division of labour:
the LLM handles ambiguity resolution and produces structured candidates, while synthesis and checking back-ends provide the authority on correctness \cite{mondal2023routerconfigs,mondal2025ambiguity,Liu2024AutoConfigRepair}.

Recent domain evidence supports this division without eliminating its limits. Network Meets ChatGPT organises intent translation, autonomous management, and network control as a layered architecture rather than a free-form command path \cite{Wang2023NetworkMeetsChatGPT}. The LUMI pipeline in Jacobs et al. uses operator confirmation, grammar-constrained intent assembly, ambiguity detection, and post-deployment monitoring to make natural-language intent more trustworthy, although its evaluated information-extraction component is not an LLM \cite{Jacobs2025TrustIBN}. A YANG-grounded LLM agent further shows that schema grounding can constrain candidate configurations on a multi-vendor OpenROADM testbed \cite{Xiao2026YANGAgent}. Together, these studies support typed intent records, clarification, and independent validation. They do not establish unrestricted autonomous configuration of production networks.

Recent work makes ambiguity explicit and studies how an LLM-driven assistant should ask questions \emph{before} generating configurations \cite{mondal2025ambiguity,mondal2023routerconfigs}.
A practical and checkable condition for \emph{mandatory clarification} is that there exist multiple distinct diffs that satisfy the current checks but induce different forwarding semantics for some traffic class.
Let \(\mathcal{S}\) be the set of diffs \(\Delta\) that pass the current policy and invariant checks.
We have an ambiguity witness when:
\begin{equation}
\begin{aligned}
\exists\, \Delta_1,\Delta_2 \in \mathcal{S},\ \Delta_1 \neq \Delta_2:
\quad
\exists (s,t,c) \ \text{s.t} \\
\textsf{Path}_{x\oplus \Delta_1}(s,c)
\neq
\textsf{Path}_{x\oplus \Delta_2}(s,c).
\end{aligned}
\end{equation}
In practice, the most common sources of such witnesses are rule overlap and priority conflicts, where an operator’s intent does not specify precedence even though the configuration must \cite{mondal2025ambiguity,ElHassany2018NetComplete,Synet2017SIGCOMM}.
Under these conditions, the agent should not ``pick a reasonable default''.
It should ask a small typed set of disambiguating questions, commit the answers to the intent record, and only then generate diffs \cite{mondal2025ambiguity,mondal2023routerconfigs}.

Finally, synthesis is not only about producing configs but also about supporting review and post-incident learning.
Work on producing localised explanations for synthesised configurations supports this direction and fits naturally with agentic workflows where explanations are linked to specific lines, policies, and checker outputs \cite{Nazari2024LocalizedExplanationsHotNets,Liu2024AutoConfigRepair}.

\subsection{Update safety: correctness during rollout, not only at rest}
Update safety is not only about the final configuration being correct.
Even correct end states can be reached through unsafe intermediate states, such as loops, blackholes, or policy violations during rollout, because updates are applied under asynchronous convergence and partial deployment \cite{Reitblatt2012Abstractions,Reitblatt2011ConsistentUpdates,McKeown2008OpenFlow}.
Agentic NetOps should treat ``apply change'' as a protocol, not a single tool call \cite{Reitblatt2012Abstractions,Alipourfard2019RiskBasedPlanning,AlFares2023ChangeManagement}.

Let \(x_t\) denote the realised network state during rollout, as partial deployments and convergence events take effect. Transient safety requires every intermediate state to satisfy the relevant invariants:
\begin{equation}
\forall t \in [0,T]: \mathcal{I}(x_t)=1,
\end{equation}
where \(\mathcal{I}\) combines the required reachability, isolation, waypointing, and loop-freedom checks. This separates final-state correctness from rollout correctness, which is the key lesson of consistent updates and staged operations \cite{Reitblatt2012Abstractions,Alipourfard2019RiskBasedPlanning,Vanbever2014HotSwap}. 


\subsection{Change risk scoring and autonomy gating}
NetOps practitioners assess risk using criteria that align with autonomy decisions, such as the number of affected devices, novelty of the change, the adequacy of verification procedures, and the history of similar changes causing incidents \cite{Alipourfard2019RiskBasedPlanning,AlFares2023ChangeManagement,Wu2023ChangeInducedIncidents}.
Agentic systems benefit from making that reasoning explicit because it links verification coverage to permissioning and gating, rather than treating coverage as an afterthought \cite{Xu2023NetCov,autio2024genai}.

Let \(\Delta\) be a candidate diff (or change bundle).
A minimal risk score that is easy to compute and easy to review is:
\begin{equation}
\begin{aligned}
\textsf{Risk}(\Delta)
={}& \alpha\,\textsf{Blast}(\Delta)
+\beta\,\textsf{Novelty}(\Delta) \\
&+\gamma\,\bigl(1-\textsf{Coverage}(\Delta)\bigr).
\end{aligned}
\end{equation}
where \(\textsf{Blast}(\Delta)\) approximates potential impact, \(\textsf{Novelty}(\Delta)\) measures distance from known-safe patterns, and \(\textsf{Coverage}(\Delta)\) measures how well tests and checks exercise the configuration lines and behaviours touched by \(\Delta\).
Recent coverage work shows why this is operationally important: ``we have tests'' is not the same as ``we have assurance for the lines we have just changed'' \cite{Xu2023NetCov,Ye2020Hoyan}.
In practice, when coverage is low for the touched lines, autonomy should be reduced rather than expanded \cite{Xu2023NetCov,Alipourfard2019RiskBasedPlanning}.

This connects directly to the earlier ladder-of-autonomy and gate definitions.
A simple execution gate can include a risk threshold:
\begin{equation}
\begin{aligned}
g_k(\Delta,E,\Pi,\mathcal{I})
= \mathbf{1}\Big[
&\ \textsf{PolicyCheck}(\Delta,\Pi)=1 \\
&\wedge\ \textsf{Verify}(\Delta,\mathcal{I})=1 \wedge\ \textsf{Risk}(\Delta)\le \theta_k
\Big].
\end{aligned}
\end{equation}
This makes "validation barriers" an important principle. The scope of coverage and the risks decide if an agent can carry out a plan, not just if they can have a major effect \cite{fogel2015batfish,Khurshid2013VeriFlow,Kazemian2013NetPlumber,Xu2023NetCov,autio2024genai}.

\subsection{Troubleshooting: from symptoms to hypotheses with reproducible traces}
Network troubleshooting is typically a search on an incomplete set of evidence. This is because signals are often noisy, telemetry data may be missing, and symptoms can deviate from the root cause due to retries, buffering, and control plane convergence. Therefore, practical application relies on a standardized workflow: collecting discriminative observations to narrow down the fault, and then validating hypotheses through targeted testing \cite{Handigol2014PacketHistories}.
The same idea carries over to agentic NetOps.
The agent should be rewarded for producing a reproducible evidence trail \(E\), not for writing a smooth account of events.
Recent dialogue-driven diagnosis in data centre networks points in the same direction: success depends on workflow discipline, tool sequencing, and explicit escalation when confidence remains low \cite{Wang2024NetAssistant}.

Cross-layer cases are still the most difficult.
Application symptoms may originate from routing policy, transport behaviour, middlebox policy, or transient forwarding states during updates.
Agentic NetOps systems should therefore represent hypotheses at several layers and test them with independent signals.
A practical rule is to maintain competing hypotheses and require at least one disconfirming test before committing to any change with a broad blast radius.

\subsection{Causal inference for explainable NetOps diagnosis}
\label{subsec:causal-netops}

Causal inference provides a useful counterweight to purely LLM-centred NetOps diagnosis. Many network failures are not isolated events; they propagate through topology, protocol state, control-plane dependencies, service dependencies, and time-ordered event streams \cite{Yan2012GRCA,Kobayashi2019CausalNetworkLogs}. A causal diagnosis method aims to recover this propagation structure, so that the output is not only a ranked root cause but also an explanation path linking symptoms to candidate causes \cite{Kobayashi2018MiningCausality,Kobayashi2019CausalNetworkLogs}.

Earlier NetOps work already reflects this idea. G-RCA models service-quality problems in large IP networks using dependency relationships, temporal and spatial event correlation, and reasoning logic, showing that operator-facing RCA benefits from explicit dependency models rather than opaque prediction alone \cite{Yan2012GRCA}. Work on mining causes of network events from log data uses causal inference to go beyond co-occurrence and recover likely causal relationships among network events \cite{Kobayashi2018MiningCausality}. Later work further combines causal inference with protocol-layer and topology knowledge, reducing spurious causal edges and producing information that is more useful for troubleshooting \cite{Kobayashi2019CausalNetworkLogs}.

A closely related cellular-network line of work studies causal inference for estimating or explaining the operational effect of radio-access parameter and software changes. For example, causal models have been used to estimate the impact of handover-parameter adjustments and other cellular-network configuration interventions, including transmission-power and cell-offset changes, on subsequent service and performance time series \cite{Hua2023CSDnet,Zhang2023DCDN,Li2024ParameterChangeCellular}. This is especially relevant to agentic NetOps because it shows that some root-cause and change-impact questions can be answered through structured causal models before an LLM is asked to reason over large telemetry and runbook contexts.

For agentic NetOps, these methods should be treated as diagnostic back-ends rather than competitors to the agentic framing. A useful division of labour is: causal inference constructs or updates a fault graph; the LLM turns operator intent into queries, requests additional evidence, explains the causal path, and prepares a reviewable mitigation proposal \cite{Yan2012GRCA,Kobayashi2018MiningCausality,Kobayashi2019CausalNetworkLogs}. This hybrid design has two advantages. First, it improves explainability, because the diagnosis is grounded in a causal path rather than a free-form narrative. Second, it can reduce repeated reasoning over raw telemetry and runbook context, because the causal graph narrows the candidate space before the LLM is asked to summarise, compare, or prepare an action proposal \cite{Kobayashi2018MiningCausality,Kobayashi2019CausalNetworkLogs}.

\subsection{Repair as an iterative workflow}
There is growing interest in combining localisation, repair, and validation into an integrated loop.
Automatic Configuration Repair is illustrative: it frames misconfiguration handling as localise--fix--validate, which matches how operators actually de-risk changes during incidents \cite{Liu2024AutoConfigRepair}.
This maps well to agentic design because it encourages iterative, checkable steps and makes rollback and validation central rather than decorative.

\subsection{Putting it together: why NetOps is the sharpest testbed}
NetOps makes agent-safety claims concrete because it offers: (i) checkable invariants, (ii) a known failure mode of transient rollout unsafety, and (iii) a principled link between test coverage and autonomy. A conservative autonomy policy then follows, with broad read access, narrow write access, and closed-loop behaviour only for low-impact actions with strong verification support. This is consistent with the NetOps literature on outage causes and the limits of informal reasoning in complex control planes \cite{Turner2010CalFaultLines,Gill2011DCFailures,XuZhou2015ConfigErrorsSurvey,Brown2023BatfishLessons}.

\begin{figure}[t]
\centering
\includegraphics[width=.7\linewidth]{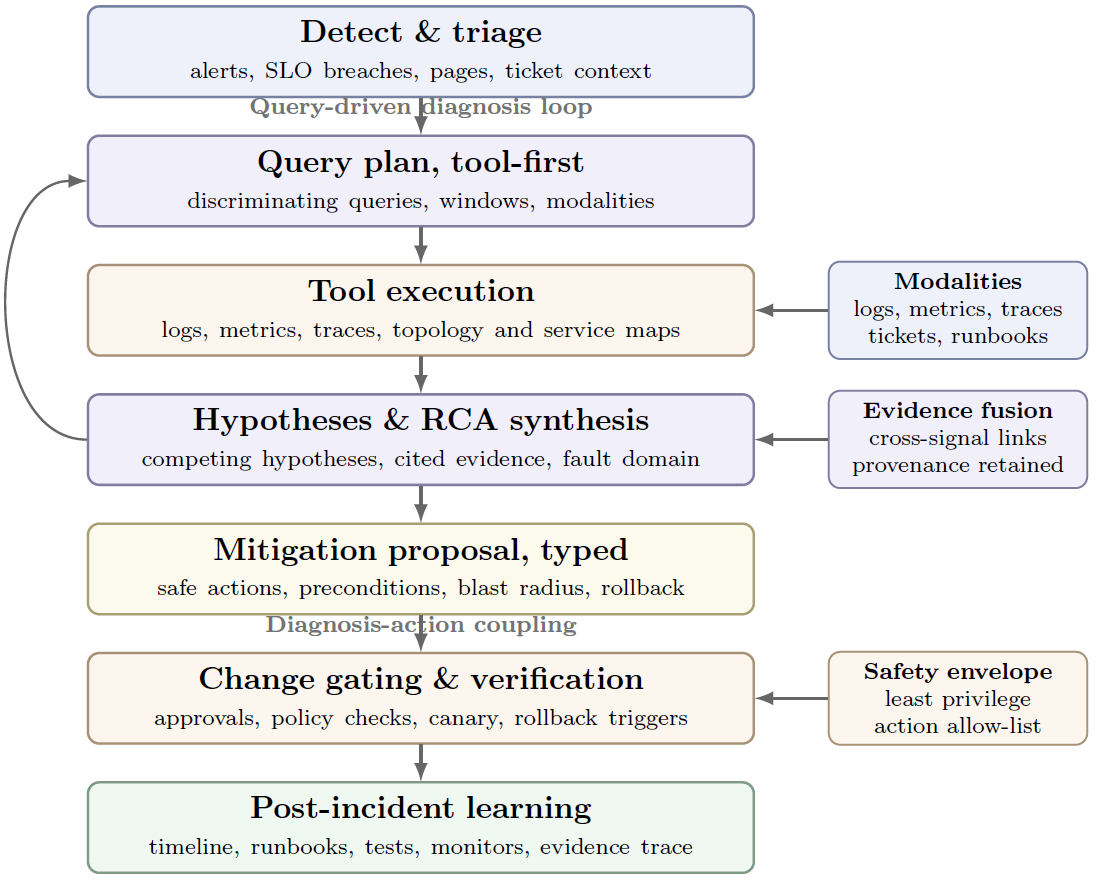}
\caption{Workflow view of agentic AIOps. The central loop is query-driven diagnosis: select the next discriminating query, gather evidence through tools, revise competing hypotheses, and prepare mitigation proposals that remain subject to policy checks, approval gates, and rollback-ready verification}

\label{fig:agentic-aiops-loop}
\end{figure}

\section{Agentic AIOps}
\label{sec:agentic-aiops}

AIOps has a richer public literature than NetOps because software services produce abundant observability data and incident work is already mediated through telemetry stores, dashboards, incident histories, and runbooks \cite{Li2022Observability,OpenTelemetrySpec,Chen2020IncidentManagement}.
What changes with agentic LLMs is not simply that a system can describe an incident; it is that the system can conduct the investigation as a structured workflow: selecting the next discriminating query, interpreting partial results, maintaining competing hypotheses, and drafting an action proposal tied to evidence, approvals, and rollback \cite{xpert2024,chen2024autosys,Wang2024RCAgentCIKM,LasCasas2024LLexus}.

Two background facts from the systems and SRE literature explain why AIOps is a natural home for agents.
First, modern production diagnosis is fundamentally query-driven, and good teams treat incident response as repeated hypothesis testing under partial observability \cite{Beyer2016SRE,Mace2015PivotTracing,Sillito2020FailuresFixes}.
Second, tracing and analysis stacks in large organisations are designed to support rapid, iterative diagnosis at scale, because static dashboards are not enough \cite{Kaldor2017Canopy,Sigelman2010Dapper,Fonseca2007XTrace,Li2022Observability}.
Figure~\ref{fig:taxonomy} shows the main roles, their tools, and needed safeguards. Agentic AIOps is practical only when it follows operational routines: tool-driven investigation, not open-ended explanation. Figure~\ref{fig:agentic-aiops-loop} maps the AIOps lifecycle and marks where tool-based RCA and mitigation planning must tie to clear gates and rollback \cite{Mace2015PivotTracing,Kaldor2017Canopy,xpert2024,chen2024autosys,Wang2024RCAgentCIKM,Chen2020IncidentManagement,autio2024genai}.

\paragraph{AIOps invariants}
AIOps can be related to NetOps verification by treating mitigation safety as formal conditions over the current service state. Let \(s_t\) denote the observed state at time \(t\), including SLOs, error budget, dependency health, replica status, rollout state, and tenant impact. A candidate action \(a_t\) is admissible only when the relevant service conditions hold:
\begin{equation}
\phi_{\mathrm{SLO}}(s_t,a_t) \quad \Pr[L_{p99}(s_{t+1}) \leq L_{\max}] \geq 1-\epsilon ,
\end{equation}
\begin{equation}
\phi_{\mathrm{budget}}(s_t,a_t) \quad B_{\mathrm{err}}(t+1) \geq B_{\min},
\end{equation}
\begin{equation}
\phi_{\mathrm{blast}}(s_t,a_t) \quad |\mathcal{U}_{\mathrm{affected}}(a_t)| \leq \beta_{\max},
\end{equation}
\begin{equation}
\phi_{\mathrm{dep}}(s_t,a_t) \quad \mathrm{Reachable}(G_{\mathrm{dep}}^{t+1}, S_{\mathrm{critical}})=\mathrm{true}.
\end{equation}
These conditions cover latency, error budget, blast radius, and dependency health. They are proposed examples of gate clauses, not claims that all reviewed AIOps systems implement probabilistic SLO checking. A benchmark must define the forecast horizon, estimator, uncertainty interval, and threshold for $\phi_{\mathrm{SLO}}$; deterministic limits remain preferable where available. After an agent suggests an action, the surrounding system checks the applicable service-level requirements before proceeding.

\subsection{A causal framing for RCA under partial observability}
Root-cause analysis in AIOps is rarely a single-step prediction problem.
A more faithful model is iterative ranking and hypothesis testing over a dependency structure under incomplete sensing and changing context \cite{Chen2002Pinpoint,Ikram2022CausalDiscovery,ChainOfEvent2024FSE,Zhu2024HeMiRCA}.

Let the service dependency graph be
\begin{equation}
G_s=(\mathcal{S},\mathcal{E}),
\end{equation}
where nodes \(u\in\mathcal{S}\) are components (services, databases, queues, hosts), and edges encode dependencies (RPC, data, or resource coupling).
During an incident, the agent obtains multi-modal evidence through tools: metrics, traces, logs, and recent change context.
A compact way to express RCA is as a ranking problem:
\begin{equation}
\textsf{score}(u) = f\!\big(\text{metrics},\text{traces},\text{logs},\text{change}(u)\big),
\end{equation}
where the agent returns a top-$k$ list with evidence for each candidate. This framing matches production debugging as an iterative process: issue a discriminating query, refine the candidate set, and choose the next test \cite{Mace2015PivotTracing,Kaldor2017Canopy,Chen2002Pinpoint,Bhagwan2018Orca}. Recent systems strengthen this approach by learning or constructing event-based causal graphs that yield interpretable RCA candidates instead of relying only on persuasive correlation \cite{Ikram2022CausalDiscovery,ChainOfEvent2024FSE,Li2021TraceRCAIWQoS,Zhu2024HeMiRCA}.
For agentic AIOps, the architectural lesson is direct: a diagnosis cannot be justified by language alone. The agent must use tools to collect evidence and perform discriminating tests that rule out competing explanations \cite{chen2024autosys,Wang2024RCAgentCIKM,phiri2025auditable}.

\subsection{Log understanding and anomaly detection with LLMs}

Logs remain the most common operational signal and one of the least forgiving \cite{He2017Drain,Du2017DeepLog,Zhao2021PracticalLogAD,Ma2025PractitionerExpectationsLogAD}.
They are verbose, uneven across systems, and liable to drift as code, dependencies, and deployment habits change.
Accordingly, AIOps pipelines rarely consume raw logs directly.
They usually introduce at least three intermediate steps: parsing or templating to turn free text into structured events; aggregation to form service-, host-, or request-level sequences and distributions; and detection or scoring to identify anomalies that warrant escalation \cite{He2017Drain,He2019LogParsingToolsBenchmarks,Du2017DeepLog,Zhu2023Loghub}.

\paragraph{What LLMs change, and what they do not}
A sensible way to place LLMs in log-based AIOps is to treat them as semantic adapters, rather than as replacements for classical detectors.
Deep sequence models such as DeepLog showed that event-sequence modelling can detect anomalies without hand-written rules \cite{Du2017DeepLog}.
In real deployments, however, this earlier promise depends on rather plain conditions: good log quality, stable parsing, manageable workload shift, and templates that do not change faster than the detector can learn them \cite{Zhao2021PracticalLogAD,He2019LogParsingToolsBenchmarks,Zhu2023Loghub}.

Practitioner studies sharpen the point.
Operators need methods that remain useful under drift, produce evidence-linked outputs, and degrade safely when signals are missing \cite{Ma2025PractitionerExpectationsLogAD}.
LLMs can help here by mapping heterogeneous log language into more stable schemas, condensing long bursts into incident-focused summaries, and suggesting the next query or filter needed to test a hypothesis \cite{xpert2024,LasCasas2024LLexus,chen2024autosys}.
This is useful only when the system constrains the output format and preserves provenance \cite{qi2024attributionrag,autio2024genai}.
If an LLM summary cannot be traced back to log lines and timestamps, it becomes a familiar operational nuisance: neat enough to circulate, and unsafe when someone has to act on it \cite{NCSC2025Integrity,phiri2025auditable}.
\paragraph{Parsing and benchmarking remain foundational}
Because parsing errors cascade into every downstream step, log parsing and its evaluation remain first-class background.
LogPai’s work on tools and benchmarks for automated log parsing helped standardise evaluation practice and made clear that parser choice can dominate downstream quality \cite{He2019LogParsingToolsBenchmarks}.
LogHub consolidated commonly used datasets into a shared collection, enabling more reproducible comparisons across systems and domains \cite{Zhu2023Loghub}.
For agentic AIOps, the implication is direct: if the agent is trained or evaluated on cleanly parsed logs, but deployed on noisy and drifting templates, autonomy will be miscalibrated \cite{Zhao2021PracticalLogAD,Ma2025PractitionerExpectationsLogAD}.
This argues for explicitly reporting (i) parsing quality, (ii) drift sensitivity, and (iii) behaviour when parsers disagree \cite{He2019LogParsingToolsBenchmarks,Zhu2023Loghub}.

\subsection{Incident triage and RCA with tool-grounded agents}
Incident response has an implicit structure that is easy to state and hard to operationalise:
detect, triage, diagnose, mitigate, and learn \cite{Chen2020IncidentManagement,Sillito2024LearningFromFailure}.
Agentic systems add value in the gaps between these stages, where humans spend time translating between representations: ticket text to telemetry queries, telemetry results to hypotheses, hypotheses to mitigations, mitigations to change records \cite{xpert2024,chen2024autosys,Wang2024RCAgentCIKM,LasCasas2024LLexus}.

\paragraph{From natural language to telemetry queries}
A high-leverage capability is converting incident descriptions into concrete telemetry queries.
Xpert studies the role of query writing in a large-scale incident workflow and proposes LLM-supported query recommendations to reduce time-to-signal \cite{xpert2024}.
This is a key agentic building block because it reframes “LLM reasoning” as evidence acquisition.
If the system cannot reliably propose the next useful query, it will default to narrative \cite{xpert2024,Yu2024MonitorAssistant,yao2025taubench}.

\paragraph{Evidence fidelity beats narrative plausibility}
Recent LLM-based RCA systems increasingly connect explanations to collected diagnostic evidence. RCACopilot illustrates this approach by combining predefined incident handlers with a language model that predicts root-cause categories and generates explanations \cite{chen2024autosys}. The study evaluates the combined pipeline on real incidents, while the diagnostic collection component has a longer production history than the LLM prediction component; these should not be conflated. The transferable lesson is that an explanation becomes operationally useful when it records what was checked and why \cite{chen2024autosys,qi2024attributionrag,phiri2025auditable}.

\paragraph{Agents and tool-augmented RCA}
A parallel line of work frames RCA as an autonomous, tool-augmented agent problem.
RCAgent uses tool-augmented action trajectories for data collection and analysis, and evaluates across multiple RCA targets (root cause, solution, evidence, ownership) \cite{Wang2024RCAgentCIKM}.
For AIOps, the point is that RCA quality depends on tool sequencing and stopping rules, not only on language \cite{Wang2024RCAgentCIKM,Jha2025ITBench,yao2025taubench}.
An agent that cannot stop when evidence is insufficient will consume budget on noisy queries and eventually ``decide'' from fragile cues \cite{Wang2024RCAgentCIKM,autio2024genai}.

The updated evidence base separates three forms of RCA evidence. Large-scale task evaluations assess generated causes or mitigations against incident records, including more than 40,000 incidents in Ahmed et al., 100,000 production incidents in the in-context learning study, and cross-lifecycle grounding in X-Lifecycle \cite{Ahmed2023CloudIncidents,Zhang2024ICLRCA,Goel2024XLifecycle}. Tool-grounded systems instead evaluate how an agent gathers and tests evidence. Roy et al. study ReAct-style agents with operational retrieval tools, while Flow-of-Action constrains multi-agent RCA with standard operating procedures and an explicit stopping agent \cite{Roy2024RCAAgents,Pei2025FlowOfAction}. OpenRCA provides a separate benchmark object with 335 failures from three enterprise systems and more than 68 GB of logs, metrics, and traces. Its best reported agent configuration resolves only a small fraction of complete cases, which is evidence against equating fluent investigation with reliable RCA \cite{Xu2025OpenRCA}.

\subsection{Mitigation planning under a constrained action envelope}
Once an agent proposes mitigations (restart, failover, throttling, config toggle), the system becomes a socio-technical control loop.
The failure mode is no longer wrong answer but wrong change at the wrong time \cite{Beyer2016SRE,Humble2010ContinuousDelivery,Basiri2019ChaosEngineering,Widder2021TrustAutomation}.
Safe autonomy therefore begins by formalising what actions are even on the table \cite{autio2024genai,phiri2025auditable}. A useful example is queue-aware streaming intrusion detection \cite{bilal2026gate}, where online evidence accumulation triggers bounded mitigation rather than open-ended configuration change. Such designs illustrate the kind of controlled action surface that agentic NetOps and AIOps systems should expose to LLM planners.
The update adds two bounded execution studies. Cao et al. evaluate a GPT-4-based agent over 150 Linux administration tasks inside a Docker sandbox, and Sarda et al. generate and execute Ansible remediation playbooks for microservice faults \cite{Cao2024LinuxAgents,Sarda2024AutoRemediation}. These studies justify classifying some systems as write-limited planners or bounded executors. Their sandboxes, task envelopes, and scripted interfaces remain part of the result, so neither study supports broad closed-loop production autonomy.
The distributed nature of modern services constrains AIOps remediation. Restarts, failovers, replica changes, traffic shifts, and rollbacks can involve multiple components, and their safety depends on consistency, replication, coordination, and recovery rules \cite{Lamport1978Time,Schneider1990StateMachine}. Stateless front-end failovers are generally low-risk. Replaying logs or changing leaders, however, can disrupt ordering, persistence, or client-visible consistency. AIOps systems must therefore account for state, arbitration, and leader-election rules and avoid remedial actions that could violate the required protocols.
Let \(\mathcal{M}\) be the permitted mitigation set, partitioned into low-risk and high-risk actions:
\begin{equation}
\mathcal{M} = \mathcal{M}^{\text{low}} \cup \mathcal{M}^{\text{high}},
\end{equation}
where \(\mathcal{M}^{\text{low}}\) includes actions with well-tested rollback paths, while \(\mathcal{M}^{\text{high}}\) includes changes that alter correctness assumptions or have long-tailed side effects (schema migrations, broad policy changes, or permanent configuration edits). This organisation-specific partition must be encoded in policy \(\Pi\), not left to prompt text \cite{saltzer1975protection,autio2024genai}. Mitigation selection can then be stated as a proposed constrained optimisation template:
\begin{equation}
a^\star = \arg\min_{a\in \mathcal{M}} \Big( \textsf{Impact}(a) + \lambda\,\textsf{Risk}(a) \Big)
\quad \text{s.t} \quad g(a,E,\Pi)=1,
\end{equation}
where \(E\) is the collected evidence trace and \(g(\cdot)\) is the non-bypassable gate that enforces policy, approvals, and required checks.
The loss terms and $\lambda$ are benchmark-specific and require empirical calibration; hard gate failure is not offset by a lower weighted objective. The separation is explicit: the agent proposes, while the surrounding system decides what is permitted and when \cite{acto_sosp23,conveyor_osdi23,autio2024genai,phiri2025auditable}.

\subsection{Workflow artefacts as adversarial inputs}
AIOps depends heavily on tickets, runbooks, on-call notes, postmortems, and chat transcripts \cite{Sillito2024LearningFromFailure}. These artefacts record valuable local practice but remain untrusted input channels: they may be stale, incomplete, or maliciously crafted. Prompt injection and integrity threats therefore belong to the primary design surface rather than to a peripheral ``LLM application'' checklist \cite{NCSC2025Integrity,pedro2025prompt,Chismon2025PromptInjection,greshake2023indirect,Liu2023HouYi}. Operational artefacts must not directly authorise privileged actions \cite{Chismon2025PromptInjection,autio2024genai}. A tool-first design may use them as leads, but must confirm or reject those leads against independent operational evidence before proposing a change \cite{chen2024autosys,Wang2024RCAgentCIKM,qi2024attributionrag}. Evaluations should test this distinction; benchmarks containing only clean, authoritative incident descriptions are likely to overstate safe autonomy \cite{Jha2025ITBench,yao2025taubench}.

\subsection{Datasets and benchmarks for agentic AIOps}
Progress is limited by proprietary incident corpora, heterogeneous telemetry infrastructures, and weak or ambiguous ground truth \cite{Li2022AIOpsDatasets,notaro2021mgt}. Public datasets support reproducibility, but agentic evaluation requires elements they often omit: labelled incidents or adjudicated causal structures, realistic tool interfaces, policy constraints, and operationally meaningful success criteria \cite{Jha2025ITBench,yao2025taubench,autio2024genai}.

\paragraph{Logs: shared corpora and realism}
LogHub makes evaluation repeatable, but that does not guarantee realism \cite{Zhu2023Loghub}. In real systems, log anomaly detection encounters missing events, changing templates, and unclear labels \cite{Zhao2021PracticalLogAD}. Practitioner studies suggest benchmarks should reward robustness along with actionable outcomes, not accuracy alone \cite{Ma2025PractitionerExpectationsLogAD}. For agentic AIOps, benchmarks should cover drift, partial observability, noisy metadata, incorrect tickets, and outdated runbooks \cite{Zhao2021PracticalLogAD,Ma2025PractitionerExpectationsLogAD,yao2025taubench}.

\paragraph{Traces and multi-modality}
Trace-based RCA benchmarks expose causal structure more directly. TraceRCA localises microservice root causes using traces and related signals \cite{Li2021TraceRCAIWQoS}. TADBench targets trace anomaly detection. This matters when agents must decide whether an anomaly requires escalation or mitigation \cite{Lin2025TADBenchTSC}. Causal and event-graph RCA provide richer ground truth. They show not only which service failed but also which event chain mattered and which tests ruled out alternatives \cite{Ikram2022CausalDiscovery,ChainOfEvent2024FSE,Zhu2024HeMiRCA,Wang2024MRCA}.

\paragraph{Agentic benchmark requirements}
A practical benchmark acts as an environment:
\begin{equation}
\mathcal{B} = (\mathcal{D}, \mathcal{T}, \Pi, \mathcal{A}, \mathcal{M}),
\end{equation}
where \(\mathcal{D}\) provides multi-modal observations, \(\mathcal{T}\) exposes tool calls, \(\Pi\) encodes policy and approvals, \(\mathcal{A}\) defines the permitted action envelope, and \(\mathcal{M}\) defines metrics including accuracy, time-to-diagnosis, cost, and safety violations. Without \(\mathcal{T}\) and \(\Pi\), evaluation reduces to story quality \cite{Jha2025ITBench,yao2025taubench,liu2023agent,li2023apibank}.

Recent IT automation benchmarks use real tasks and interfaces rather than static text alone \cite{Jha2025ITBench,yao2025taubench}. AIOps evaluation still lacks several elements: tool-call traces, intermediate hypotheses, action-outcome simulation, and penalties for unsafe actions even when the final diagnosis is correct \cite{Li2022AIOpsDatasets,autio2024genai,phiri2025auditable}. As argued in Section~\ref{sec:llm-foundations}, safe autonomy can be evaluated only when tools, gates, and action envelopes form part of the benchmark \cite{autio2024genai,Jha2025ITBench,yao2025taubench}.

Three recent benchmarks illustrate this progression. OpenRCA evaluates long-context, multi-modal diagnosis over enterprise telemetry \cite{Xu2025OpenRCA}. AIOpsLab deploys cloud environments, injects faults, exposes telemetry and shell interfaces, and evaluates detection, localisation, RCA, and mitigation \cite{Chen2025AIOpsLab}. TeleEval-OS covers 13 telecommunications operations-scheduling subtasks spanning ticket creation, resolution, closure, and operational assessment \cite{Wang2025TeleEvalOS}. Their task units and interfaces differ and do not support a single leaderboard. They are better compared by task level, tool surface, available evidence, action envelope, and recovery obligation.

\begin{table*}[!h]
\centering
\scriptsize
\setlength{\tabcolsep}{2.5pt}
\renewcommand{\arraystretch}{1.08}
\caption{Representative evidence and model-level attribution mapped to the review schema. Panel A distinguishes operational LLM systems, enabling analytical frameworks, and benchmark components; Panel B reports backbone/scale, prompting or agent strategy, retrieval, adaptation, and tools without attributing compound-system results to the model alone \cite{Chen2020IncidentManagement,notaro2021mgt,Jha2025ITBench}}
\label{tab:systems}
\begin{tabularx}{\textwidth}{@{}
>{\RaggedRight\arraybackslash}p{0.105\textwidth}
>{\RaggedRight\arraybackslash}p{0.105\textwidth}
>{\RaggedRight\arraybackslash}p{0.105\textwidth}
>{\RaggedRight\arraybackslash}p{0.135\textwidth}
>{\RaggedRight\arraybackslash}p{0.125\textwidth}
>{\RaggedRight\arraybackslash}p{0.135\textwidth}
>{\RaggedRight\arraybackslash}X
@{}}
\toprule
System & Domain / task & Evidence object / autonomy & Tool surface & Evaluation type & Safety controls & Core survey use \\
\midrule

DeepLog \cite{Du2017DeepLog,Zhao2021PracticalLogAD} &
AIOps / log anomaly detection &
Enabling analytical framework / read-only component &
log streams, parsers, anomaly scores &
offline classification; log-sequence prediction &
thresholding, drift checks, operator review &
Establishes a baseline for log evidence extraction. Realism depends on parser quality, drift handling, and whether alerts support later diagnosis or mitigation \cite{Zhao2021PracticalLogAD,Ma2025PractitionerExpectationsLogAD,He2019LogParsingToolsBenchmarks}. \\

\addlinespace
Xpert \cite{xpert2024,Chen2020IncidentManagement} &
AIOps / triage and diagnosis &
Operational LLM system / tool-grounded analyst &
telemetry queries, DSLs, incident context &
tool-use workflow; query recommendation &
bounded tool budget, schema-constrained queries, feedback loops &
Shifts evaluation from answer generation to evidence acquisition. Quality depends on the DSL, retrieved evidence, and reproducibility of the query trail \cite{xpert2024,yao2025taubench}. \\

\addlinespace
RCACopilot \cite{chen2024autosys,qi2024attributionrag} &
AIOps / RCA &
Operational LLM system / tool-driven analyst &
diagnostic data collection, logs, traces, category prediction &
workflow RCA; explanation quality &
evidence linking, provenance, auditability &
Shows that RCA should be evaluated as a workflow, not a single answer. Explanations are useful only when grounded in collected evidence \cite{chen2024autosys,phiri2025auditable}. \\

\addlinespace
RCAgent \cite{Wang2024RCAgentCIKM,Jha2025ITBench} &
AIOps / RCA agents &
Operational LLM system / tool-augmented analyst or planner &
multi-step tools, telemetry, hypothesis refinement &
agent trajectory; task completion; tool use &
termination rules, tool budgets, unsafe-action penalties &
Shows the need to score trajectories, not only final diagnoses. Evaluation should include safety requirements, cost accounting, and failure handling under partial observability \cite{Wang2024RCAgentCIKM,yao2025taubench,autio2024genai}. \\

\addlinespace
TraceRCA \cite{Li2021TraceRCAIWQoS,Zhu2024HeMiRCA} &
AIOps / trace-based RCA &
Benchmark and analytical framework / evidence construction &
distributed traces, causal or temporal paths &
offline localisation; benchmark comparison &
causal consistency, supported edges, uncertainty checks &
Supports causal structure and localisation. Higher autonomy evaluation still needs agent traces, counterfactual tests, and action outcomes \cite{Ikram2022CausalDiscovery,ChainOfEvent2024FSE}. \\

\addlinespace
TADBench \cite{Lin2025TADBenchTSC,Li2022Observability} &
AIOps / trace anomaly detection &
Benchmark / escalation evidence &
traces, anomaly scores, service dependencies &
benchmark suite; anomaly detection under drift &
drift testing, escalation thresholds, partial-observability checks &
Helps assess when anomalies justify further investigation. Benchmarks should include drift, partial observability, and penalties for unsafe or unsupported escalation \cite{Lin2025TADBenchTSC,yao2025taubench}. \\

\bottomrule

\end{tabularx}
\end{table*}

\begin{table*}[!h]
\ContinuedFloat
\centering
\scriptsize
\setlength{\tabcolsep}{2.8pt}
\renewcommand{\arraystretch}{1.08}
\caption[]{Table~\ref{tab:systems} (continued), Panel B: model-level synthesis for representative systems. Results belong to the compound system unless component ablations isolate the backbone, prompting, retrieval, adaptation, tool, and control layers}
\begin{tabularx}{\textwidth}{@{}
>{\RaggedRight\arraybackslash}p{0.08\textwidth}
>{\RaggedRight\arraybackslash}p{0.14\textwidth}
>{\RaggedRight\arraybackslash}p{0.18\textwidth}
>{\RaggedRight\arraybackslash}p{0.16\textwidth}
>{\RaggedRight\arraybackslash}p{0.15\textwidth}
>{\RaggedRight\arraybackslash}p{0.13\textwidth}
>{\RaggedRight\arraybackslash}X@{}}
\toprule
System & Backbone / scale & Prompting or agent strategy & Retrieval / grounding & Fine-tuning or adaptation & Tools & Attribution caution \\
\midrule
DeepLog & LSTM sequence model; not an LLM & Next-event prediction over parsed log sequences & Parsed log templates and execution paths & Trained on normal-operation sequences & Log parser and anomaly scorer & Analytical baseline; it isolates log modelling from language-agent orchestration. \\
Xpert & GPT-3.5 and GPT-4 variants; parameter counts not disclosed & Few-shot in-context KQL generation with post-rectification & Historical incidents and associated KQL records & Prompt-based main system; fine-tuned code models are comparators & KQL generation and validation pipeline & Quality reflects model, demonstrations, retrieval corpus, DSL constraints, and rectification. \\
RCACopilot & GPT-3.5-turbo and GPT-4 with an 8K context in the reported study & Few-shot chain-of-thought category prediction after diagnostic summarisation & Predefined incident handlers collect diagnostic data; similar historical cases supply demonstrations & In-context main system; a fine-tuned GPT-3.5 comparator is reported & Incident handlers and diagnostic collectors & The evaluated effect belongs to the handler--retrieval--prompt--model pipeline, not GPT-4 alone. \\
RCAgent & Locally deployed Vicuna-13B-v1.5-16K & ReAct-style controller, expert agents, JSON repair, and trajectory self-consistency & Tool observations, snapshot-key memory, and RAG over log chunks & No task-specific backbone fine-tuning reported; domain knowledge enters through tools and prompts & Cloud RCA diagnostic tools & Ablations are needed to separate backbone capacity from tool design, context management, and stabilisation. \\
\bottomrule
\end{tabularx}
\end{table*}

\begin{table*}[!h]
\ContinuedFloat
\centering
\scriptsize
\setlength{\tabcolsep}{2.8pt}
\renewcommand{\arraystretch}{1.08}
\caption[]{Table~\ref{tab:systems} (continued), Panel B continued: model-level synthesis for representative systems}
\begin{tabularx}{\textwidth}{@{}
>{\RaggedRight\arraybackslash}p{0.08\textwidth}
>{\RaggedRight\arraybackslash}p{0.14\textwidth}
>{\RaggedRight\arraybackslash}p{0.18\textwidth}
>{\RaggedRight\arraybackslash}p{0.16\textwidth}
>{\RaggedRight\arraybackslash}p{0.15\textwidth}
>{\RaggedRight\arraybackslash}p{0.13\textwidth}
>{\RaggedRight\arraybackslash}X@{}}
\toprule
System & Backbone / scale & Prompting or agent strategy & Retrieval / grounding & Fine-tuning or adaptation & Tools & Attribution caution \\
\midrule
OWL \cite{Guo2024OWL} & LLaMA2-13B base; OWL-13B after adaptation & Operations-specific instruction following with Homogeneous Markov Context Extension & Owl-Instruct and operations-domain corpora; no retrieval layer evaluated & Supervised tuning with a mixture of LoRA adapters across domains and tasks & No live operational tools; evaluated as a domain model & Results on Owl-Bench, log parsing, and anomaly detection establish model adaptation, not operational authority. \\
AIOpsLab \cite{Chen2025AIOpsLab} & GPT-4 and GPT-3.5 shell agents plus ReAct and FLASH variants; parameter counts not disclosed & Interactive agents plan across detection, localisation, RCA, and mitigation levels & Logs, metrics, traces, and live cloud state exposed through the benchmark interface & No benchmark-specific backbone fine-tuning reported & Shell, Kubernetes, telemetry APIs, fault injection, and workload control & Reported behaviour belongs to each agent--interface--scenario combination; the framework also evaluates tool cost and invalid API use. \\
OpenRCA \cite{Xu2025OpenRCA} & Claude 3.5 Sonnet, GPT-4o, Gemini 1.5 Pro, Mistral Large 2, Command R+, and Llama 3.1 & Sampling baselines and a ReAct-style controller--executor RCA agent with code generation & More than 68 GB of logs, metrics, and traces from 335 failures across three systems & No task-specific fine-tuning reported & Python execution over benchmark telemetry & Low complete-case accuracy shows that model choice, telemetry access, generated code, and error recovery jointly determine performance. \\
\bottomrule
\end{tabularx}
\end{table*}


\section{Evaluation and benchmarking}
\label{sec:evaluation}

Many agentic operations papers state their main claims too loosely. The relevant test is correctness under operational constraints, not correctness alone \cite{Zhao2021PracticalLogAD,Sculley2015HiddenTechnicalDebt,Breck2017MLTestScore,Northcutt2021LabelErrors,Ribeiro2020CheckList}. Evaluation should therefore examine the workflow: what evidence was collected, whether the agent stopped appropriately, and whether its actions obeyed policy and rollback rules \cite{Wang2024RCAgentCIKM,Ma2024AgentBoard,Gioacchini2024AgentQuest,Xie2024OSWorld}. Reliable systems leave an auditable trace, behave conservatively under uncertainty, and recover when execution departs from plan \cite{Humble2010ContinuousDelivery,Reitblatt2012Abstractions,Basiri2019ChaosEngineering,Beyer2016SRE}.

\begin{table*}[t]
  \centering
  \footnotesize
  \setlength{\tabcolsep}{3pt}
  \renewcommand{\arraystretch}{1.08}
  \caption{Task taxonomy for agentic NetOps and AIOps, stated at the tool and action level. The final column links each task to the contract components most strongly stressed by the evaluation \cite{Ma2024AgentBoard,Gioacchini2024AgentQuest,Xie2024OSWorld}}
  \label{tab:tasks}
  \begin{tabularx}{\textwidth}{@{}
    >{\RaggedRight\arraybackslash}p{0.13\textwidth}
    >{\RaggedRight\arraybackslash}p{0.13\textwidth}
    >{\RaggedRight\arraybackslash}p{0.17\textwidth}
    >{\RaggedRight\arraybackslash}p{0.27\textwidth}
    >{\RaggedRight\arraybackslash}X
    @{}}
    \toprule
    Task & Inputs & Tools (surface) & Success criteria (operational) & Contract components stressed \\
    \midrule

    Incident triage \& routing &
    alerts, tickets, short context &
    ticketing, CMDB/service map, on-call rota \cite{Chen2020IncidentManagement,LasCasas2024LLexus} &
    correct ownership/team, low time-to-first-meaningful-action, no privacy leaks \cite{Chen2020LinkedIncidents,NCSC2025Integrity} &
    mainly \(\mathcal{R}_k,\mathcal{B}_k\): enough fresh evidence for routing, under time and privacy budgets \\

    \addlinespace
    Evidence acquisition (querying) &
    incident description, symptoms &
    log search, metrics DSL, trace queries \cite{xpert2024,Guo2024CToolEval} &
    discriminating queries early, bounded tool budget, reproducible query trail \cite{xpert2024,Ma2024AgentBoard,Gioacchini2024AgentQuest} &
    mainly \(\mathcal{T}_k,\mathcal{R}_k,\mathcal{B}_k\): permitted queries, useful evidence, bounded tool cost \\

    \addlinespace
    Root cause analysis (RCA) &
    logs/metrics/traces, change history &
    correlation, dependency graph, repo metadata \cite{chen2024autosys,Wang2024RCAgentCIKM,Ikram2022CausalDiscovery,ChainOfEvent2024FSE} &
    correct root cause in top-\(k\), evidence-linked explanation, calibrated uncertainty \cite{chen2024autosys,Wang2024RCAgentCIKM,Guo2017Calibration} &
    mainly \(\mathcal{R}_k,\mathcal{B}_k\): sufficient evidence, calibrated uncertainty, and bounded diagnostic cost \\

    \addlinespace
    Remediation planning &
    runbooks, policy, failure domain info &
    change API wrappers, feature flags, runbook retriever \cite{chen2024autosys,Gioacchini2025AutoPenBench} &
    safe plan as diffs/typed steps, explicit preconditions, rollback trigger defined \cite{chen2024autosys,Gioacchini2025AutoPenBench} &
    mainly \(\mathcal{R}_k,\mathcal{G}_k,\mathcal{U}_k\): evidence-backed plan, policy gates, and rollback-ready steps \\

    \addlinespace
    Net intent-to-config synthesis &
    intent, inventory, topology &
    config generators, compilers, verifiers \cite{fogel2015batfish,Beckett2017Minesweeper,mondal2025ambiguity} &
    correct diff, invariants hold under failures, ambiguity resolved before synthesis \cite{fogel2015batfish,Beckett2017Minesweeper,mondal2025ambiguity} &
    mainly \(\mathcal{T}_k,\mathcal{G}_k,\mathcal{U}_k\): write-capable tooling, invariant gates, and controlled change artefacts \\

    \addlinespace
    Change safety \& rollout control &
    proposed diff, risk context &
    sandbox, canary controller, rollback hooks \cite{Reitblatt2012Abstractions,Alipourfard2019RiskBasedPlanning,Xie2024OSWorld} &
    safe intermediate states, bounded blast radius, automatic rollback on guardrail breach \cite{Reitblatt2012Abstractions,Alipourfard2019RiskBasedPlanning} &
    mainly \(\mathcal{G}_k,\mathcal{U}_k,\mathcal{B}_k\): gate compliance, staged rollout, rollback, blast-radius and latency budgets \\

    \bottomrule
  \end{tabularx}
\end{table*}
Table~\ref{tab:tasks} sets out the terminology by defining the tasks, their inputs, tool surfaces, operational success criteria, and the contract components most strongly stressed by each task. Table~\ref{tab:metrics-glossary} then makes the metrics explicit, and Figure~\ref{fig:eval-report-card} gives a minimal evaluation report card that every agentic operations paper should include.

\subsection{Metrics glossary and a consistent scoring vector}
Operational success is multi-objective and should be represented as a reporting vector rather than collapsed into a single aggregate score \cite{Ma2024AgentBoard,Guo2024CToolEval,Xu2025CRAB}.
A compact proposed template that covers both diagnosis and action is:
\begin{equation}
\begin{aligned}
\mathbf{m}=\big(&\textsf{RCA@}k,\ \textsf{MRR},\ \textsf{Violations},\\
&\textsf{RollbackProfile},\ \textsf{ToolCost},\ \textsf{Latency}\big).
\end{aligned}
\label{eq:score-vector}
\end{equation}

The key is that some dimensions are “hard” constraints (policy violations should be near zero), while others are trade-offs (tool cost versus time-to-signal). The entries are not commensurate: a paper should not compensate for a safety violation by averaging it with higher task accuracy.
For hybrid causal--LLM systems, evaluation should also report where computation is spent: causal-graph construction or update, telemetry querying, retrieval, and LLM calls. This distinction matters because a causal model may amortise diagnostic cost across incidents, while an LLM-heavy design may repeatedly consume context and compute for each investigation. Explainability should likewise be scored separately from answer fluency, for example by checking whether the system returns a causal path, the evidence supporting each edge, and the tests that would falsify the proposed cause.

\paragraph{Metric reporting protocol}
Every metric should identify (i) the autonomy rung and unit of analysis---incident, proposal, rollout, or trace; (ii) the ground truth or adjudication process; (iii) the unit, observation horizon, and direction of improvement; (iv) whether it is a hard constraint or trade-off; (v) aggregation by mean, median, percentile, or macro/micro averaging; (vi) uncertainty through bootstrap or repeated-run confidence intervals where dependence permits; (vii) important strata such as incident severity, change class, model version, and missing-telemetry condition; and (viii) known gaming paths and missing-data handling. Safety-event counts should be reported both as rates and absolute numerators and denominators. Rare events require exposure-based denominators and conservative intervals rather than a claim of zero risk.

\begin{table}[t]
  \centering
  \scriptsize
  \setlength{\tabcolsep}{2pt}
  \renewcommand{\arraystretch}{1.05}
  \caption{Operational metric templates for agentic NetOps/AIOps. Each metric requires a declared rung, ground truth, unit, aggregation rule, direction, and uncertainty analysis; no single metric is sufficient \cite{Guo2017Calibration,Ribeiro2020CheckList,Northcutt2021LabelErrors}}
  \label{tab:metrics-glossary}
  \begin{tabularx}{\linewidth}{@{}
    >{\RaggedRight\arraybackslash}p{0.20\linewidth}
    >{\RaggedRight\arraybackslash}p{0.40\linewidth}
    >{\RaggedRight\arraybackslash}X
    @{}}
    \toprule
    Metric & Definition (what to compute) & Why it matters \\
    \midrule
    \textsf{MTTD} (proxy) &
    Elapsed seconds from a benchmark-defined incident start to the first adjudicated correct triage signal or discriminating query; lower is better &
    Measures time-to-signal, but requires a fixed start event and adjudicated query utility \cite{xpert2024,Chen2020IncidentManagement} \\

    \textsf{MTTR} (proxy) &
    Elapsed seconds to verified mitigation in replay or sandbox, or to an approved plan in a human-gated study; lower is better, with the endpoint named &
    Avoids conflating plan approval, change execution, and restored service \cite{Humble2010ContinuousDelivery,Beyer2016SRE,Reitblatt2012Abstractions} \\

    \textsf{RCA@}$k$ &
    Indicator that any adjudicated causal contributor appears in the top-$k$ list; report a macro average and the multi-cause matching policy; higher is better &
    Reflects shortlist inspection while permitting interacting causes and partial labels \cite{Li2021TraceRCAIWQoS,Wang2024MRCA} \\

    \textsf{MRR}/nDCG &
    Ranking quality against an adjudicated relevance ordering for causal contributors or next-step recommendations, aggregated per incident, with higher values better &
    Requires fixed relevance grades and a candidate-set policy, preventing longer unbounded lists from appearing better \cite{xpert2024,Wang2024RCAgentCIKM,Ma2024AgentBoard} \\

    \textsf{PolicyViolations} &
    Number and exposure-normalised rate of attempted or executed forbidden tools, scopes, or actions; zero is the hard target &
    Attempted and executed violations must be separated and reported explicitly \cite{saltzer1975protection,Guo2024CToolEval,Gioacchini2025AutoPenBench} \\

    \textsf{RollbackProfile} &
    Trigger recall on harmful canaries, false-rollback rate on healthy canaries, completion rate, and trigger-to-recovery latency &
    No scalar direction is valid: successful detection and poor proposals can both increase raw rollback frequency \cite{Reitblatt2012Abstractions,Basiri2019ChaosEngineering} \\

    \textsf{UnnecessaryActions} &
    Actions beyond an adjudicated minimal sufficient plan, or actions whose paired replay or no-action counterfactual shows no benefit, with lower values better &
    Requires a declared counterfactual oracle and fixed action granularity, otherwise the count is not identifiable \cite{Widder2021TrustAutomation,Chen2020IncidentManagement} \\

    \textsf{ToolCost} &
    $\sum_i c(\tau_i)$ over tool calls $\tau_i$ (tokens, API cost, retries) &
    Budgets bind autonomy in practice \cite{li2023apibank,patil2024gorilla,Guo2024CToolEval} \\

    \textsf{Latency} &
    End-to-end wall-clock time (tool time included), reported with percentiles &
    Determines operational usability \cite{Xie2024OSWorld,Kapoor2024OmniACT} \\

    \textsf{Abstention quality} &
    Coverage--risk curve and conditional stop rates under benchmark-labelled sufficient and insufficient evidence &
    Separates safe caution from paralysis; neither universal abstention nor universal action is rewarded \cite{Ma2024AgentBoard,Xu2025CRAB} \\

    \textsf{ECE} &
    $\textsf{ECE}=\sum_b \frac{|B_b|}{n}\big|\textsf{acc}(B_b)-\textsf{conf}(B_b)\big|$, using a declared confidence source, binning rule, and outcome label; lower is better &
    Self-reported verbal confidence is not automatically calibrated; compare it with elicited probabilities or repeated-sample estimates \cite{Guo2017Calibration} \\

    \textsf{Drift robustness} &
    Score degradation under controlled drift: missing telemetry, template shifts, stale runbooks &
    Drift is the default case in operations \cite{Zhao2021PracticalLogAD,Sculley2015HiddenTechnicalDebt,Breck2017MLTestScore,Ribeiro2020CheckList} \\
   \textsf{CausalPathQuality} &
Fraction of adjudicated path edges supported by evidence, plus coverage of required disconfirming tests; higher is better. &
Distinguishes a checkable causal explanation from plausible narrative \cite{Kobayashi2018MiningCausality,Kobayashi2019CausalNetworkLogs,Ikram2022CausalDiscovery}. \\

\textsf{ComputeBreakdown} &
Cost split across telemetry queries, causal-graph construction/update, retrieval, and LLM calls, reported in native units and normalised monetary cost where possible. &
Prevents improvements from hiding a shift to more expensive repeated reasoning \cite{Kobayashi2019CausalNetworkLogs,Liang2022HELM,Guo2024CToolEval}. \\
    \bottomrule
  \end{tabularx}
\end{table}

\paragraph{Metric-specific instantiation and applicability}
The protocol above applies to every row in Table~\ref{tab:metrics-glossary}. MTTD and MTTR use seconds per incident or rollout, with benchmark-defined start and end events, censoring rules, tail percentiles, and uncertainty across independent incidents. RCA@$k$, MRR, nDCG, and CausalPathQuality use an incident-level adjudicated causal set, relevance grades, or evidence-edge oracle. They require macro or stratified aggregation and fixed candidate-set rules to prevent gaming through longer lists. PolicyViolations, RollbackProfile, and UnnecessaryActions use proposal or rollout units. Their oracles are the policy decision, paired harmful and healthy canaries, and an adjudicated minimal sufficient action set or paired no-action replay. Violations remain hard constraints rather than compensable utility terms. ToolCost, Latency, and ComputeBreakdown use trace or incident units and are interpreted as lower-is-better only conditional on matched task success. Abstention quality, ECE, and StopScore require independently labelled outcomes, a declared confidence source, and evidence-sufficiency labels, with coverage and missed-action rates reported separately. Drift robustness uses paired clean and perturbed cases and reports score degradation with incident-level uncertainty. Read-side outcome metrics apply from R1 upward, tool and trace metrics from R2 upward, and write, rollout, and rollback metrics only to R3--R4 studies that exercise the corresponding authority. Where the source does not expose the required oracle or unit, the metric is not estimable and should be reported as such.

A proposed stop-decision summary is:
\begin{equation}
\textsf{StopScore}=\Pr(\text{stop}\mid \text{insufficient evidence})-\Pr(\text{stop}\mid \text{sufficient evidence}).
\label{eq:stopscore}
\end{equation}
Its range is $[-1,1]$, and higher is better only when the benchmark defines evidence-sufficiency labels independently of the agent. It should be reported with both conditional rates, confidence intervals, and sensitivity to the sufficiency adjudication; otherwise, the difference can conceal poor performance in one class. Under those conditions, it penalises premature conclusion while rewarding stopping or escalation when evidence is genuinely insufficient \cite{Wang2024RCAgentCIKM,Ma2024AgentBoard}.

\subsection{A standard evaluation report card (what every paper should report)}
\begin{figure}[t]
\centering
\includegraphics[width=.8\linewidth]{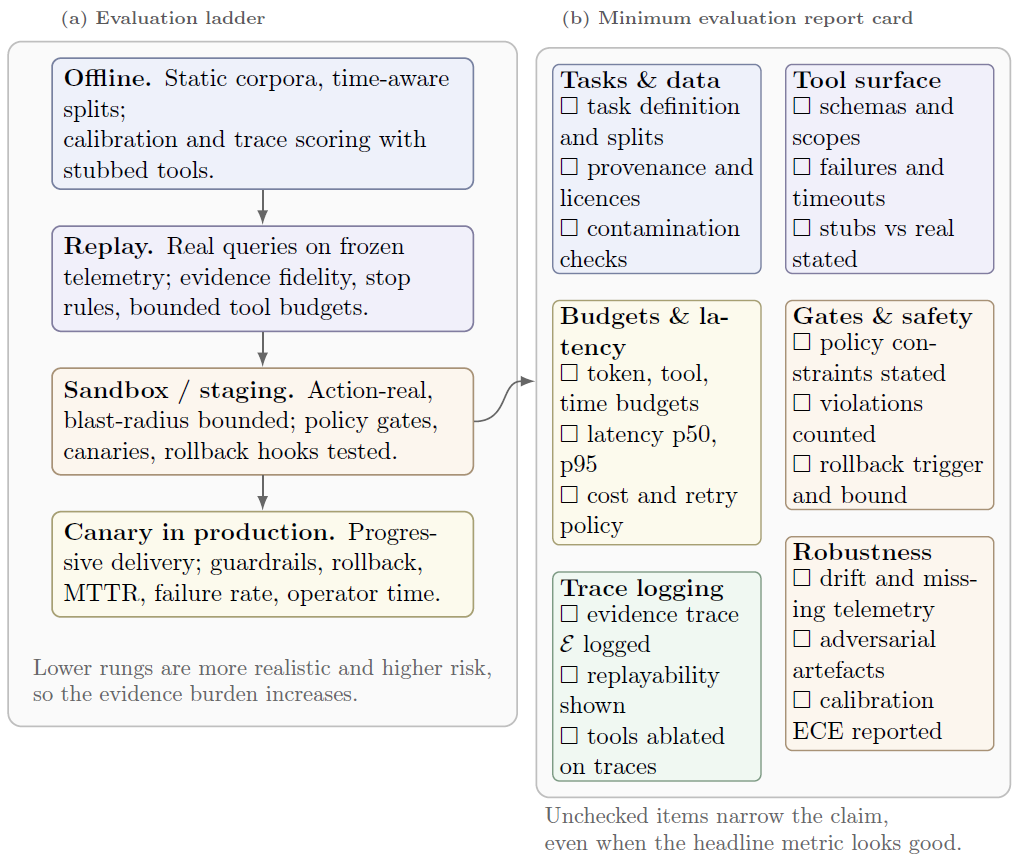}
\caption{Evaluation realism and matched reporting burden. The left side orders study environments from offline corpora to canary-in-production evaluation; the right side lists the minimum tasks/data, tool, budget, gate, trace, and robustness information needed for claims at each level. This figure concerns evidence strength, not agent architecture}

\label{fig:eval-report-card}
\end{figure}

Many results are hard to compare because papers omit the operational interface.
Figure~\ref{fig:eval-report-card} is a minimal report card: datasets and splits, tool surfaces, budgets, gates, trace logging, drift tests, and cost and latency \cite{Ribeiro2020CheckList,Northcutt2021LabelErrors,Ma2024AgentBoard}.
It is intentionally boring, because boring is what lets the field make scientific progress \cite{Breck2017MLTestScore,Sculley2015HiddenTechnicalDebt,Ribeiro2020CheckList}.

\subsection{Worked example: scoring an agent trace (process quality is observable)}
Let an agent produce a trace $E=(\tau_1,\ldots,\tau_n)$ where each $\tau_i$ is a tool call, a retrieved artefact, or a proposed action.
A practical trace score decomposes into: (i) discriminating value, (ii) policy compliance, and (iii) budget use \cite{Ma2024AgentBoard,Guo2024CToolEval,Gioacchini2024AgentQuest}.
One minimal form is:
\begin{equation}
\textsf{TraceScore}(E)
=\sum_{i=1}^{n}\Big(\textsf{Gain}(\tau_i)-\lambda c(\tau_i)-\mu \mathbb{I}\{\tau_i\ \text{violates policy}\}\Big),
\label{eq:tracescore}
\end{equation}
paired with outcome metrics such as RCA@k and the rollback profile. This is a proposed within-benchmark utility, not a validated cross-benchmark score. The benchmark must define and normalise $\textsf{Gain}(\tau_i)$, for example as reduction in oracle-labelled hypothesis entropy or agreement with a gold evidence trace; fix $\lambda$ and $\mu$ before testing; and publish sensitivity analyses. Hard policy violations should also be reported separately and should not become acceptable merely because accumulated gain is large. The useful property of the template is that policy violations and budget overruns are not hidden behind a fluent final paragraph \cite{Ma2024AgentBoard,Gioacchini2025AutoPenBench}.

\subsection{Worked benchmark instantiations: one NetOps task and one AIOps task}
\paragraph{NetOps change and rollback}
Consider 40 repeated sandbox rollouts of an ACL diff against an emulated topology: 20 seeds contain an injected canary regression and 20 are healthy. The benchmark fixes the intended reachability and isolation predicates, the allowed tool set, a maximum of 12 read calls, the canary duration, and a 60-second recovery deadline. If rollback triggers in 18 harmful cases and 2 healthy cases, the rollback profile reports trigger recall (18/20=0.90), false-rollback rate (2/20=0.10), completion among triggered rollbacks, and the distribution of trigger-to-restoration latency. A rollback rate of \(20/40=0.50\) is not meaningful on its own. Proposal validity, verifier rejection, gate-bypass attempts, intermediate-state violations, and final recovery should be reported separately. Rates should include binomial intervals, and latency should be given by median and tail percentiles across seeds.

\paragraph{AIOps diagnosis and stopping}
Consider 100 incident replays from a microservice testbed, with 50 cases containing sufficient evidence for an adjudicated causal set and 50 with one required modality removed. If the agent stops or escalates in 42 insufficient-evidence cases and 6 sufficient-evidence cases, Equation~\eqref{eq:stopscore} gives \(42/50-6/50=0.72\). The two conditional rates are also needed to distinguish appropriate caution from unnecessary abstention. Each replay should report RCA@\(k\) under a multi-cause matching rule, time to first discriminating query, causal-path evidence precision, tool cost, and abstention coverage. Mitigations are tested only in a sandbox. Correct diagnosis does not authorise live execution. Bootstrap intervals are computed by incident, with repeated stochastic runs nested within each incident.
These examples make the templates measurable once the task, oracle, permissions, perturbations, horizon, and aggregation rule are fixed. They define illustrative protocols, not new experimental results.

\subsection{Offline evaluation: datasets, metrics, and protocols}
Offline evaluation is necessary, but it must be designed to avoid over-claiming \cite{Ribeiro2020CheckList,Northcutt2021LabelErrors,Ma2024AgentBoard}.

\paragraph{(i) Leakage and contamination}
Incident corpora contain near-duplicates (recurring alerts, repeated signatures, repeated playbooks).
At minimum, splits should be time-aware and duplicate-aware, with overlap checks reported.
This matters even more for LLM systems because pretraining contamination can silently inflate benchmark performance, and contamination checks are now part of mainstream LLM evaluation practice \cite{Zheng2023Judging,Jimenez2024SWEbench,Northcutt2021LabelErrors}.

\paragraph{(ii) Metrics must match operational intent}
For RCA, top-$k$ accuracy is useful, but ranking metrics (MRR, nDCG) often match engineering practice better.
For action planning, report policy violations, unnecessary actions, and rollback completeness.
For confidence gating and abstention, calibration matters, because wrong-but-confident agents create the worst operational incidents \cite{Guo2017Calibration,Ma2024AgentBoard}.
LM-PACE provides domain-specific evidence for this requirement by estimating confidence for GPT-4 cloud-incident RCA predictions through retrieval-grounded prompting and calibration \cite{Zhang2024LMPACE}. The result should be interpreted within its evaluated incident distribution and predictor family. Expected calibration error is informative only when binning, sample size, confidence definition, and shift conditions are reported, and calibration does not convert an unsupported diagnosis into authority to act.

\paragraph{(iii) Tool interfaces and traces are first-class}
Agentic ops evaluation should not treat tools as an implementation detail.
Benchmarks for tool use and long-horizon control provide useful templates for reporting tool schemas, trajectories, and budgets \cite{zhou2024web,li2023apibank,patil2024gorilla,Guo2024CToolEval,Koh2024VisualWebArena,Kapoor2024OmniACT}.
A pragmatic protocol is to run offline evaluation in a stubbed tool environment with deterministic responses, then score both outcome and trace (Equation~\ref{eq:tracescore}) \cite{Gioacchini2024AgentQuest,Ma2024AgentBoard}.

\paragraph{Robustness under drift and partial observability}
Operational signals drift continuously.
Benchmarking should include stress variants: missing telemetry, delayed telemetry, contradictory signals, and evolving templates \cite{Zhao2021PracticalLogAD,Ribeiro2020CheckList,Northcutt2021LabelErrors}.
If an agent becomes more autonomous under these stressors, it is usually a sign that the evaluation is too forgiving.

\paragraph{LLM-as-a-judge is not enough}
LLM graders can help with structure and readability, but they are not a substitute for tool-grounded correctness and safety scoring.
Use LLM judges as an auxiliary signal, never as the only metric \cite{Zheng2023Judging,Ribeiro2020CheckList}.
\begin{figure}[t]
\centering
\includegraphics[width=0.6\linewidth]{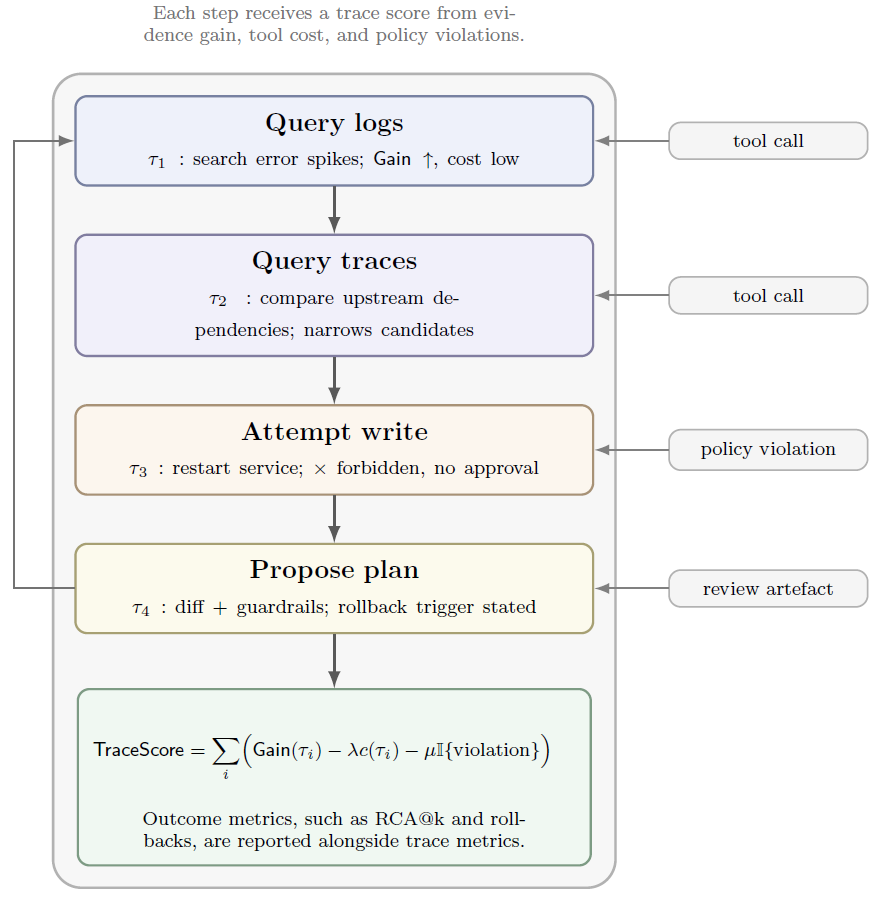}
\caption{Process quality is scorable. A trace is evaluated by discriminating value (did the query narrow hypotheses), policy compliance (did it attempt forbidden actions), and rollout discipline (were guardrails and rollback triggers specified)}
\label{fig:trace-scoring}
\end{figure}

\subsection{Benchmarks and Evaluation Realism}
\label{subsec:benchmark-realism}

The benchmark used for an agentic operations system determines the strength of the claim that can be made about it.
For NetOps and AIOps, a benchmark is not realistic merely because it contains operational vocabulary.
It must expose the agent to the same sequence of constraints that appears in practice: incomplete evidence, bounded tool access, stale or conflicting artefacts, policy checks, rollout constraints, and recovery obligations.
For this reason, benchmark realism should be treated as a ladder rather than as a single property.

At the first level, \emph{static question answering} tests whether a model can answer questions about logs, tickets, incidents, configurations, or operating procedures.
This is useful for measuring background knowledge and local reasoning, but it does not test whether the agent can gather evidence, choose tools, respect permissions, or avoid unsafe action.
A high score at this level therefore supports only a weak operational claim: the model can produce plausible operational text under a fixed prompt.

At the second level, \emph{tool-use trace replay} evaluates the sequence of tool calls made by the agent.
Here the output is not only the final answer, but also the trace: which tools were called, in what order, with which parameters, and whether each call reduced uncertainty.
This level is better matched to agentic NetOps and AIOps because many operational tasks are solved by asking the right next query, not by producing the right sentence.
Trace replay also makes it possible to score tool budgets, redundant calls, missing evidence, provenance, and stopping behaviour \cite{Ma2024AgentBoard,Gioacchini2024AgentQuest,Guo2024CToolEval,yao2025taubench}.

At the third level, \emph{sandbox execution} places the agent inside a controlled operational environment.
The agent may inspect logs, metrics, traces, tickets, configurations, or topology state, while proposed changes are applied only to a simulator, digital twin, replay environment, or validation pipeline.
This level is necessary for testing planner-executor systems because a proposed mitigation may look safe in text but fail under policy, reachability, dependency, or rollout constraints.
For NetOps, this level matches configuration analysis, invariant checking, and safe-update testing. For AIOps, it matches frozen incident replay over logs, traces, and change histories.

At the fourth level, \emph{live testbed evaluation} runs the agent against a controlled but active system. This setting is stronger than sandbox replay because telemetry is produced by a running stack, tool responses may change over time, and the agent must operate under partial observability. Such testbeds are useful for evaluating diagnosis under drift, tool failures, noisy alerts, and ambiguous symptoms. They also support fault injection and chaos-style tests, where the benchmark can measure whether the agent detects, investigates, and contains the fault without widening the blast radius \cite{Xie2024OSWorld,Gioacchini2025AutoPenBench}.

At the highest level, \emph{canary and rollback-aware evaluation} tests whether the agent can support bounded execution. The benchmark must specify the approval path, the canary condition, the guardrail metric, the rollback trigger, and the maximum acceptable response time after a violation. This level is essential once \(\mathcal{T}_k^{\text{execute}}\neq\emptyset\), because action quality cannot be separated from recovery quality. An agent that applies a correct final change, but cannot detect a bad canary or prepare a rollback, has not yet demonstrated safe autonomy.

ITBench is an important step in this direction because it evaluates agents on real-world-style IT automation scenarios rather than isolated question answering. It contains 94 scenarios across SRE, CISO, and FinOps, and its early results report low success rates for current agents, including 13.8\% for SRE tasks and 25.2\% for CISO tasks \cite{Jha2025ITBench}. These results support the central claim of this survey: current agents may look capable in demonstrations, but operational competence requires workflow-level evaluation. The relevant benchmark question is therefore wider than answer correctness. It is whether the agent trace satisfies the contract \(\mathcal{C}_k\): the evidence is sufficient, the tool use is bounded, the gates are respected, the rollout is controlled, and the rollback and audit path is complete.

\subsection{Online evaluation: sandboxes, canaries, and rollback tests}
Online evaluation becomes necessary once an agent proposes actions, because action quality depends on the behaviour of the system being changed \cite{Basiri2019ChaosEngineering,Xie2024OSWorld,Kapoor2024OmniACT}. A useful first stage is incident replay, where telemetry queries are real but actions are simulated. This fits NetOps pre-change verification pipelines \cite{fogel2015batfish,Beckett2017Minesweeper}, and AIOps replay over frozen trace and log stores when the agent also leaves an auditable evidence trail \cite{chen2024autosys,Wang2024RCAgentCIKM,Ma2024AgentBoard}.

A later stage is then a controlled rollout. A change may be correct in its final state yet unsafe while it is being applied, so the agent should stage the change, run a canary, validate the result, and only then widen the rollout \cite{Reitblatt2012Abstractions,Alipourfard2019RiskBasedPlanning,Xie2024OSWorld}.

\paragraph{Guardrail-triggered rollback as an evaluation primitive}
An evaluation that cannot tell whether the agent would have rolled back is not yet measuring safe autonomy.
A stronger protocol specifies the guardrails, the telemetry used to detect breaches, the rollback action, and the allowed response time \cite{Basiri2019ChaosEngineering,Gioacchini2025AutoPenBench}. Rollback-aware learning further indicates that reversibility should be incorporated into the action model \cite{sorstkins2025learning}.

\paragraph{Fault injection and chaos-style tests}
Agents should be evaluated on controlled fault scenarios, because real incidents are rare, costly, and unevenly observed.
Chaos engineering provides a disciplined way to inject failures and test whether detection and mitigation workflows behave as expected \cite{Basiri2019ChaosEngineering}.

\subsection{Benchmarks for RCA and operations}

Against the realism ladder above, existing RCA and operations benchmarks can be read as partial tests of the contract: log and trace datasets mainly stress \(\mathcal{R}_k\), tool-use benchmarks stress \(\mathcal{T}_k\) and \(\mathcal{B}_k\), while sandbox and rollout benchmarks begin to test \(\mathcal{G}_k\) and \(\mathcal{U}_k\).
It shapes the scientific claim itself, as discussed earlier \cite{Ma2024AgentBoard,Gioacchini2024AgentQuest,Xie2024OSWorld}. For AIOps, useful anchors include shared log corpora and trace-based RCA datasets \cite{Zhu2023Loghub,Li2021TraceRCAIWQoS}. In NetOps, configuration analysis and verification form the foundation. This ecosystem allows operators to check network invariants and test changes in a repeatable way, which helps ensure that updates do not introduce unexpected problems \cite{fogel2015batfish,Beckett2017Minesweeper,Xu2023NetCov}. Agentic evaluation introduces two requirements often missing from existing benchmarks. These requirements reflect the practical demands created by increasingly workflow-centred evaluation. First, the tool interface should be clearly defined. Second, it should be able to acquire intermediate supervision information, such as by recovering evidence traces from task setup methods \cite{Ma2024AgentBoard,Guo2024CToolEval,Gioacchini2024AgentQuest,Xu2025CRAB}. When these are absent, a plausible narrative can too easily stand in for a sound investigation.

\subsection{Cost, latency, and operational usability}
In practice, agentic systems often face limitations due to excessive tool invocations, retries, and the use of verbose context prompts. These factors are typically more critical than the performance of any single model invocation. Therefore, evaluations should encompass not only end-to-end latency but also the number of tool invocations, failure rate, token usage, cost estimates, and the time required for operator involvement \cite{Xie2024OSWorld,Kapoor2024OmniACT,Koh2024VisualWebArena}. Usability should be judged through the artefacts operators actually use: diffs, evidence trails, guardrail monitors, and rollback plans \cite{Beyer2016SRE,Humble2010ContinuousDelivery,Ma2024AgentBoard}.

\section{Security, privacy, safety, and governance}
\label{sec:security}

The introduction of agentic NetOps and AIOps systems has altered the risk profile of NetOps and AIOps. Large language models are no longer limited to generating text; they can now interact with the environment in a more direct way. Agents can now coordinate access to privileged tools, propose and even implement changes, and choose which evidence to rely on. This is a stark contrast to their previous, more passive role. This creates a classic confused-deputy problem: an attacker need not compromise the tool if the attacker can influence an authorised agent to misuse it \cite{saltzer1975protection,greshake2023indirect,Zhan2025AdaptiveIPI,Xia2025SafeToolBench}. 
A sound engineering response begins with architecture. This choice fixes how the system will behave in a live setting and sets the conditions for later hardening. In practice, operational artifacts should be treated as potentially adversarial from the outset. The system should enforce least privilege at every tool boundary, and all consequential operations should pass through independent approval gates that the model cannot bypass. As complexity increases, these controls reduce the risk that a plausible instruction becomes an unsafe action \cite{chen2025struq,Chen2025DefensePI,Zhang2025MoEPI}.

\begin{figure}[t]
\centering
\includegraphics[width=0.85\linewidth]{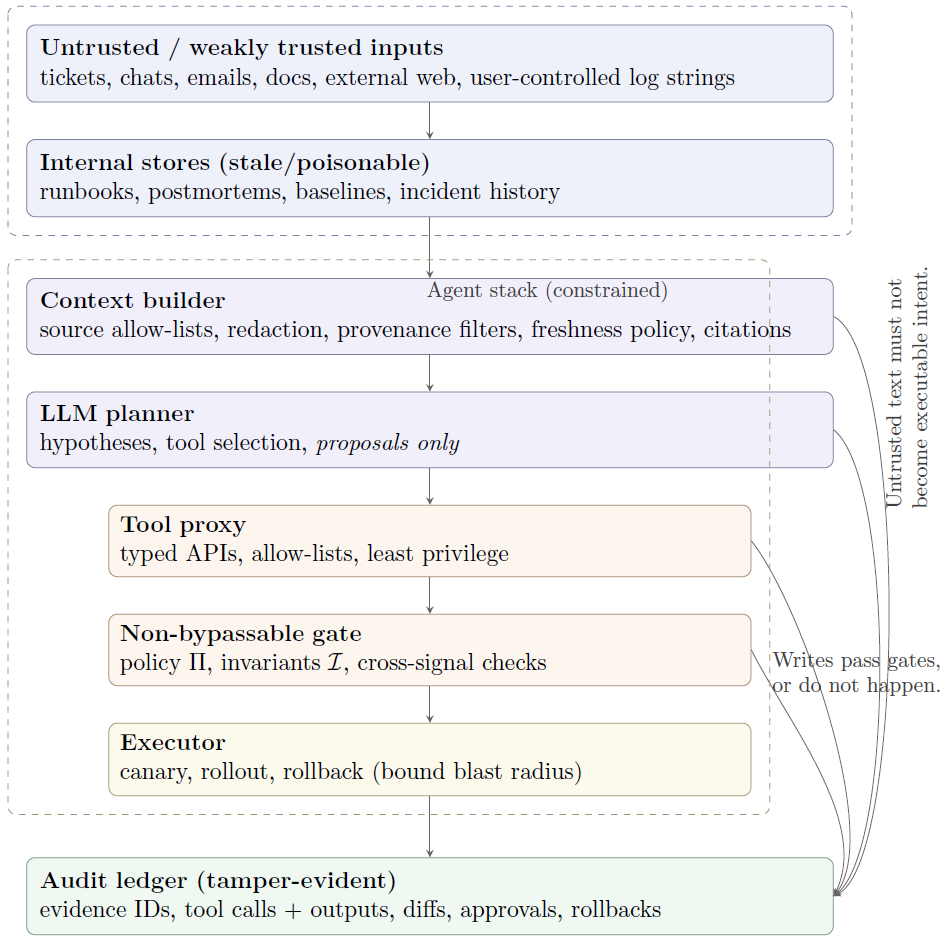}
\caption{Adversarial trust-boundary view. Untrusted artefacts can inject instructions or distort evidence; least-privilege tools, policy and invariant gates, cross-signal integrity checks, and audit trails constrain propagation into action. This figure models attack paths, unlike the operational workflow and architecture figures}
\label{fig:threatmodel}
\end{figure}

Figure~\ref{fig:threatmodel} presents the trust boundaries explicitly. Table~\ref{tab:threat-control} maps threats to controls and measurable tests. Section~\ref{sec:security-eval} gives a concrete evaluation protocol for security claims in agentic operations systems. Security is treated here as a set of enforceable controls and falsifiable tests, rather than as a catalogue of possible hazards \cite{Xia2025SafeToolBench,Zhan2025AdaptiveIPI}.

\subsection{Attacker model, assets, and trust boundaries}
A usable threat model for agentic operations should name assets, attacker capabilities, and the concrete surfaces where those capabilities apply \cite{greshake2023indirect,Xia2025SafeToolBench}.

\paragraph{Assets at risk}
Beyond confidentiality of secrets in logs and tickets, the primary assets are:
(i) \emph{integrity of operational truth} (what is happening in the system),
(ii) \emph{integrity of changes} (what was applied, where, and why),
and (iii) \emph{availability of the workflow} (can on-call staff still diagnose and act).
In NetOps, a wrong config or unsafe rollout can create wide blast-radius outages.
In AIOps, the more common failure is accelerating a bad mitigation, or delaying a correct mitigation through repeated refusal or mis-triage \cite{pasquini2025aioopsdoom,shafran2025machineRAG}.

\paragraph{Attacker capabilities and goals}
We model an attacker by the tuple
\begin{equation}
    \Theta_{\mathrm{adv}}=(\mathcal{C},\mathcal{G},\mathcal{S}),
\end{equation}
where \(\mathcal{C}\) is the set of capabilities, \(\mathcal{G}\) is the set of goals, and \(\mathcal{S}\) is the set of reachable surfaces.
Capabilities commonly include:
(1) writing or influencing text that enters prompts (tickets, chat, wiki pages),
(2) inserting or modifying documents in retrieval stores (poisoning or jamming),
(3) influencing telemetry strings (user-controlled payloads that appear in logs),
and (4) inducing unsafe tool use by shaping the agent’s context.
Goals include unsafe action, denial of workflow (refusal storms), and data exfiltration.
These are now well evidenced in prompt-injection and RAG-security work \cite{greshake2023indirect,zou2025poisonedrag,shafran2025machineRAG,Zhan2025AdaptiveIPI,Chen2025BackdoorPI}.

\paragraph{Trust boundaries in agentic ops}
Operationally, it helps to split the system into four zones:
\begin{enumerate}
  \item \textbf{Untrusted inputs:} tickets, external docs, user-influenced log lines, chat transcripts.
  \item \textbf{Semi-trusted evidence stores:} internal runbooks, postmortems, baselines; high value, but stale or poisonable.
  \item \textbf{Privileged tools:} change APIs, controller interfaces, restart and drain actions, ticketing actions.
  \item \textbf{Verification and governance:} policy-as-code, validators, simulators, approvals, tamper-evident auditing.
\end{enumerate}
The design requirement is simple: zone (4) must be non-bypassable by the model \cite{saltzer1975protection,Xia2025SafeToolBench}.

The scarcity of domain-specific security evaluations is itself material. AIOpsDoom directly tests telemetry manipulation in an operational-agent setting, while much of the remaining evidence is transferred from general tool-integrated agents. For example, Gasmi et al. compare function-calling and Model Context Protocol deployments across 3,250 attack scenarios and show that architectural choice changes the attack surface \cite{pasquini2025aioopsdoom,Gasmi2026AgentSecurity}. This supports explicit testing of both model-facing and system-facing attack paths. It does not supply a NetOps or AIOps production-safety estimate.

\subsection{Threats mapped to controls and measurable tests}
Table~\ref{tab:threat-control} links the most common threat classes to concrete controls and to tests that can be run in evaluation.

\begin{table}[t]
\centering
\footnotesize
\setlength{\tabcolsep}{3pt}
\renewcommand{\arraystretch}{1.08}
\caption{Threat-to-control matrix for agentic operations. Controls must be enforceable at the tool boundary, and they should come with measurable tests (attack success, unsafe tool calls, bypass attempts, and refusal storms)}
\label{tab:threat-control}
\begin{tabularx}{\linewidth}{@{}
  >{\RaggedRight\arraybackslash}p{0.16\linewidth}
  >{\RaggedRight\arraybackslash}X
  >{\RaggedRight\arraybackslash}X
  >{\RaggedRight\arraybackslash}p{0.19\linewidth}
  @{}}
\toprule
Threat class & Typical surface and goal & Primary controls (enforceable) & Measurable tests \\
\midrule

Prompt injection (direct / indirect) &
Injected instructions in tickets, runbooks, or chats that steer tool use or bypass policy &
Instruction--data separation; context-builder sanitisation; tool allow-lists; non-bypassable gates &
Attack success rate; unsafe tool-call rate; bypass attempts blocked \cite{greshake2023indirect,chen2025struq,Zhan2025AdaptiveIPI,Chen2025BackdoorPI,Chen2025DefensePI,Zhang2025MoEPI} \\

RAG poisoning (targeted) &
Corrupt runbook snippets or ``known fix'' documents that induce a wrong change proposal &
Provenance and allow-lists; signed or reviewed documents; retrieval filters by owner and freshness; evidence citation &
Targeted answer-flip rate; source-trust violation rate; retrieval provenance coverage \cite{zou2025poisonedrag} \\

RAG jamming / denial &
Blocker documents induce refusal loops, slow the workflow, or stop incident handling altogether &
Refusal budget; fallback retrieval; document-level risk scoring; safe degraded mode &
Refusal-storm rate; time-to-first-meaningful-action; false-positive refusal rate \cite{shafran2025machineRAG} \\

Telemetry manipulation (integrity) &
Single-channel dishonest signals steer diagnosis and mitigation &
Cross-signal consistency checks; independent measurements; ``trust but verify'' for each claim &
Cross-signal inconsistency detection rate; wrong-mitigation rate under adversarial signals \cite{pasquini2025aioopsdoom} \\

Tool misuse and excessive agency &
Wrong tool, wrong arguments, unsafe sequencing, or broad blast radius &
Least privilege by tool scope; typed tools; staged execution; approvals for writes &
Forbidden-call rate; argument-safety violations; rollback completeness and timeliness \cite{Xia2025SafeToolBench} \\

Data exfiltration &
Prompt or tool output coerces disclosure of secrets or memorised data &
Pre-model redaction; role-based retrieval; output filtering; tight audit on data access &
Secret-leak rate; policy violations on data access; retention and deletion compliance \cite{carlini2021extracting,Song2024SecureSQL,Wang2024Raccoon} \\

\bottomrule
\end{tabularx}
\end{table}

\subsection{Telemetry integrity via cross-signal consistency}
In operations, failures of integrity often matter more than failures of confidentiality. An agent may still act wrongly, even when it is grounded, if the evidence on which it relies is false or incomplete \cite{pasquini2025aioopsdoom}. Telemetry provides evidence, but it does not offer certainty. The state observed by an agent can be skewed due to a partial view \cite{Li2022Observability,Zhao2021PracticalLogAD,Sculley2015HiddenTechnicalDebt,Breck2017MLTestScore}. In adversarial settings, compromised devices may inject misleading logs, counters, or protocol events \cite{pasquini2025aioopsdoom,NCSC2025Integrity}. Agentic systems should therefore attach provenance, coverage, and uncertainty to telemetry before using it for diagnosis or mitigation.

A basic safeguard is to check whether metrics, traces, and logs tell a consistent story. Let \(\phi_m=\Phi_m(o^{\text{metrics}})\), \(\phi_t=\Phi_t(o^{\text{traces}})\), and \(\phi_\ell=\Phi_\ell(o^{\text{logs}})\) denote the corresponding feature views. 
\begin{equation}
\operatorname{Consistent}
= \mathbf{1}\!\left\{d(\phi_m,\phi_t,\phi_\ell)\le \delta\right\}.
\end{equation}

Here, \(d(\cdot)\) may be a maximum normalised discrepancy or a learned disagreement score, and \(\delta\) is fixed on clean validation data before adversarial testing. This proposed check does not establish truth: correlated sensors may agree and still be wrong. It is an integrity trigger for additional confirmation or reduced autonomy when modalities disagree \cite{pasquini2025aioopsdoom}.

\subsection{Non-bypassable policy gates and the propose--commit split}

Every write action should pass through a gate that the model cannot bypass:
\begin{equation}
g(a,E,\Pi)
= \mathbf{1}\!\left\{\Pi(a)=1 \wedge \mathcal{I}(x\oplus a)=1
\wedge \mathsf{Integrity}(E)=1\right\}.
\end{equation}

Here, \(a\) is the proposed action or diff, \(E\) is the evidence trace, \(\Pi\) is the access and change policy, \(\mathcal{I}(x\oplus a)\) checks the candidate system state, and \(\mathsf{Integrity}(E)\) checks evidence provenance and consistency. This distinction is important: state invariants are not functions of an evidence trace. The model may propose a change, but an independently enforced gate decides whether execution is permitted. The gate's assurance remains limited to the scope and soundness of its policy, invariant, and integrity checks.

\subsection{Privacy, retention, and compliance as system properties}

Operational data frequently includes secrets, customer identifiers, architectural information, location traces, traffic patterns, and service-usage metadata. Agentic systems heighten exposure risks, as such data may be incorporated into retrieval contexts, tool outputs, memory, and audit logs \cite{Song2024SecureSQL,Wang2024Raccoon}. Even anonymized telemetry can remain linkable through rare-event patterns \cite{Narayanan2008DeAnonymization}.

Accordingly, privacy controls must be enforced prior to model access. These include data minimization, role- and incident-scoped retrieval, tenant isolation, secret scanning, redaction, retention limits, and auditable access justifications \cite{nist2023airmf}. Post-generation filtering is inadequate because the model has already processed sensitive information \cite{Song2024SecureSQL}. As models may reproduce memorized content, operational agents should, by default, avoid persistent memory and treat each incident as a distinct retention boundary \cite{carlini2021extracting,Song2024SecureSQL,Wang2024Raccoon}.

\subsection{Governance as artefacts: approvals and audit integrity}
Governance is most effective when expressed as enforceable artefacts within the operating process, rather than policy text alone. Policy checks, approvals, and audit records should identify the on-call operator, the authority for each affected domain, mandatory escalation points, and post-incident responsibility. Human approval may fail through fatigue, automation bias, misplaced trust, time pressure, or unclear ownership \cite{Bainbridge1983Ironies,Parasuraman1997HumansAutomation,Lee2004TrustAutomation,Widder2021TrustAutomation}. Higher-autonomy systems should therefore expose the approval path, the evidence and uncertainty under review, and avoid repeated low-information requests that promote routine approval.

\paragraph{Human performance across autonomy rungs}
At the copilot and analyst rungs, evaluation should determine whether the agent reduces search effort without narrowing attention too early. Junior operators may benefit from procedural guidance but be less able to detect plausible errors. Experienced operators may detect subtler errors, although verbose explanations increase interruption cost. At the planner--executor rung, approval depends on whether \(\mathcal{R}\) exposes change scope, failed checks, uncertainty, canary conditions, and rollback triggers under realistic time pressure. Closed-loop operation adds review fatigue, loss of situation awareness, and delayed intervention after long periods of correct operation.

\paragraph{Deskilling, handover, and responsibility}
Longitudinal studies should examine whether routine delegation weakens diagnostic skill, knowledge of system dependencies, or the ability to operate during agent failure. Periodic manual exercises, replay-based training, and audits requiring reconstruction of causal and change rationale may limit this effect. Handover quality should be judged by whether another operator or team can recover the current hypothesis, unresolved uncertainty, actions taken, ownership boundary, and safe next step. Cross-team incidents also require an authority matrix because networking, security, application, and infrastructure teams may control different tools and policies. Acceptance, modification, rejection, override, and emergency bypass should be logged separately. Responsibility remains with the deployment owner, including the authority granted, gate design, monitoring, and recovery.

\paragraph{Status of the human-factors evidence}
The update identifies human evaluation and operator confirmation, but no study whose primary contribution is longitudinal human-factors evaluation of an operational LLM agent. Incident-owner assessments in large cloud RCA studies examine perceived correctness or readability, while LUMI requires operator confirmation of interpreted intent \cite{Ahmed2023CloudIncidents,Zhang2024ICLRCA,Goel2024XLifecycle,Jacobs2025TrustIBN}. These observations do not measure automation bias, review fatigue, deskilling, handover quality, or accountability over sustained use. The human-factors agenda therefore remains weakly supported by direct evidence.


\paragraph{Audit trails that support forensics}
For agentic ops, the minimum audit record includes: policy version, retrieved evidence identifiers and provenance, tool calls with arguments and outputs, proposed and executed diffs, and verification results that permitted action.
If audit logs are not integrity-protected, post-incident accountability collapses.
Tamper-evident logging is a natural fit for this requirement \cite{crosby2009tamper}.

\subsection{Security evaluation protocols}
\label{sec:security-eval}
Security claims should be evaluated with explicit attacker inputs and explicit tool surfaces, not only with benign accuracy tests.
A practical protocol is:
(i) define the permitted tool set and policy \(\Pi\),
(ii) define attacker surfaces \(\mathcal{S}\) (tickets, docs, telemetry strings),
(iii) run injected and poisoned variants alongside clean variants,
and (iv) report both task outcomes and security outcomes \cite{Wang2024Raccoon,Zhan2025AdaptiveIPI,Xia2025SafeToolBench,Song2024SecureSQL}.

We recommend reporting at least:
attack success rate, unsafe tool-call rate, bypass attempts blocked, refusal-storm rate (false positives that halt the workflow), and cost/latency impact under attack \cite{shafran2025machineRAG,Wang2024Raccoon,Xia2025SafeToolBench}.


\section{Open problems and research agenda}
\label{sec:open-problems}

The field of agentic NetOps and AIOps sits at an awkward intersection.
On one side, the tool surfaces (telemetry queries, configuration validators, deployment APIs, ticketing systems) are mature and operationally meaningful.
On the other side, LLM agents are still difficult to bound, hard to evaluate under realistic drift, and easy to mislead when they treat untrusted text as instruction \cite{Zhan2024InjecAgent,Wang2025AgentVigil,Jia2025TaskShield}.
This section focuses on unresolved problems and transforms them into contract satisfaction questions. 

\paragraph{Priority and test design}
The agenda follows the dependencies in the available evidence. \emph{Immediate priorities} are benchmark design and reporting. Existing systems can already support better evidence if studies declare the autonomy rung, expose tool interfaces and traces, report model configuration, use matched clean and adversarial cases, and distinguish hard safety constraints from performance trade-offs. Baselines should include a non-agentic method, an LLM without tools, and, where applicable, the same agent without the control under study. \emph{Medium-term priorities} are architectural: compositional safety, calibrated stopping, causal--LLM interfaces, non-bypassable policy enforcement, and verification of write-side actions. \emph{Long-term priorities} concern sustained operation, including cross-team accountability, retention of operator skill, interoperable audit records, calibration under drift, recovery from correlated failures, and governance of repeated autonomous action.

Each study should specify the NetOps or AIOps domain, operational environment or dataset, baselines, unit of analysis, autonomy rung, and measurable success and failure criteria. It should pose a testable question and, where meaningful, compare a non-agentic method, an LLM without tools, and the same agent without the control under study. The main empirical question is whether reported benefits and safeguards survive system change, partial observability, tool failure, and adversarial input.

\subsection{Problem framing: autonomy levels and operational assurance contracts}

A recurring mistake is to treat autonomy as binary. In practice, operations teams already use graded autonomy through access control, approval procedures, change windows, and progressive rollout. A deployable research agenda should therefore treat autonomy as an explicit variable and evaluate systems at several rungs. General agent frameworks distinguish reasoning from action, but in operational settings this separation must be enforceable and auditable, rather than stated only in prompts \cite{yao2023react,schick2023toolformer,Zhan2024InjecAgent,Jia2025TaskShield}.

This survey expresses this requirement through the assurance contract \(\mathcal{C}_k\) introduced in Section~\ref{sec:agentic-patterns}. Future work should specify the autonomy rung, permitted tool surface \(\mathcal{T}_k\), required evidence \(\mathcal{R}_k\), gates \(\mathcal{G}_k\), rollout and audit protocol \(\mathcal{U}_k\), and budget constraints \(\mathcal{B}_k\). The central question is therefore whether an agent's execution trace satisfies this contract under drift, uncertainty, and adversarial input. Deployable autonomy can then be assessed directly through missing evidence, budget overruns, gate violations, unsafe actions, and incomplete audit trails \cite{Zeng2025AIRepr,Kim2025BenchmarkProfiling}.

\subsection{Verification, guarantees, and compositional safety across layers and time}
The central technical challenge for safe autonomy is that operations properties are layered and time-dependent.
A NetOps change can satisfy final-state reachability and still create transient loops or blackholes during rollout. An AIOps mitigation may reduce the error rate while still violating tail-latency SLOs and triggering cascading retries. The research problem is therefore not only verification, but \emph{compositional safety} across layers and across the rollout timeline.

\paragraph{Network-level invariants remain the clearest foothold}
NetOps has a strong tradition of proactive verification and safe-update reasoning. Batfish and Minesweeper show how real configurations can be analysed against protocol semantics and invariants before deployment \cite{fogel2015batfish,Beckett2017Minesweeper}. Complementary systems emphasise fast online detection of policy violations, together with scalable reasoning about forwarding behaviour and network changes \cite{Khurshid2013VeriFlow,Kazemian2012HSA,Kazemian2013NetPlumber}. Work on consistent updates formalises the central risk: correct end states can still be reached through unsafe transient states, so deployment must be treated as a protocol \cite{Reitblatt2012Abstractions}.

\paragraph{Service-level safety needs testable operational semantics}
For AIOps, safety targets are naturally expressed as SLO constraints, error budgets, and blast-radius limits. They often require monitored rollouts and realistic failure scenarios. Large-scale systems experience shows that production failure modes are subtle, and validation must include representative workloads and fault patterns \cite{Veeraraghavan2016Kraken,sun2022automatic}. Benchmark suites such as DeathStarBench provide reproducible microservice workloads that can support end-to-end evaluation of diagnosis and mitigation policies under stress \cite{Gan2019DeathStarBench}.

\paragraph{Compositional safety in a shared invariant interface}
To avoid re-stating definitions already introduced in the assurance sections, the research agenda reuses the same state-invariant and rollout-safety semantics. The compositional target combines network-state constraints, service-level constraints, and rollout safety across intermediate states. A study should therefore report pre-change verification, rollout monitoring, and how often each layer catches a violation that the others miss \cite{Reitblatt2012Abstractions,fogel2015batfish}.

\subsection{Benchmark realism and reproducibility for contract-based evaluation}

Agentic evaluation becomes weak when reduced to narrative scoring. Benchmarks should record tool use, intermediate hypotheses, budgets, and safety constraints. Without such records, disciplined investigation cannot be distinguished from plausible guessing. General agent benchmarks capture multi-step interaction and tool use, but operational settings require stricter safety and audit criteria \cite{liu2023agent,zhou2024web,qin2023tool,Hu2025CompileAgent}. SWE-bench makes a related point: evaluation changes when tasks involve real artefacts and operational constraints rather than isolated question answering \cite{Jimenez2024SWEbench,Hu2025CompileAgent}.

A contract-based benchmark can score each instance by whether \(\tau \models \mathcal{C}_k\), with partial credit for satisfied clauses. This is consistent with established evaluation practice, where robustness, calibration, and efficiency matter in high-stakes deployment \cite{Liang2022HELM,Zeng2025AIRepr,Kim2025BenchmarkProfiling}.

\subsection{Hybrid causal--LLM operations}
\label{subsec:hybrid-causal-llm}

An important open problem is how to combine causal diagnostic models with LLM agents without weakening either. Causal models provide structured explanations, reduce the hypothesis space, and lower repeated diagnostic cost once a dependency graph is available \cite{Yan2012GRCA,Kobayashi2018MiningCausality,Kobayashi2019CausalNetworkLogs,Li2022CIRCA}. LLM agents can coordinate tools, interpret operator intent, summarise evidence, and draft mitigations for review \cite{xpert2024,chen2024autosys,Wang2024RCAgentCIKM,Jha2025ITBench,yao2025taubench}. A useful division of labour is therefore to let causal inference narrow and explain the fault space, while the LLM gathers evidence, interacts with operators, and prepares bounded mitigation proposals \cite{Kobayashi2019CausalNetworkLogs,Li2022CIRCA,Wang2024RCAgentCIKM}.

Several questions remain. Causal graphs must remain valid under topology, workload, and service drift \cite{Zhao2021PracticalLogAD,Sculley2015HiddenTechnicalDebt,Breck2017MLTestScore}. Agents also need criteria for deciding when causal evidence is sufficient to support a mitigation proposal \cite{Kobayashi2018MiningCausality,Kobayashi2019CausalNetworkLogs,Li2022CIRCA}. Evaluation should therefore extend beyond root-cause ranking to the causal path, supporting evidence, and cost of reaching the diagnosis \cite{Yan2012GRCA,Kobayashi2018MiningCausality,Kobayashi2019CausalNetworkLogs,Liang2022HELM,Guo2024CToolEval,Jha2025ITBench,yao2025taubench}.

\subsection{Security and integrity as contract clauses, not add-ons}

The main security risk in agentic operations is not privacy alone, but confused-deputy behaviour. Untrusted tickets, runbooks, or telemetry may influence an agent with access to privileged tools. Prompt injection and tool hijacking have proved effective in tool-integrated systems, while prompt-only defences remain fragile \cite{Liu2023HouYi,Liu2024PromptInjectionBenchmark,Deng2024MasterKey,Zhan2024InjecAgent,Wang2025AgentVigil}. Evidence poisoning poses a related threat to retrieval and telemetry pipelines. Provenance, allow-lists, and integrity checks must therefore form part of the control plane \cite{Carlini2024Poisoning}.

Structured query interfaces and strict separation of untrusted input are useful safeguards, although difficult to enforce across realistic tool chains \cite{chen2025struq,Jia2025TaskShield,Wen2025InstructionDetection}. In the contract model, these controls belong in \(\mathcal{G}_k\): non-bypassable gates, provenance checks, audit integrity, and explicit bounds on unsafe tool calls, bypass attempts, and excessive blocking.

\subsection{Human factors, accountability, and organisational fit}

Agentic operations will fail in practice if they ignore how incident response actually works. On-call work is collaborative, interrupt-driven, and full of handoffs. This makes artefacts central: evidence bundles, proposed diffs, checks, ownership records, and rollback triggers. Empirical studies show that coordination and information foraging dominate real incident work \cite{Sillito2020FailuresFixes}, while trust in automation remains contextual and can be weakened by inconsistent or opaque behaviour \cite{Widder2021TrustAutomation}.
In contract terms, the question is which \(\mathcal{G}_k\) artefacts support fast, correct human intervention and post-incident learning, rather than merely producing persuasive prose \cite{Zeng2025AIRepr}.

\subsection{Standardisation and minimal interoperable schemas}
Interoperability is an enabling constraint.
Without shared schemas for tool calls, evidence pointers, diffs, approvals, rollbacks, and audit logs, systems remain bespoke and results remain hard to compare.

There is a practical opportunity to align agent evidence collection with OpenTelemetry as a common substrate for logs, metrics, and traces \cite{OpenTelemetrySpec}.
For networking management, intent and telemetry standards and typed management APIs (such as gNMI) provide scaffolding for safe, auditable tool calls \cite{RFC9315Intent,RFC9232Telemetry,gNMISpec}.
These do not solve agent safety, but they reduce accidental diversity so that traces can be replayed and compared \cite{MCP2025Spec}.

\paragraph{Minimal schema set}
A single universal standard is unlikely in the near term. A more practical route is a small set of interoperable schemas that preserve local system design while making traces checkable, replayable, and auditable.
\begin{itemize}
  \item \textbf{Tool call:} tool name, version, typed arguments, timestamp, result pointer, and error code.
  \item \textbf{Evidence pointer:} source ID, provenance, trust tier, freshness timestamp, and citation anchors such as query IDs, offsets, or row IDs.
  \item \textbf{Diff:} structured change representation, including configuration diffs, feature-flag toggles, or deployment patches, with scope and target inventory.
  \item \textbf{Approval:} approver role, decision, and reviewed bundle, including the diff, checks, and rollout plan.
  \item \textbf{Rollback:} triggers, time bounds, rollback actions, and post-rollback verification queries.
  \item \textbf{Audit log:} hash-chained events linking prompts, policies, retrieved evidence, tool calls, diffs, approvals, and outcomes.
\end{itemize}
Together, these schemas provide the minimum structure needed to assess \(\tau\models\mathcal{C}_k\)  and replay traces in reference sandboxes \cite{MCP2025Spec}.

\subsection{Continual operation: drift, learning, governance, and cost}
Agentic NetOps and AIOps systems should be considered as a continuous process, not as a one-time deployment. Models and prompts can age, tool APIs may change, disappear, or acquire new behaviour, the signals underneath the system also drift, and all this sometimes happens quietly. Agent systems should therefore be managed as operational services, with routine evaluation, and where needed, retraining or recalibration. Governance frameworks such as the NIST AI Risk Management Framework provide a useful vocabulary for mapping risks, measuring performance, and monitoring systems over time. This vocabulary fits well with contract-based deployment and phased release practices, since both require the system boundary and its obligations to be made explicit \cite{nist2023airmf}.

General evaluation frameworks that track several dimensions and publish comparable artefacts are a useful starting point.
For NetOps and AIOps, however, the record needs to be more operationally specific. An agent card specifies tool scopes \(\mathcal{T}_k\), contract clauses, approval rules, rollback conditions, and monitors safety performance over time \cite{Liang2022HELM,Kim2025BenchmarkProfiling,Zeng2025AIRepr}. Table~\ref{tab:open-problems} links each research question to a corresponding failure mode and evaluation method. In all instances, empirical evidence is required; assertions of increased autonomy must clearly define permitted actions and demonstrate that the system fulfills these criteria.

\begin{table}[t]
\centering
\footnotesize
\setlength{\tabcolsep}{3pt}
\renewcommand{\arraystretch}{1.08}
\caption{Prioritised research questions for agentic NetOps/AIOps. The surrounding text assigns immediate, medium-term, or long-term horizons; each row pairs a contract-aligned question with a dominant failure mode and a measurable evaluation handle}
\label{tab:open-problems}
\begin{tabularx}{\linewidth}{@{}
  >{\RaggedRight\arraybackslash}p{0.16\linewidth}
  >{\RaggedRight\arraybackslash}X
  >{\RaggedRight\arraybackslash}p{0.18\linewidth}
  >{\RaggedRight\arraybackslash}p{0.24\linewidth}
  @{}}
\toprule
Category & Research question (contract view) & Dominant failure mode & Evaluation handle \\
\midrule

Safe autonomy \& guarantees &
How to choose \((\mathcal{R}_k,\mathcal{G}_k,\mathcal{B}_k)\) so that \(\tau \models \mathcal{C}_k\) is achievable yet non-trivial? &
Refusal storms or over-permissive execution &
\(\Pr[\tau \models \mathcal{C}_k]\), gate-violation rate, refusal rate under missing telemetry \cite{zhou2025meshagent,mondal2025ambiguity,Jia2025TaskShield} \\

Compositional safety &
How to bind \(\mathcal{I}_{\text{net}}, \mathcal{I}_{\text{slo}}, \mathcal{I}_{\text{rollout}}\) into \(\mathcal{U}_k\) and make it non-bypassable? &
Unsafe transient states despite a ``correct'' final state &
Pre-change vs.\ runtime catch rates; false-assurance rate \cite{Reitblatt2012Abstractions} \\

Benchmark realism &
What benchmark artefacts are minimally sufficient to score traces, not only outcomes? &
Offline over-claim; brittle tool use under drift or budgets &
Trace metrics (tool efficiency, stop quality) + outcome metrics under drift variants \cite{liu2023agent,zhou2024web,qin2023tool,Hu2025CompileAgent,Zeng2025AIRepr} \\

Security \& integrity &
How to reduce confused-deputy tool misuse under injection and poisoning attacks? &
Unsafe tool call that appears tool-grounded &
Attack success rate, unsafe tool-call rate, bypass attempts, over-blocking rate \cite{Liu2024PromptInjectionBenchmark,Deng2024MasterKey,Carlini2024Poisoning,chen2025struq,Zhan2024InjecAgent,Wang2025AgentVigil} \\

Human factors &
What review bundles and uncertainty signals improve intervention quality? &
Over-trust or under-trust; poor handover &
Handover success, intervention appropriateness, time-to-key-evidence \cite{Sillito2020FailuresFixes,Widder2021TrustAutomation,Zeng2025AIRepr} \\

Interoperability &
What minimal schema set enables replay, auditing, and comparison across stacks? &
Bespoke systems that cannot be replayed consistently &
Schema coverage, portability across back-ends, replay fidelity \cite{OpenTelemetrySpec,RFC9315Intent,gNMISpec,MCP2025Spec} \\

Continual ops \& cost &
How to maintain safety and calibration as tools, signals, and prompts drift? &
Silent degradation; rising tool cost; confident wrong actions &
Safety trends over time, drift-trigger accuracy, cost/latency per incident \cite{nist2023airmf,Liang2022HELM,Kim2025BenchmarkProfiling} \\

\bottomrule
\end{tabularx}
\end{table}

\section{Conclusion} 
\label{sec:conclusion} 
In NetOps and AIOps, agents are no longer confined to producing operational documentation. They are increasingly involved in operational work itself, where reliability, security, and accountability remain active constraints. In network and cloud operations, this work follows a controlled path from evidence to action. An agent may read artefacts, query tools, form a diagnosis, suggest a mitigation, and, in limited cases, carry it out. These steps need to remain separate. The evidence synthesised here suggests that operational reliability depends critically on the control system surrounding the agent, particularly as authority approaches the write boundary. Safer agent design therefore depends on clear authority boundaries, restricted tool access, and checks that remain visible during ordinary operation.

The evidence synthesis must nevertheless be read at the correct level. The six analytical pillars describe the full LLM-enabled NetOps and AIOps system, while the four evidential statuses constrain the claims that each record may support. Within the 28 directly audited LLM-facing records, only 13 are operational systems and none demonstrates broad closed-loop control. Evidence is strongest for read-oriented assistance and tool-grounded diagnosis and remains substantially thinner for write-capable operation, domain-specific security, and longitudinal human factors. Our qualitative synthesis therefore indicates, rather than statistically estimates, a capability--assurance gap.

Autonomy is best treated as a set of operational commitments rather than a single capability. Read-only support, evidence collection, proposal generation, limited execution, and self-healing expose different tools, risks, and duties of accountability. The sharpest distinction appears at the write boundary. An agent may recommend a configuration change, mitigation, or policy update, but it should not apply that change to a live system without checks that do not depend on the model. Small operational errors can spread. A misplaced rule or premature deployment may widen into service failure.

NetOps already has a tradition of pre-change checking, including reachability, isolation, loop freedom, and update safety. AIOps has a parallel diagnostic requirement: a proposed cause must be tied to dependencies and observed facts. Evaluation should therefore examine the workflow as well as the final answer. It should show what was investigated, which tools were used, what evidence was missing, and whether the causal path can be inspected. Policy violations, latency, rollback behaviour, robustness, and auditability matter because they describe how the system behaves under operational pressure. A sound evaluation path should move from offline, time-aware datasets to replay testing, sandbox testing, and controlled trials in live environments.

The mathematical contracts and several workflow scores in this survey are proposed conceptual or reporting templates, not theorem claims or universal metrics. Their quantities require benchmark-specific task definitions, independent oracles, calibrated thresholds, declared aggregation rules, and uncertainty reporting.

Security and governance follow the same reasoning. Operational artefacts contain local knowledge, but may be stale, sensitive, incomplete, or adversarially influenced. Trust must therefore be part of the workflow from the outset. Prompt injection, poisoned retrieval, tampered telemetry, excessive tool use, and refusal storms all indicate weak control over evidence, access, action, or recovery.

Accountability follows directly. An agent operating near a control surface must leave actions that can be explained after the event. Otherwise, it may remain useful, but it is difficult to govern. This also limits current evidence. Public benchmarks rarely represent the drift, partial observability, stale documentation, tool failure, and adversarial input found in real operations. Claims for closed-loop autonomy in NetOps and AIOps should therefore remain limited. Read-only assistance is already useful, and narrow self-healing may be feasible. Broad self-healing in large, variable systems remains substantially harder.

Human approval does not by itself resolve this problem. Automation bias, review fatigue, rubber-stamping, deskilling, and poor cross-team handover may weaken an otherwise sound approval process. Evaluation should therefore record who accepted, modified, rejected, or overrode a proposal, and whether another operator can safely continue the incident.

This survey therefore adopts a contract-based model of autonomy. At each level, the system should specify permitted tools, required evidence, mandatory checks, and the provisions for rollback and audit. Greater autonomy requires a dependable connection between evidence, permitted action, and safe recovery. Without it, autonomy demonstrated under controlled conditions may fail in operational use.

The conclusion is conditional. Operational assurance does not necessarily increase with operational authority. It must be designed, reported, and tested at each rung, and claims should not exceed the evidence available at that level of authority.

\section*{Acknowledgement}

The authors thank Hamed Haddadi for comments on earlier drafts. AI-assisted tools were used for language editing and structural refinement. The authors remain responsible for the final text.

\bibliographystyle{IEEEtran}

\bibliography{ref}

\end{document}